\documentclass[aps,prb,showpacs,twocolumn,amsmath,amssymb,superscriptaddress,citeautoscript]{revtex4}
\renewcommand{\arraystretch}{1.2}
\usepackage{hyperref}
\usepackage{subfig}
\usepackage{graphicx}
\usepackage{tabularx}
\usepackage{pdflscape}
\usepackage[justification=raggedright]{caption}
\newcolumntype{Y}{>{\centering\arraybackslash}X}
\hypersetup{
    colorlinks,
    citecolor=blue,
    filecolor=blue,
    linkcolor=blue,
    urlcolor=blue
}
\def\vk{{\bf k}}
\def\vr{{\bf r}}

\begin{document}
	
\title{Crystal-field splittings in rare-earth-based hard magnets: An {\it ab initio} approach}
\author{Pascal Delange}
\affiliation{ Centre de Physique Th\'eorique, \'Ecole Polytechnique, CNRS, Universit\'e Paris-Saclay, 91128 Palaiseau Cedex, France }
\author{Silke Biermann}
\affiliation{ Centre de Physique Th\'eorique, \'Ecole Polytechnique, CNRS, Universit\'e Paris-Saclay, 91128 Palaiseau Cedex, France }
\affiliation{Coll\`{e}ge de France, 11 place Marcelin Berthelot, 75005 Paris, France}
\author{Takashi Miyake}
\affiliation{CD-FMat, AIST, Tsukuba 305-8568, Japan}
\author{Leonid Pourovskii}
\affiliation{ Centre de Physique Th\'eorique, \'Ecole Polytechnique, CNRS, Universit\'e Paris-Saclay, 91128 Palaiseau Cedex, France }
\affiliation{Coll\`{e}ge de France, 11 place Marcelin Berthelot, 75005 Paris, France}
\affiliation{Materials Modeling and Development Laboratory, National University of Science and Technology ``MISIS'', 119049 Moscow, Russia}

\date{\today}

\begin{abstract}
We apply the first-principles density functional theory + dynamical mean field theory framework to evaluate the crystal-field splitting on rare-earth sites in hard magnetic intermetallics. An atomic (Hubbard-I) approximation is employed for local correlations on the rare-earth 4$f$ shell and self-consistency in the charge density is implemented. We reduce the density functional theory self-interaction contribution to the crystal-field splitting by properly averaging the 4$f$ charge density before recalculating the one-electron Kohn-Sham potential. Our approach is shown to reproduce the experimental crystal-field splitting in the prototypical rare-earth hard magnet SmCo$_5$. Applying it to $R$Fe$_{12}$ and $R$Fe$_{12}X$ hard magnets ($R=$Nd, Sm and $X=$N, Li), we obtain in particular a large positive value of the crystal-field parameter $A_2^0\langle r^2\rangle$ in NdFe$_{12}$N resulting in a strong out-of-plane anisotropy observed experimentally. The sign of  $A_2^0\langle r^2\rangle$ is predicted to be reversed by substituting N with Li, leading to a strong out-of-plane anisotropy in SmFe$_{12}$Li. We discuss the origin of this strong impact of  N and Li interstitials on the crystal-field splitting on rare-earth sites. 
\end{abstract}


\maketitle

\section{Introduction}
Permanent magnets are a key component of modern electronic devices, ranging from electric motors to medical imaging. 
An important breakthrough in the quest for high-performance permanent magnets was the discovery of rare-earth intermetallic magnets, starting with SmCo$_5$ in 1966\cite{strnat1970}. 
Since its discovery in 1982, the champion of hard magnets has been Nd$_{2}$Fe$_{14}$B\cite{herbst1991}. 
More recently, rare-earth iron-based hard magnets $R$Fe$_{12}X$ with the ThMn$_{12}$ structure such as NdFe$_{12}$N have been under renewed scrutiny\cite{korner2016,suzuki2016,harashima2015,hirayama2015,miyake2014}. 
The underlying reason is the high price and strategical importance of rare-earths and cobalt, and the ongoing research effort to find good permanent magnets with reduced rare-earth concentration\cite{coey2011}.
The ThMn$_{12}$ structure has a reduced ratio of rare-earth vs transition metal compared to Nd$_{2}$Fe$_{14}$B, but nevertheless conserves strong hard magnetic properties (large magnetization and Curie temperature, and strong anisotropy) when doped with light elements such as nitrogen\cite{suzuki2016,harashima2015,hirayama2015}.

The main physical ingredients for a rare-earth hard magnet are the high magnetic anisotropy energy provided by rare-earth ions combined with the high magnetization and Curie temperature from the transition metal sublattice, typically composed of Fe or Co atoms\cite{coey1996,gutfleisch2011,coey2011}.  
The 3$d$ transition metal atoms carry little anisotropy; because of their rather small spin-orbit coupling, their magnetization direction is essentially fixed by that of the rare-earth ion through an exchange coupling.  
The majority of rare-earth elements, especially heavy rare-earth elements, are very expensive. 
Moreover, the magnetic moment of heavy rare-earth is normally anti-parallel to the transition-metal one reducing the net magnetization\cite{coey2011}.
Hence, one advantage of new compounds like $R$Fe$_{12}X$ is a reduced rare-earth concentration. 
In turn, a higher Fe concentration is favorable for achieving a large magnetization, which is another advantage of $R$Fe$_{12}X$ compounds.
However, this reduced rare-earth concentration means each rare-earth ion must contribute a strong magnetic anisotropy to keep the overall magnetic hardness.  
The preferred magnetization direction (in-plane or out-of-plane) of a given rare-earth ion is determined by the interplay between the crystal-field (CF) splitting and spin-orbit (SO) interaction. 
To the first order, the crystalline magnetic anisotropy energy reads:
$$E_A \approx K_1 \textrm{sin}^2\theta$$
where $\theta$ is the angle between the magnetization and the easy axis, and 
\begin{equation}
K_1 = -3J(J-\frac{1}{2})\alpha_J A_2^0 \langle r^2\rangle n_R
\label{CF_to_K}
\end{equation}
where $J$ is the total angular momentum for the rare-earth 4$f$ shell, $n_R$ the concentration of rare-earth atoms, $\alpha_J$ the corresponding Stevens factor and $A_2^0 \langle r^2\rangle$ is the lowest-order crystal-field parameter (CFP). 
Additional small doping of light elements is
also found to strongly modify the anisotropy by affecting the rare-earth CF splitting \cite{matsumoto2016,harashima2016,harashima2015}. 
They also modify the structural stability: doping B makes the Nd$_2$Fe$_{14}$B phase more stable, while interstitial nitrogen has only a minor effect in structural stability.

It follows that the CF splitting on the rare-earth 4$f$ shell  is a crucial quantity defining the magnetic hardness of rare-earth intermetallics. The theoretical search for new rare-earth hard magnets thus requires a reliable approach to calculating CFP.
The importance of crystal-field effects  for the optical, magnetic, and other properties of solids has been recognized long ago, and semi-empirical models of the CF Hamiltonian, such as the point charge model\cite{hutchings1964} and the superposition model\cite{newman1989}, have been developed since the 60's. While they provide an inexpensive and physically transparent description of CF parameters, their predictive power is limited as they require experimental input to determine the actual values.  Experimental information is readily available for large band-gap rare-earth insulators, where the CFP can be extracted from measurements of dipole-forbidden optical transitions between $f$-states\cite{brecher1967}. In the case of rare-earth intermetallics, where the $f-f$ transitions are hidden by the optical response of conduction electrons, inelastic neutron spectroscopy can be used to determine CFP\cite{buschow1974,sankar1975,givord1979,tie1991,tils1999}, but its results are more ambiguous as one needs to sort out the contributions of phonons and the effect of inter-site exchange interactions.

{\it Ab initio} calculations do not rely on experimental input and can have truly predictive power. First-principles techniques for computing the CF parameters\cite{coehoorn1991,daalderop1992,steinbeck1994,novak1994RNi5,steinbeck1996,novak1996,Ning2007,hu2011,novak2013,miyake2014} can be separated into two main approaches. The first one \cite{daalderop1992,steinbeck1994,novak1994RNi5,steinbeck1996,novak1996,Ning2007,miyake2014} consists in extracting the nonspherical Kohn-Sham potential $V_{lm}$ and the 4$f$ charge density $\rho_{4f}$ around the rare-earth site and then computing the corresponding crystal-field parameter.
As the density functional theory (DFT) is not able to fully capture the physics of partially-filled localized 4$f$ shells, one imposes their localization by treating the 4$f$ orbitals as semi-core states. 
The non-spherical 4$f$ charge density $\rho_{4f}(\boldsymbol{r})$ of the rare-earth ion includes an unphysical contribution to the CFP stemming from the local-density-approximation (LDA) self-interaction error. 
This is usually corrected by spherically-averaging the 4$f$ charge density, but then approximations have to be made for the long-range ``tails'' of $\rho_{4f}(\boldsymbol{r})$.

The importance of excluding the self-interaction of the non-spherical part of the partial 4$f$ charge density to obtain proper crystal-field energies was first recognized by Brooks \emph{et al.} in a publication aimed at calculating the spin Hamiltonian parameters of rare-earth compounds\cite{brooks1997density}.

In the second, more recent, approach the 4$f$ states are represented by Wannier functions\cite{hu2011,haverkort2012,novak2013}, while the charge density and, correspondingly, the Kohn-Sham potential are generated by self-consistent DFT calculations with 4$f$ states treated as semi-core. 
An additional {\it ad hoc} parameter is used to correct the charge transfer energy between 4$f$ and conduction bands.

One may also mention recent work on determining the CFP of lanthanides and transition metals using quantum-chemical methods, in particular, in order to understand the properties of magnetic molecules.
Such approaches employ, for instance, the complete active space self-consistant field method\cite{liddle2015improving} or multireference second order perturbation theory\cite{ungur2017ab}.
Here, however, we choose to focus on perfect crystals rather than on molecules.

Overall,  {\it ab initio} calculations of CFP for rare-earth ions are a formidable theoretical problem, due to generally small values of those CFP and their extreme sensitivity to computational details. 
The main weak point of previously proposed DFT-based approaches is that they are not able to correctly treat the localized valence 4$f$ states. 
Hence, the charge density is derived under the drastic approximation of treating them as fully localized core states, spherically-averaged inside the atomic sphere. 
The DFT+U method provides a more realistic treatment for the 4$f$ density in the limit of strong ordered magnetism. 
However, it is usually not able to capture the true quasi-atomic (multiplet) nature of rare-earth shells in the paramagnetic or partially-polarized state. 
The DFT+U calculations can nevertheless be used to estimate the CFP by
converging them to the on-site density matrix corresponding to a given atomic wave function. The CF splitting can be then evaluated from  the difference in DFT+U total energy  between such  calculations for relevant CF states. This method in fact makes use  of the (usually inconvenient) tendency of DFT+U to remain in a local energy minimum instead of converging to the ground state density.
 Zhou \emph{et al}\cite{zhou2011crystal} employed this approach together with an orbital-dependent self-interaction correction\cite{zhou2009obtaining}, to obtain total energies for different orbital occupancies in UO$_2$ and deduce its CFP.

In this work, we propose an approach to {\it ab initio} CFP calculations based on self-consistent DFT+Dynamical Mean-Field Theory (DFT+DMFT)\cite{anisimov97,lichtenstein1998} treating the local many-body problem for the 4$f$ shell in the quasi-atomic (Hubbard-I) approximation. 
While this approach of using DFT+DMFT with the Hubbard I approximation, which we may call DFT+Hub-I, is rather simple and computationally efficient, it was shown to capture not only the 4$f$ multiplet structure in the paramagnetic state\cite{lichtenstein1998,lebegue2005electronic,lebegue2006multiplet,Pourovskii2009,Locht2016} and in the ferromagnetic state\cite{graanas2012charge}, but also the 4$f$--conduction band exchange interaction and the resulting exchange splitting of the Fermi-surface\cite{Pourovskii2009}. 
This scheme also provides a rather natural way of averaging the 4$f$ partial density to reduce the self-interaction error from the CF Hamiltonian.  
We validate it by applying it to the well-known hard magnet SmCo$_5$, for which the crystal-field splitting has been measured in multiple experiments\cite{buschow1974,sankar1975,givord1979,tie1991,tils1999}. 
We then apply our method to much less investigated new hard magnets of the  $R$Fe$_{12}X$ family, computing their CFP for different rare-earth elements (Sm or Nd) and considering N and Li interstitials. 
Our calculations predict the hypothetical  SmFe$_{12}$Li compound to possess a strong axial anisotropy and, possibly, interesting hard-magnetic properties. 

The paper is organized as follows: in Sec.~\ref{sbsec:cf_theory} we introduce basic notions as well as relevant notations of the CF theory. 
Our first-principles computational approach is presented in more details in Sec.~\ref{sbsec:cf_calc}. 
Our results for the DFT+Hub-I electronic structure and CFP for  the $R$Fe$_{12}(X)$ hard magnets  are presented in Secs. \ref{sbsec:res_es} and \ref{sbsec:res_cf}, respectively. 
In Sec.~\ref{sec:discussion} we analyze the shape 4$f$ Wannier functions (WF) in real space and employ a projective approach to evaluate the WF localization and the contribution of hybridization effects to CFP.

\section{Method}
\subsection{Crystal-field parameters: notation and symmetry }\label{sbsec:cf_theory}

We start by introducing crystal-field parameter notations. The local Hamiltonian for a rare-earth  ion with a partially-filled 4$f$ shell subject to the exchange field created by the transition-metal sublattice and to a crystal-field potential reads
\begin{equation}\label{H}
\hat{H}=\hat{H}_{\mathrm{1el}}+\hat{H}_U=\hat{E}_0+\lambda \sum_{i}s_i l_i  +2\mu_B B_{\mathrm{ex}}\hat{S}_a+\hat{H}_{cf}+\hat{H}_U
\end{equation}
where the one-electron part of the Hamiltonian corresponds to the first four terms on the right-hand side, namely, a uniform shift, spin-orbit, exchange-field, and crystal-field terms. 
$\hat{S}_a$ is the in-plane or ouf-of-plane spin operator, corresponding to the case where $B_{\mathrm{ex}}$ is along $x$ or $z$, respectively.
$\hat{H}_U$ represents the electron-electron Coulomb repulsion term of the many-body Hamiltonian. The crystal-field term $\hat{H}_{cf}$ is defined as the non-spherically symmetric part of the one-electron Hamiltonian. The corresponding non-spherical part $V_{ns}(\boldsymbol{r})$ of the one-electron potential can be expanded into spherical harmonics as follows
\begin{equation}\label{V_cf}
V_{ns}(\boldsymbol{r})=\sum_{k=1}^{\infty}\sum_{q=-k}^{k} A_k^q(r) Y_{kq}(\hat{\boldsymbol{r}}).
\end{equation}
where $Y_{kq}(\hat{\boldsymbol{r}})$ is the spherical harmonic function with total angular moment $k$ and projected angular moment $q$. 
The matrix elements of $V_{ns}(\boldsymbol{r})$ between 4$f$ orbitals define $\hat{H}_{cf}$. Due to the properties of the spherical harmonics, only $A_k^q$  for $k \le 2l$, i.e.  $k \le 6$ in the case of  an $f$ shell, can contribute to  $\hat{H}_{cf}$ . In the point-charge CF calculations  $A_k^q(r)$ is  reduced to
$A_k^q r^k$. While we do not assume this form for $A_k^q(r)$ in the present formalism we still employ the now standard notation  $\langle A_k^q(r)\rangle \equiv A_k^q\langle r^k\rangle$. For historic reasons, several conventions exist for the parametrization of $\hat{H}_{cf}$, leading to a rather confusing variety of definitions for the crystal-field parameters. Using the so-called Stevens operator equivalents\cite{stevens1952}, $\hat{H}_{cf}$ is decomposed as follows

\begin{equation}
\hat{H}_{cf}=\sum_{kq} A_k^q\langle r^k\rangle  \Theta_k(J) \hat{O}_k^q 
\end{equation}
where $\hat{O}_k^q$ is the Stevens operator equivalent, $A_k^q\langle r^k\rangle$, as explained above, is the standard notation for the crystal-field parameter for given $k$ and $q$. $\Theta_k(J)$ is the Stevens factor for a given ground state multiplet defined by the quantum number $J$.  
$\Theta_k(J)$ for $k=$2, 4, and 6 are often designated by $\alpha_J$, $\beta_J$, and $\gamma_J$, respectively. 
The Stevens operator equivalents are more convenient for analytical calculations and somewhat outdated, but they are still extensively used in the literature. 
For numerical calculations it is more convenient to express  $\hat{H}_{cf}$  in terms of Wybourne's\cite{wybourne1965} spherical tensor operators $\hat{C}_k^q$:
\begin{equation}\label{H_cf_Stev}
\hat{H}_{cf} = \sum_{kq} B_k^q \hat{C}_k^q
\end{equation}
where $\hat{C}_k^q$ are defined by $$C_k^q(\hat{\boldsymbol{r}}) = \sqrt{4\pi/(2k+1)}Y_{kq}(\hat{\boldsymbol{r}})$$
Moreover, the CFP can be made real by employing the Hermitian combination of Wybourne's operators $\hat{T}_k^q$ defined by 
$$\hat{T}_k^0 = \hat{C}_k^0 \textrm{   and   } \hat{T}_k^{\pm|q|} = \sqrt{\pm1} \left[ \hat{C}_k^{-|q|}  \pm (-1)^{|q|} \hat{C}_k^{|q|} \right]$$
$\hat{H}_{cf}$ can then be expressed as 
\begin{equation}\label{H_cf}
\hat{H}_{cf} = \sum_{kq} L_k^q \hat{T}_k^q
\end{equation}
with a set of real parameters $L_k^q$. $L_k^q$ are linked to the Stevens CFP $A_k^q\langle r^l\rangle$ by a set of positive prefactors $\lambda^q_k = A_k^q\langle r^k\rangle / L_k^q$. For a more extensive discussion of CFP conventions see, e.g., Refs.~\onlinecite{newman1989,mulak2000,newman2007}.

The number of a priori non-zero CF parameters $A_k^q\langle r^k\rangle$ is constrained by the point-group symmetry of a given rare-earth site. In particular, in the presence of inversion symmetry, $V_{ns}(\hat{\boldsymbol{r}})=V_{ns}(-\hat{\boldsymbol{r}})$, only  $A_k^q\langle r^k\rangle$ for even $k$ can be nonzero (cf. Eq.~\ref{V_cf}). Other point-group symmetries further reduce the number of relevant $A_k^q\langle r^k\rangle$.
As a consequence, the crystal-field on Sm 4$f$ in SmCo$_5$ can be fully described with only four CF parameters: $A_2^0\langle r^2\rangle$,  $A_4^0\langle r^4\rangle$, $A_6^0\langle r^6\rangle$, and $A_6^6\langle r^6\rangle$. In the case of the $R$Fe$_{12}X$ family, the relevant parameters are  $A_2^0\langle r^2\rangle$,  $A_4^0\langle r^4\rangle$, $A_4^4\langle r^4\rangle$, $A_6^0\langle r^6\rangle$, and $A_6^4\langle r^6\rangle$ .

In our calculations, we extract the set of parameters $L_k^q$ (or $A_k^q\langle r^k\rangle$), as well as $B_{\mathrm{ex}}$ and $\lambda$ by a least-square fit of {\it ab initio} $\hat{H}_{\mathrm{1el}}$ (using the usual Frobenius norm) obtained within DFT+Hub-I (see Eq.~\ref{H1el_HI} in the next section). Note that one may assign a spin label to the CF parameters in  Eqs. \ref{H_cf_Stev} and \ref{H_cf}, hence, allowing for different CF potentials for spin up and down electrons. We found that this improves the fit for spin-polarized $\hat{H}_{\mathrm{1el}}$. 

\subsection{Calculational approach}\label{sbsec:cf_calc}

We employ the DFT+Hub-I approach\cite{aichhorn2016} based on the TRIQS library\cite{Parcollet2015} and the full potential linearized augmented planewave Wien-2k\cite{Wien2k} band structure code in conjunction with the projective Wannier-orbitals construction\cite{Amadon2008,Aichhorn2009}. The charge-density self-consistency\cite{lechermann2006,Pourovskii2007} is implemented as described in Ref. \onlinecite{Aichhorn2011}.  The Hubbard-I impurity solver is provided by the TRIQS library.

The Wannier orbitals representing the rare-earth 4$f$ states are constructed from the Kohn-Sham bands within the window $[-\omega_{\mathrm{win}},\omega_{\mathrm{win}}] = \left[-2,2\right]$~eV relative to the Fermi level. The choice of the half-window size $\omega_{\mathrm{win}}$ is the only significant parameter in our calculations (indeed, the choice of Hubbard $U$ and Hund's coupling $J$ has limited impact on the results, as we demonstrate in Appendix \ref{app_U_J}). In order to construct a complete orthonormal basis of Wannier orbitals one needs to choose $\omega_{\mathrm{win}}$ large enough to include at least all 4$f$-like Kohn-Sham bands. 
Wannier orbitals constructed with a ``small window''  leak\cite{aichhorn2016} to neighboring sites due to hybridization between 4$f$ states and conduction band states.
A larger window results in more localized Wannier orbitals consisting almost exclusively of the corresponding 4$f$ partial waves inside the rare-earth atomic sphere \cite{Aichhorn2009,aichhorn2016}, as discussed in in Sec.~\ref{sec:discussion} and Appendix~\ref{app_window_size} below. DFT+Hub-I studies of rare-earth wide-gap insulators show a rather strong sensitivity of calculated CFP to the window size; less-localized small window Wannier 4$f$ orbitals result in a better agreement with experimental CFP\cite{pourovskii_unpub}.  In the present case of rare-earth intermetallics we find a rather weak dependence of CFP to variations of $\omega_{\mathrm{win}}$  within the reasonable range from 2 to 8 eV, see Appendix~\ref{app_window_size}. Hence, we employ $\omega_{\mathrm{win}}=$2~eV in our calculations throughout.  

In the Hubbard-I approximation the hybridization function is neglected and solving of the DMFT impurity problem is reduced to the diagonalization of the atomic Hamiltonian (\ref{H}).  The one-electron part $\hat{H}_{\mathrm{1el}}$ of (\ref{H}) is then given by\cite{Pourovskii2007}
\begin{equation}\label{H1el_HI}
\hat{H}_{\mathrm{1el}}=-\mu+\langle H^{ff} \rangle -\Sigma_{\mathrm{DC}}
\end{equation}
where $\mu$ is the chemical potential, $\langle H^{ff} \rangle$ is the Kohn-Sham Hamiltonian projected to the basis of 4$f$ Wannier orbitals and summed over the Brillouin zone, $\Sigma_{\mathrm{DC}}$ is the double counting correction term for which we employ the fully-localized-limit (FLL) form\cite{czyzyk1994} that is known to work best for localized states such as 4$f$ orbitals. 
In our calculations, we evaluate the FLL double-counting using the occupancy of the DMFT local Green's function, which comes out to be close to the nominal 4$f$ occupancy of the corresponding 3+ rare-earth ion. If the nominal occupancy is used in FLL DC  instead one obtains almost the same CFP, with differences no larger than 10 to 20~K.
We carry out DFT+Hub-I iterations until convergence in the total energy with precision $10^{-5}$~Ry is reached and then extract the CFP from Eq.~\ref{H1el_HI} as described in the previous section.

Self-consistent DFT+Hub-I calculations produce a non-spherical one-electron Kohn-Sham potential (\ref{V_cf}), that includes several non-spherical contributions acting on 4$f$ states: the long-range electrostatic (Madelung) interaction, as well as the local-density-approximation (LDA) exchange-correlation potential due to the conduction electrons and 4$f$ states themselves. 
This last ``intra-4$f$ shell'' contribution to the exchange-correlation potential should be removed within DFT+Hub-I, since the on-site interaction $H_U$ between 4$f$ states is already treated explicitly within DMFT.
Hence, the ``intra-4$f$ shell'' contribution in the one-electron part $\hat{H}_{\mathrm{1el}}$ of Eq.~\ref{H} due to LDA is counted twice and should be removed by a double-counting correction.
Moreover, this contribution includes the LDA self-interaction error for localized states directly impacting CFP: for low-lying CF levels, the self-interaction error will be larger than for less occupied excited CF states.  

In order to reduce the self-interaction error in the CFP we enforce uniform occupancy of all states within the 4$f$ ground state multiplet in our self-consistent DFT+Hub-I calculations. 
To that end, we define the imaginary-frequency atomic (Hubbard-I) Green's function at the fermionic Matsubara frequency $\omega_n=(2n+1)\pi T$, where $T$ is the temperature, as follows:
\begin{align}\label{eq_Gat}
\begin{split}
G^{\mathrm{at}}_{ab}(i\omega_n)=\frac{1}{M}\sum_{\gamma \in \mathrm{GSM} \atop \delta \notin \mathrm{GSM}} \biggl( & \frac{\langle \gamma|f_a|\delta\rangle \langle \delta|f_b^\dagger|\gamma\rangle}{i\omega_n-E_{\gamma}+E_{\delta}}  \\
  &+   \frac{\langle \delta|f_a|\gamma\rangle \langle \gamma|f_b^\dagger|\delta\rangle}{i\omega_n+E_{\gamma}-E_{\delta}} \biggr)
 \end{split}
\end{align}
where the eigenstates $|\gamma\rangle$ and $|\delta\rangle$ with eigenenergies $E_{\gamma}$ and $E_{\delta}$ are obtained by diagonalization of Eq.~\ref{H} and belong to the ground-state multiplet (GSM) and excited  multiplets respectively, $a$ and $b$ label 4$f$ orbitals, $M$ is the degeneracy of the GSM. 
In other words, to obtain Eq.~\ref{eq_Gat} we substitute the standard Boltzmann weight $X_{\gamma}=e^{-E_{\gamma}/T}/Z$, where $Z$ is the partition function, with the uniform weight  $\tilde{X}_{\gamma}=1/M$ for the GSM and $\tilde{X}_{\delta}=0$ for exited multiplets in the spectral representation of the Green's function
\footnote{We consider the case of $T$ being much lower than the inter-multiplet splitting, hence, the contribution of excited multiplets into the partition function $Z$ can be neglected}\footnote{An equivalent approach would be to make the one-electron Hamiltonian that serves as an input to the Hubbard I solver spherically symmetric: both approaches should be equivalent and nearly equally simple to implement.}.
In practice, the degeneracy  of the ground state multiplet $M$ is chosen to be the same as for the corresponding free ion, hence, it is given by Hund's rules. Therefore, $M=10$ for Nd and $M=6$ for Sm.
The self-energy thus obtained is then plugged back into the self-consistency cycle.
This leads to a spherically-averaged contribution from the 4$f$ orbitals, both inside and outside the rare-earth atomic sphere, while non-spherical contributions from other states are taken into account. 
We verified the validity of this method by calculating the density matrix from the local Green's function and transforming it to the relativistic basis of one-electron $J=\frac{5}{2}$ and $J=\frac{7}{2}$ orbitals.
With the averaging, the resulting density matrix is made of two identity blocks with deviations of the order of few percent, to be compared with over 50\% without the averaging.

Conceptually speaking, our approach amounts to replacing Eq.~\ref{H1el_HI} by 
\begin{equation}
\begin{aligned}\label{H1el_HI_double_counting}
\hat{H}_{\mathrm{1el}}=-\mu+\langle H^{ff} \rangle - \Sigma_{\mathrm{DC}}  &-  v_{\mathrm{KS}}\left[ \rho_{spd} + \rho_{4f} \right] \\
 & + v_{\mathrm{KS}}\left[ \rho_{spd} + \bar{\rho}_{4f} \right]
\end{aligned}
\end{equation}
where $v_{\mathrm{KS}}\left[ \rho\right]$ is the Kohn-Sham potential ($v_{\mathrm{KS}} = v_{\mathrm{Hartree}} + v_{xc}$ ) evaluated from the total electronic density $\rho$ and then projected to the basis of 4$f$ Wannier orbitals. 
$\rho_{4f}$ designates the projected electronic density belonging to the rare-earth's 4$f$ orbitals, $\bar{\rho}_{4f}$ is the same density, spherically averaged, and $\rho_{spd}$ designates all the remaining density, belonging to all atoms' $s,p$ and $d$ orbitals.

The same approach is used in the spin-polarized DFT+Hub-I calculations: in this case  the exchange splitting is also removed within the GSM. 
We found, however, that this averaging is not sufficient, since the value of the exchange field within our DFT+Hub-I iterations may become larger than the inter-multiplet splitting. 
Hence we also directly remove the 4$f$ spin polarization from the resulting DFT+Hub-I density matrix. For a given $\vk$-point the ``averaged'' density matrix $\tilde{N}^{\vk}$ in the Bloch basis  reads:
\begin{equation}\label{Nk_av}
\tilde{N}^{\vk}=N^{\vk}+\frac{1}{2}P^{\dagger}(\vk)\left(T n^{ff}(\vk) T^{\dagger}- n^{ff}(\vk)\right)P(\vk)
\end{equation}
where $N^{\vk}$ is the density matrix in the Bloch basis calculated as described in Refs.~\onlinecite{Aichhorn2011} and~\onlinecite{aichhorn2016}, $P(\vk)$ is the projector\cite{Aichhorn2009,aichhorn2016} between the Wannier and Bloch spaces, $n^{ff}(\vk)$ is the density matrix in the Wannier basis, $T$ is the time-reversal operator. The averaged density matrix $\tilde{N}^{\vk}$ is then used to recalculate the electron density at the next DFT iteration as described in Ref. \onlinecite{Aichhorn2011}.
The contribution of 4$f$ states to the spin density and local-spin-density-approximation (LSDA) exchange field is thus suppressed. The resulting exchange field is due to the polarization of the transition-metal sublattice, as expected for hard magnetic rare-earth intermetallics. In contrast, direct spin-polarized  DFT+Hub-I calculations without the averaging would lead to a large unphysical exchange field on rare-earth sites due to the magnetization density of 4$f$ electrons themselves. 

In appendix \ref{SmCo5}, we benchmark the present method on the prototypical rare-earth hard magnet SmCo$_5$, for which several measurements of CFP exist, and show good agreement between calculated and measured CFPs. Moreover, the actual eigenstates of the Sm 4$f$ shell in SmCo$_5$ obtained within DFT+HubI are also in very good agreement with previous neutron scattering and mangetic form-factor measurements, see Appendix~\ref{WF_tables}.

\subsection{Calculational details}\label{sbsec:calc_details}

\begin{figure}
\includegraphics[width=1\linewidth]{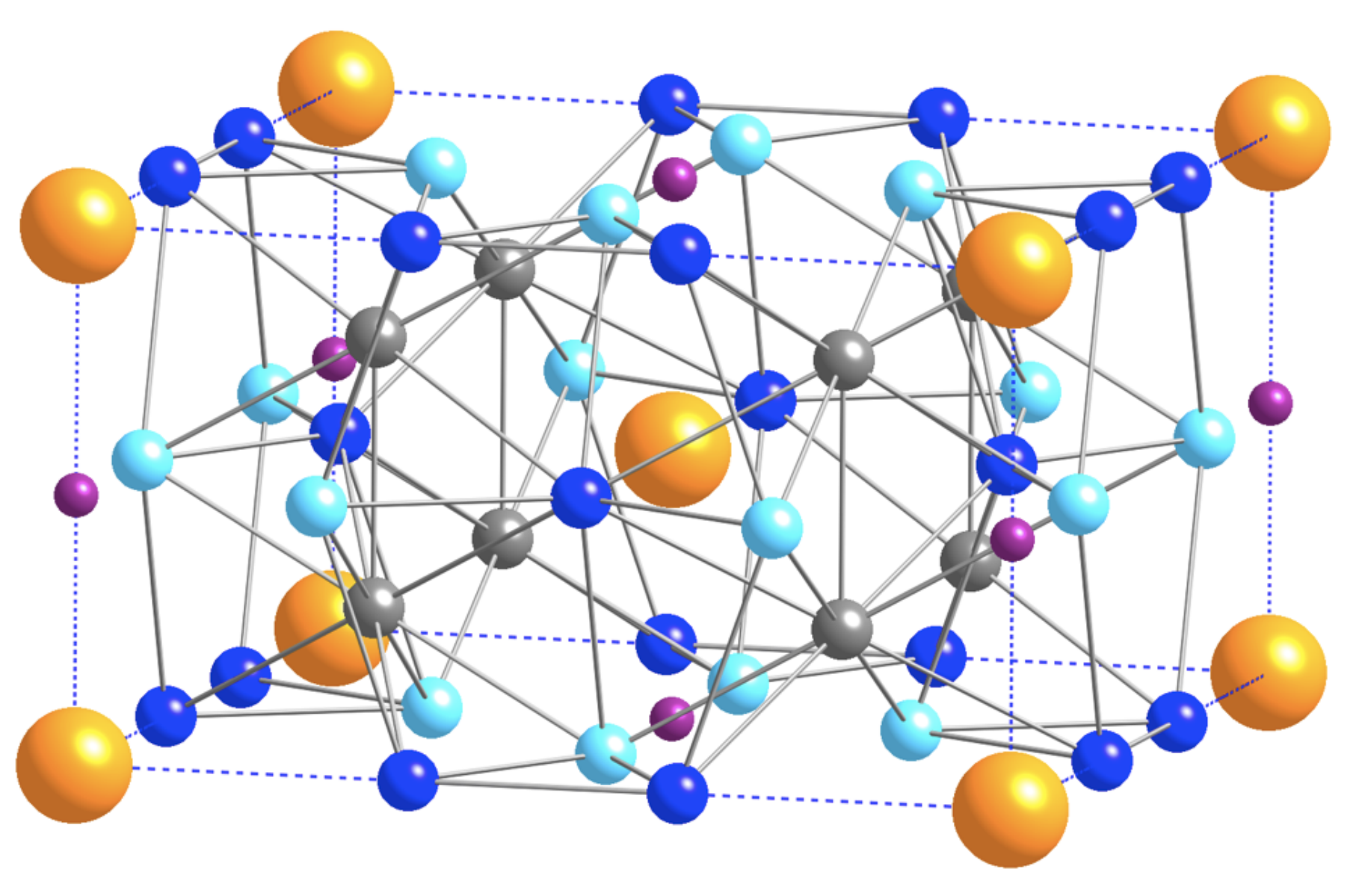}
\caption{ Conventional unit cell of $R$Fe$_{12}X$. The rare-earth $R$ sites are yellow, the three types of Fe sites are grey, light and dark blue, and dopant $X$ sites purple.}
\label{crystal_structure}
\end{figure}

The $R$Fe$_{12}X$ family has the space group $I4/mmm$, with a tetragonal primitive unit cell. 
The conventional unit cell, with twice the volume and the atoms, is orthorhombic. 
It has equivalent $R$ sites in the corner and the center at Wyckoff position 2$a$, $X$ interstitial sites between two nearest $R$ sites on Wyckoff position 2$b$, and contains 24 Fe atoms on three inequivalent sites, denoted below Fe$_1$, Fe$_2$ and Fe$_3$ at Wyckoff positions 8$j$, 8$i$ and 8$f$ respectively, as displayed in Fig.~\ref{crystal_structure}.
Calculations are done at the theoretical lattice constants for $R$Fe$_{12}X$, summarized in table \ref{tab:lattice_constants} in the conventional unit cell (from Table II of Ref. \onlinecite{harashima2015_proc} and from this work). 
The calculated lattice constants agree within 2 \% with the measured ones in the more stable NdFe$_{11}$Ti(N) and SmFe$_{11}$Ti(N) compounds.\cite{harashima2015_proc}

{\centering
\begin{tabularx}{0.6\linewidth}{|c | Y | Y |}
\cline{2-3}
\multicolumn{1}{c|}{}  & \multicolumn{2}{ c |}{Lattice constant (\AA)}  \\
\hline
Compound & $a$ & $c$ \\
\hline
NdFe$_{12}$ & 8.533 & 4.681 \\
NdFe$_{12}$N & 8.521 & 4.883 \\
NdFe$_{12}$Li & 8.668 & 4.873 \\
SmFe$_{12}$ & 8.497 & 4.687 \\
SmFe$_{12}$N & 8.517 & 4.844 \\
SmFe$_{12}$Li & 8.640 & 4.863 \\
\hline
\end{tabularx}
\captionof{table}{\label{tab:lattice_constants} Conventional unit cell lattice constants used in our calculations. $b$ = $a$, and the angles are $ \alpha = \beta = \gamma = 90^\circ$}
}

The DFT calculations are performed with spin-orbit coupling included within the second variational approach. We employ throughout the rotationally-invariant Coulomb vertex specified by Slater integrals  $F^0=U=$6.0~eV as well as $F^2=$10.13, $F^4=$6.77, and $F^6=$5.01~eV corresponding to Hund's rule coupling $J_H=$0.85~eV.  These values of $U$ and $J_H$ are in agreement with those in the literature\cite{Pourovskii2007,Nilsson2013,Locht2016}. One may notice, that while the values of $U$ and $J_H$ are important to determine the one-electron spectrum of a material, they are expected to have a rather small effect on the crystal-field parameters that we consider in this work.  \footnote{In rare-earth ions  one typically has $U \gg J_H \gg \lambda$, hence, the atomic multiplet structure is set within the $LS$-couping scheme. With the crystal-field splitting in lanthanides being typically much smaller than all those energy scales one may neglect the mixture between different multiplets, in that case CFP exhibits only very weak dependence to $U$ and $J_H$.} We discuss this dependence in Appendix \ref{app_U_J}. DFT+Hub-I calculations are carried out for the temperature of 290~K.

\section{Results} 
\subsection{DFT and DFT + Hubbard I electronic structure of $R$Fe$_{12}X$} \label{sbsec:res_es}
\begin{figure*}
\begin{minipage}[b][6cm]{0.9\linewidth}
\subfloat[NdFe$_{12}$N DFT density of states ]{\includegraphics[scale=0.35]{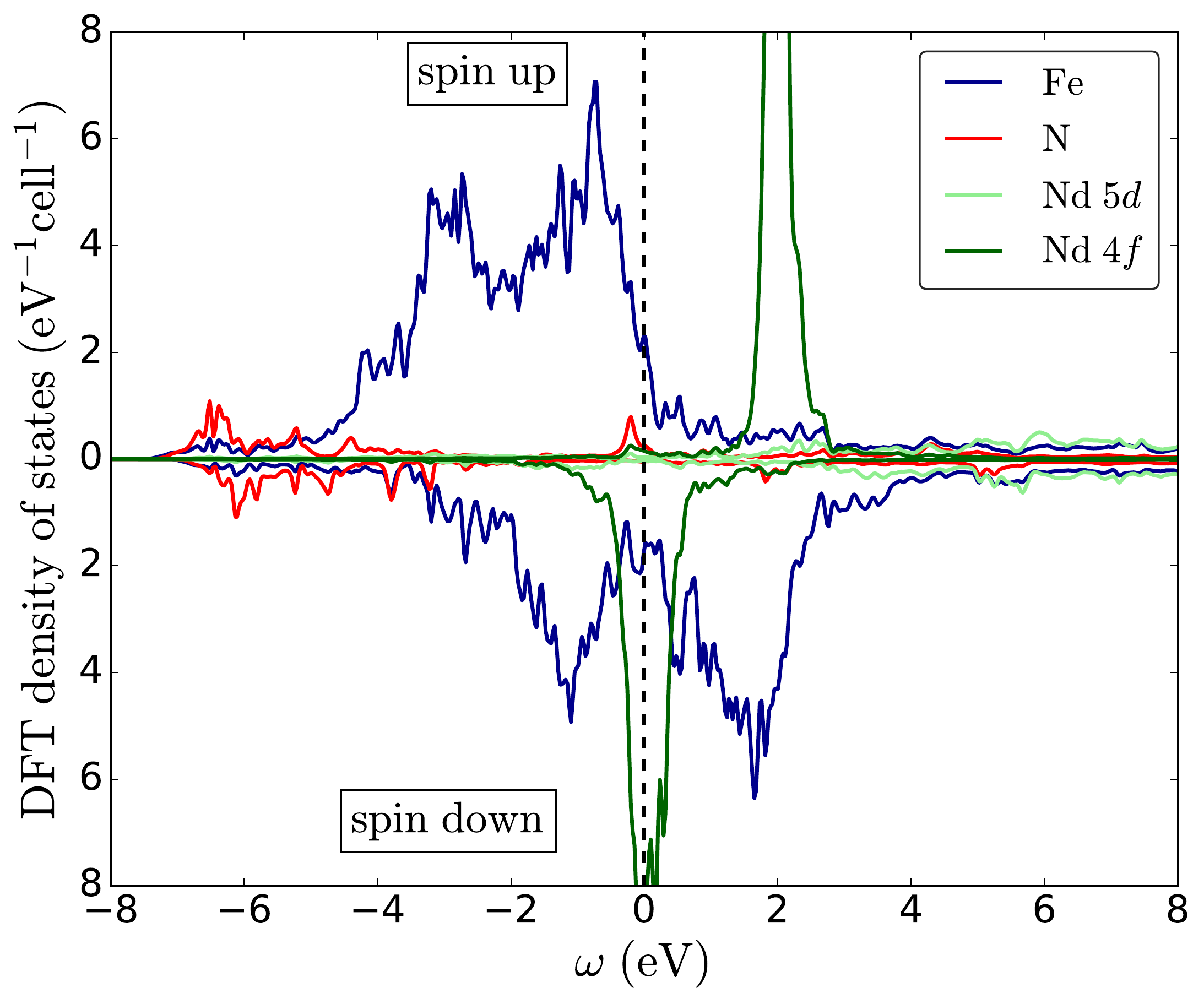}\label{sfig:LDA_DOS}}
\subfloat[NdFe$_{12}$N DFT + Hubbard I spectral function ]{\includegraphics[scale=0.35]{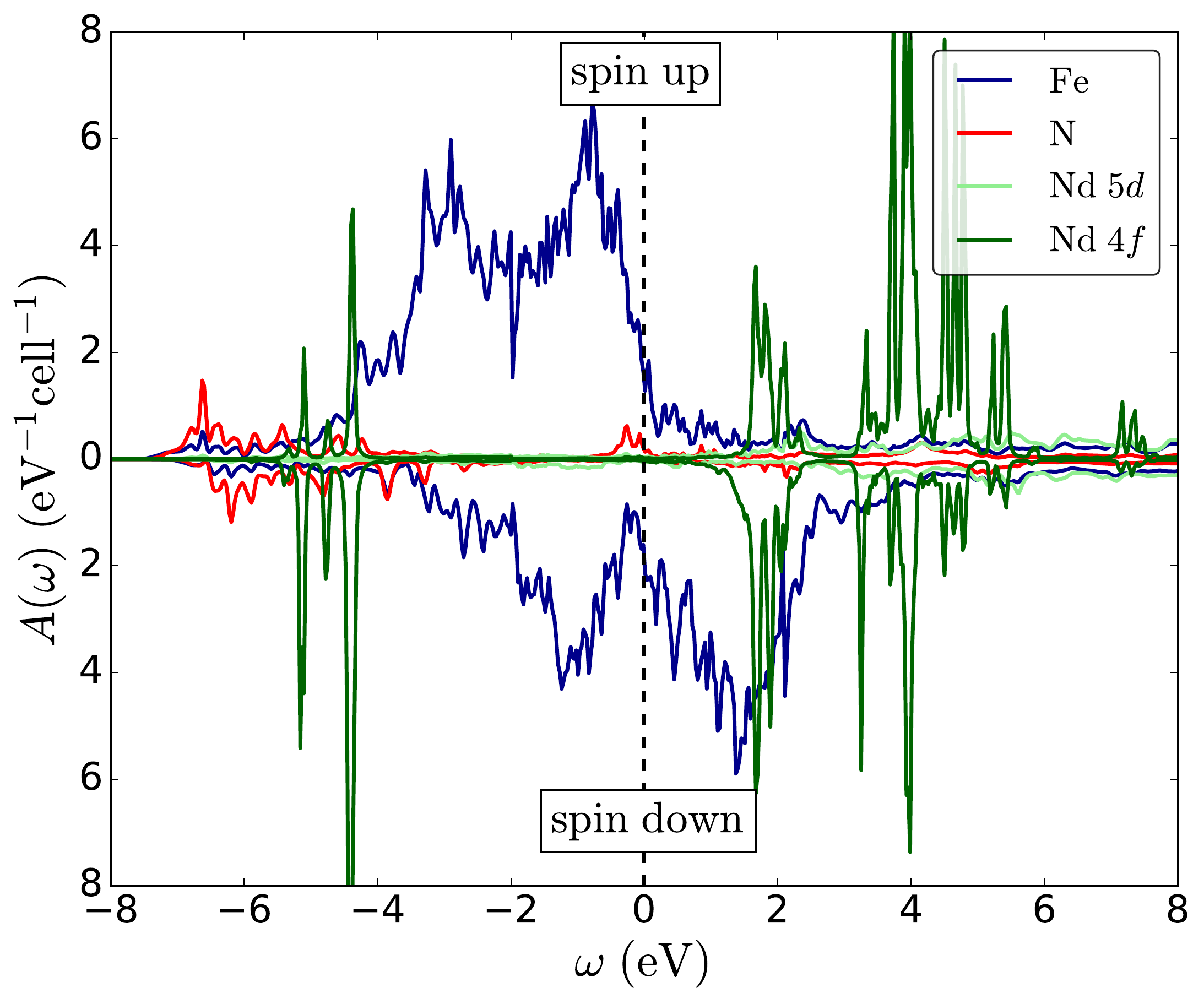}\label{sfig:hub1_spectral_fun}}
\end{minipage}
\caption{a. Atom- and orbital-resolved density of states of NdFe$_{12}$N  calculated with the spin-polarized DFT method. b. Atom- and orbital-resolved  spectral function of the same compound  obtained within self-consistent spin-polarized  DFT+Hubbard I.
For better readability we take the average over the three types of Fe atoms: the actual total Fe density of states per unit cell is three times larger.}
\label{DOS_spectral_function}
\end{figure*}

We first compare the electronic structure of $R$Fe$_{12}X$ obtained within DFT (LSDA) and DFT+Hub-I. 
A typical DFT density of states (DOS) and a DFT+Hub-I spectral function for ferromagnetic $R$Fe$_{12}X$, namely, for NdFe$_{12}$N, are shown in Fig.~\ref{DOS_spectral_function}.  
The DFT DOS of Fig.~\ref{sfig:LDA_DOS} features a strong polarization of the Fe 3$d$ band. 
N 2$p$ states are dispersive, with the bottom of the bands contributing to a peak in the DOS around -6~eV. 
The Nd 4$f$ band is fully spin-polarized and anti-ferromagnetically aligned to Fe 3$d$, with the total spin moment within the Nd atomic sphere equal to -2.77~$\mu_B$, i.e. close to the Hund's rule value of 3~$\mu_B$ for the Nd$^{3+}$ ion. 
The Nd majority-spin 4$f$ band is pinned at the Fermi level, its double-peak structure is due to spin-orbit splitting. 
This picture of 4$f$ bands pinned at the Fermi level is qualitatively incorrect and illustrates the difficulties of DFT with local or semi-local exchange-correlation functionals to correctly treat strongly-interacting localized valence states.

The spin-polarized DFT+Hub-I spectral function shown in Fig.~\ref{DOS_spectral_function} was calculated using the averaging approach described in Sec.~\ref{sbsec:cf_calc}. 
It features an almost fully-polarized Fe 3$d$ band as well as occupied and empty 4$f$ states separated, to first approximation, by $U$, thus forming  lower and upper Hubbard bands, respectively. 
The Hubbard bands are split due to the Hund's rule and spin-orbit couplings into several manifolds with characteristically sharp peaks corresponding to transitions from the ground state to different quasi-atomic multiplets upon electron addition or removal.
The 4$f$ multiplet structure in lanthanides is known to be only weakly sensitive to the crystalline environment. 
Indeed, the positions of the Hubbard bands in Fig.~\ref{sfig:hub1_spectral_fun} as well as the overall shape of the upper Hubbard band split  into two manifolds of multiplet peaks centered at about 2 and 4~eV are in agreement with photoemission and inverse-photoemission spectra of the Nd metal\cite{lang1981study}. 
One also sees that the Nd 4$f$ states in DFT+Hub-I are not fully spin-polarized, in contrast to the DFT case. 
Indeed the Nd spin moment of -1.61~$\mu_B$ obtained within DFT+Hub-I  is only about half of the Hund's rule value and is also aligned antiferromagnetically with respect to the spin moment on iron. 
The calculated Nd orbital moment is 3.40~$\mu_B$. 
It is precisely the crystal-field splitting of the Nd 4$f$ shell that prevents the full saturation of the Nd magnetization.

\subsection{crystal-field parameters and exchange fields in $R$Fe$_{12}X$}\label{sbsec:res_cf}

The calculated CF and exchange fields for Nd and Sm $R$Fe$_{12}$(N,Li) compounds are listed in Table~\ref{tab:1_12} and \ref{tab:2_12}, together with the magnetic moments on $R$ and in the full cell. 
Comparing the different materials, one sees that $R$Fe$_{12}$ has the smallest values of $A_2^0\langle r^2\rangle$ (in absolute value), while N insertion enhances  $A_2^0\langle r^2\rangle$  up to positive values of about 400 to 600~K. 
Li insertion has the opposite effect, leading to large negative  $A_2^0\langle r^2\rangle$, in particular for $R=$Nd. 
We notice some dependence of the CF parameters $A_k^q\langle r^k\rangle$ on the spin direction in the ferromagnetic phase. It is mostly weak, of the order of a few tenths of kelvin for the most important CFP $A_2^0\langle r^2\rangle$, except in NdFe$_{12}$N. It can be significant, though, for higher-order CFP. 
The magnetic state (paramagnetic of ferromagnetic) has a significant impact on $A_2^0\langle r^2\rangle$  in some compounds: one may notice larger values of $A_2^0\langle r^2\rangle$ for paramagnetic SmFe$_{12}$(N,Li) than for either spin direction  in the ferromagnetic phase. 

Finally, the total magnetization appears to be slightly reduced in Sm compounds, compared to Nd compounds: in the former, the spin magnetic moment on the rare-earth compensates the orbital magnetic moment, leading to negligible total moment, while Nd presents a total moment dominated by the orbital component, and in the same direction as the Fe sublattice magnetization.

The sign and overall magnitude of our calculated $A_2^0\langle r^2\rangle$ are in agreement with previous calculations for $R$Fe$_{12}$(N) in Ref. \onlinecite{harashima2015} using the 4$f$-in-core approach, though there are some differences in the precise values. 
We obtain a similar value for NdFe$_{12}$, a somewhat larger one for NdFe$_{12}$N, a more negative value for SmFe$_{12}$ and a smaller (positive) value for SmFe$_{12}$N.  One may notice that the results in Ref. \onlinecite{harashima2015} are quite sensitive to different  treatments of the ``tails'' of 4$f$ core orbitals: there is no such uncertainty in our approach.

\begin{table*}
	\centering
	\renewcommand{\arraystretch}{1.2}
	\setlength{\tabcolsep}{10pt}
	\caption{\label{tab:1_12} Calculated CF parameters in  ferromagnetic (FM) and paramagnetic (PM)  NdFe$_{12}$(N,Li) in kelvin. For the FM case we list the CF parameters for each spin direction. The exchange field in the FM phase (in tesla), the spin and orbital magnetic moments of the rare-earth as well as the total magnetic moment per crystal unit cell (in Bohr magneton $\mu_B$) are also listed.}
	\begin{tabular}{c| c | c c | c | c c | c | c c | }
		\cline{2-10}
		&  \multicolumn{3}{ c|}{NdFe$_{12}$} & \multicolumn{3}{ c|}{NdFe$_{12}$N} & \multicolumn{3}{ c|}{NdFe$_{12}$Li }  \\
		\cline{2-10}
		& PM & \multicolumn{2}{c|}{FM}& PM & \multicolumn{2}{c|}{FM} & PM & \multicolumn{2}{c|}{FM}  \\ 
		&        &   $\uparrow$  & $\downarrow$ &        &   $\uparrow$  & $\downarrow$ &        &   $\uparrow$  & $\downarrow$   \\
		\hline
		\multicolumn{1}{ |c|}{$A_2^0\langle r^2\rangle$} & -57   & -71  & -116  & 486  & 477 & 653  & -656 & -687 & -742 \\
		\multicolumn{1}{ |c|}{$A_4^0\langle r^4\rangle$} & -29  &  -5 & -1 & 107  & 75 & 112  & -182 & -158 & -186 \\
                 \multicolumn{1}{ |c|}{$A_4^4\langle r^4\rangle$} & -129 & -76 & -270 & 7  & -105  & -141 & -118 & -60 &  -228 \\
		\multicolumn{1}{ |c|}{$A_6^0\langle r^6\rangle$} &  52 & 62 & 54  & 51  & 32 & 63 & -24 & -17 & -31 \\
		\multicolumn{1}{ |c|}{$A_6^4\langle r^6\rangle$} &  70  & -224  & -107 &  -160  & -65 & -91  & 37 &  -6 & 96 \\
		\hline
		\multicolumn{1}{ |c|}{$B_{\mathrm{ex}}$ ($T$)} & -  &  \multicolumn{2}{c|}{265} & -  &  \multicolumn{2}{c|}{217} & -  &  \multicolumn{2}{c|}{410} \\
		\multicolumn{1}{ |c|}{Nd $M_{\mathrm{spin}} $} & -  &  \multicolumn{2}{c|}{-1.48~$\mu_B$} & -  &  \multicolumn{2}{c|}{-1.61~$\mu_B$} & -  &  \multicolumn{2}{c|}{-1.69~$\mu_B$} \\
		\multicolumn{1}{ |c|}{Nd $M_{\mathrm{orb}} $} & -  &  \multicolumn{2}{c|}{2.96~$\mu_B$} & -  &  \multicolumn{2}{c|}{3.40~$\mu_B$} & -  &  \multicolumn{2}{c|}{3.28~$\mu_B$} \\
		\multicolumn{1}{ |c|}{$M_{\mathrm{cell}}$} & -  &  \multicolumn{2}{c|}{26.39~$\mu_B$} & -  &  \multicolumn{2}{c|}{29.15~$\mu_B$} & -  &  \multicolumn{2}{c|}{27.59~$\mu_B$} \\
		\hline
	\end{tabular}
\end{table*}
\begin{table*}
	\centering
	\renewcommand{\arraystretch}{1.2}
	\setlength{\tabcolsep}{10pt}
	\caption{\label{tab:2_12} The same quantities as in Table~\ref{tab:1_12} for SmFe$_{12}$(N,Li).}
	\begin{tabular}{c| c | c c |  c | c c  |  c | c c | }
		\cline{2-10}
		&  \multicolumn{3}{c|}{SmFe$_{12}$} & \multicolumn{3}{c}{SmFe$_{12}$N}  & \multicolumn{3}{|c|}{SmFe$_{12}$Li   } \\
		\cline{2-10}
		& PM & \multicolumn{2}{c|}{FM} & PM & \multicolumn{2}{c|}{FM}  & PM & \multicolumn{2}{c|}{FM} \\
		&   &  $\uparrow$  & $\downarrow$ &        &   $\uparrow$  & $\downarrow$  &        &   $\uparrow$  & $\downarrow$  \\
		\hline
		\multicolumn{1}{ |c|}{$A_2^0\langle r^2\rangle$} & -32   & -184  & -211  &  249  & 195  & 225  & -458 & -297 & -272 \\
		\multicolumn{1}{ |c|}{$A_4^0\langle r^4\rangle$} & -11   & -21  & -18  & 99  & 78  & 70  & -116 & -68 & -71 \\
                 \multicolumn{1}{ |c|}{$A_4^4\langle r^4\rangle$} & -215  & -41 &  -136 & -122   & 22 & -91  & -124 & 61 &  -198 \\
		\multicolumn{1}{ |c|}{$A_6^0\langle r^6\rangle$} &  47 & 45  & 40  & 71  & 47 & 25  & -13 &  -2 &  -12  \\
		\multicolumn{1}{ |c|}{$A_6^4\langle r^6\rangle$} & -85  &  -95 & -58  & -184  & -97 & -82   & 44 & 30 & 38 \\
		\hline
		\multicolumn{1}{ |c|}{$B_{\mathrm{ex}}$ ($T$)} & -  &  \multicolumn{2}{c|}{232} & -  &  \multicolumn{2}{c|}{205}  & -  &  \multicolumn{2}{c|}{331} \\
		\multicolumn{1}{ |c|}{Sm $M_{\mathrm{spin}} $} & -  &  \multicolumn{2}{c|}{-3.31~$\mu_B$} & -  &  \multicolumn{2}{c|}{-2.41~$\mu_B$} & -  &  \multicolumn{2}{c|}{-3.96~$\mu_B$} \\
		\multicolumn{1}{ |c|}{Sm $M_{\mathrm{orb}} $} & -  &  \multicolumn{2}{c|}{3.29~$\mu_B$} & -  &  \multicolumn{2}{c|}{2.35~$\mu_B$} & -  &  \multicolumn{2}{c|}{3.60~$\mu_B$} \\
		\multicolumn{1}{ |c|}{$M_{\mathrm{cell}}$} & -  &  \multicolumn{2}{c|}{24.54~$\mu_B$} & -  &  \multicolumn{2}{c|}{26.83~$\mu_B$} & -  &  \multicolumn{2}{c|}{25.77~$\mu_B$} \\
		\hline
	\end{tabular}
\end{table*}

\begin{figure}
\includegraphics[width=0.9\linewidth]{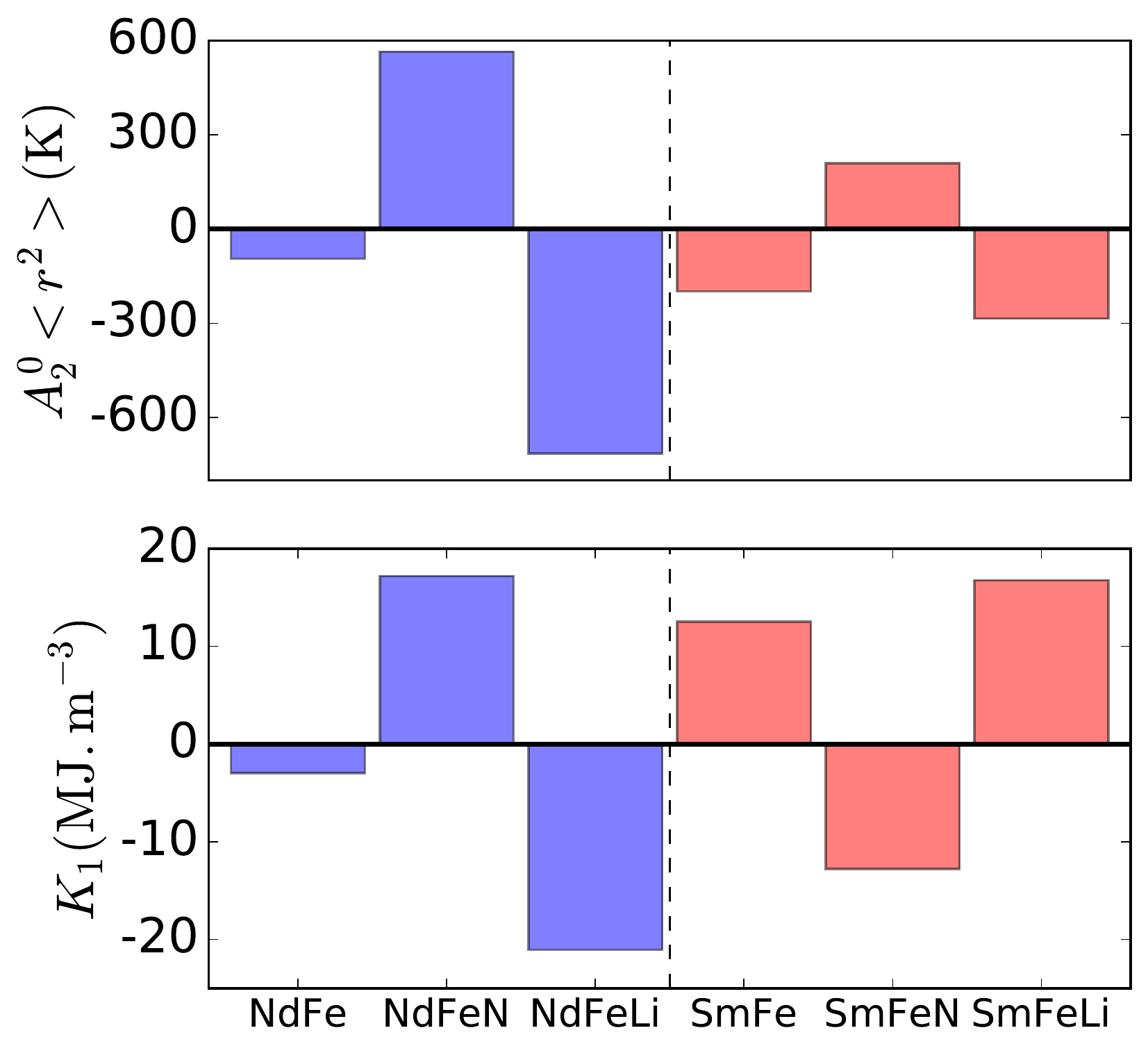}
\caption{crystal-field parameters $A_2^0\langle r^2\rangle$  (average for up and down spins in the FM phase) and anisotropy energy $K_1$ for \emph{R}Fe$_{12}$\emph{X}, with \emph{R}=Nd,Sm and \emph{X} is either empty, N or Li ($K_1$ is obtained from equation \ref{CF_to_K}). }
\label{values_chart}
\end{figure}

The lowest-order CF parameters $A_2^0\langle r^2\rangle$ and the corresponding single-ion anisotropy energies $K_1$ evaluated using Eq.~\ref{CF_to_K} are displayed in Fig.~\ref{values_chart}. 
One may see that, while NdFe$_{12}$N and NdFe$_{12}$Li exhibit larger $|A_2^0\langle r^2\rangle|$  (upper panel) than their Sm counterparts, this difference is offset by a larger Stevens prefactor of Sm in Eq.~\ref{CF_to_K}, so that the Sm and Nd-based compounds have a magnetic anisotropy coefficient $K_1$ of similar magnitude. 
An important difference between Nd and Sm is the different signs of their Stevens factors $\alpha_J$ ($\alpha_J = -7/1089$ for Nd, $\alpha_J = 13/315$ for Sm). 
Consequently, N insertion leads to a large out-of-plane anisotropy  for Nd, but in-plane anisotropy for Sm. 
Li has the opposite effect: doping Li into SmFe$_{12}$ leads to a rather large out-of-plane anisotropy of SmFe$_{12}$Li, of comparable magnitude to that of NdFe$_{12}$N.

Performing the averaging over the ground state multiplet as described in Eq.~\ref{eq_Gat} is crucial to obtain reasonable CFP: the lowest-order CFP $A_2^0\langle r^2\rangle$ is most sensitive to this.
The corresponding data without averaging for NdFe$_{12}$N are given and discussed in Appendix \ref{appendix:imp_of_averaging}.

For the sake of comparison with future experiments we list low-energy eigenvalues and eigenstates of all $R$Fe$_{12}X$ compounds in Appendix~\ref{WF_tables}. It is interesting to notice that eigenstates of the ground-state $J=5/2$ multiplet of Sm (Table~\ref{tab:atomic_levels_Sm}) are often found to exhibit  a significant admixture from exited $J=7/2$ states; the Nd $J=9/2$ states (Table~\ref{tab:atomic_levels_Nd}) contain a significantly lower admixture from the first exited multiplet.

\begin{figure}
\includegraphics[width=0.9\linewidth]{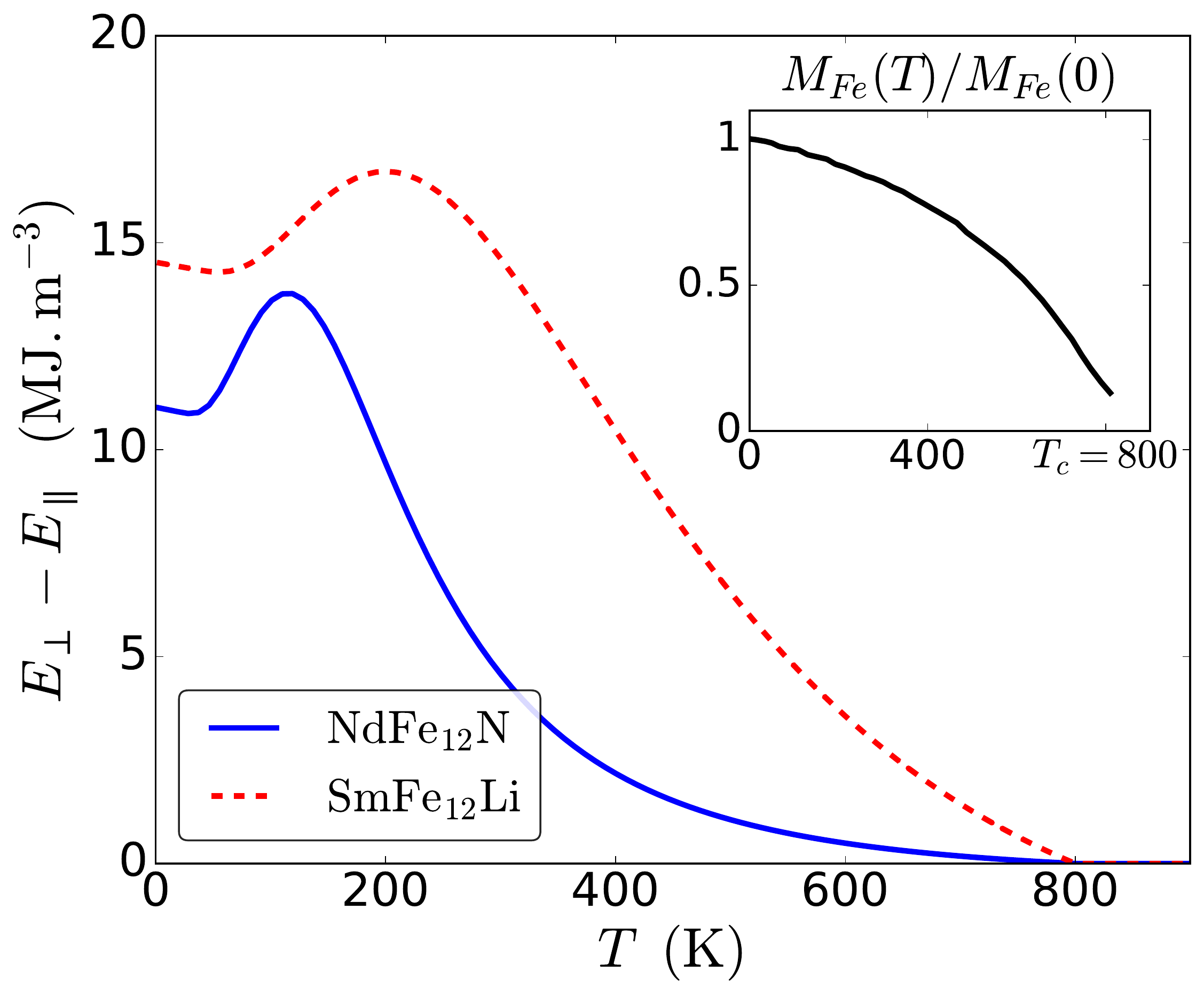}
\caption{ (Color online) Evolution with temperature of the difference in the 4$f$ shell energy $E_{\perp} - E_{\parallel}$ between the moments on $R$ and Fe aligned perpendicularly and parallel to the $z$ axis, respectively, for NdFe$_{12}$N (blue, full line) and SmFe$_{12}$Li (red, dashed line). 
Inset: magnetization fraction of the Fe sublattice in NdFe$_{12}$N, as a function of temperature from Hirayama \emph{et al}\cite{hirayama2015}.}
\label{K1_vs_temp}
\end{figure}

A last interesting point is that the exchange fields $B_{\mathrm{ex}}$ on the rare-earth are enhanced by  Li and reduced by N. 
This is useful because the exchange field, or exchange coupling between Fe and $R$, is essential for finite temperature magnetocrystalline anisotropy. 
The rare-earth-originated anisotropy becomes ineffective at high temperature, and this threshold temperature is determined by the exchange coupling $B_{\mathrm{ex}}$. 
In Fig.~\ref{K1_vs_temp}, we show the difference between the 4$f$ shell atomic energies $E_{\perp}$ and $E_{\parallel}$, computed as $ E = \mathrm{Tr}[\hat{H}e^{-\beta \hat{H}}] / \mathrm{Tr}[e^{-\beta \hat{H}}]$ with $\hat{H}$ defined in Eq.~\ref{H} and the exchange field $B_{\mathrm{ex}}$ is along the $z$ axis (along the $c$ lattice parameter) and $x$ axis (along the $a$ lattice parameter), respectively. We scale the exchange field $B_{\mathrm{ex}}$ by a coefficient $M_{Fe}(T)/M_{Fe}(0)$ at non-zero temperatures, using the measured magnetization ratio of NdFe$_{12}$N from Hirayama \emph{et al}\cite{hirayama2015}.
The energy difference plotted in Fig.~\ref{K1_vs_temp} is more general than the expression of Eq.~\ref{CF_to_K}, because it also contains higher order CFP and non-zero temperature; to compute $E_{\perp}$ and $E_{\parallel}$ we diagonalize the full Hamiltonian $\hat{H}$, without restricting ourselves to the ground state multiplet. This gives quite a different picture than Fig.~\ref{values_chart}: the strongly enhanced exchange coupling due to Li doping causes the magnetocrystalline anisotropy to persist at much higher temperatures than with N doping.

\section{Discussion: the effect of hybridization with the interstitials} \label{sec:discussion}
Let us now analyze the mechanisms determining the CFP on the rare-earth site and, in particular, the impact of the N and Li interstitials on them. We consider the NdFe$_{12}X$ ($X=$Ni, Li) compounds as example.  The N atom nominally carries three 2$p$ electrons, but in the $R$Fe$_{12}$N compounds the N 2$p$ bands are more than half-filled (Fig.~\ref{DOS_spectral_function}). To verify this we have also performed a Bader-charge analysis\cite{bader1990atoms} for NdFe$_{12}X$ and found 8.3 electrons on N resulting in an ion charge of -1.3. In contrast, the Li atom is nominally 2$s^1$, but it looses its single 2$s$ electron inside the NdFe$_{12}$ matrix, the corresponding Bader ion charge is +0.7.

\begin{figure}
\begin{minipage}[b]{0.95\linewidth}
\subfloat[NdFe$_{12}$N, $m=-3$]{\includegraphics[width=0.5\linewidth]{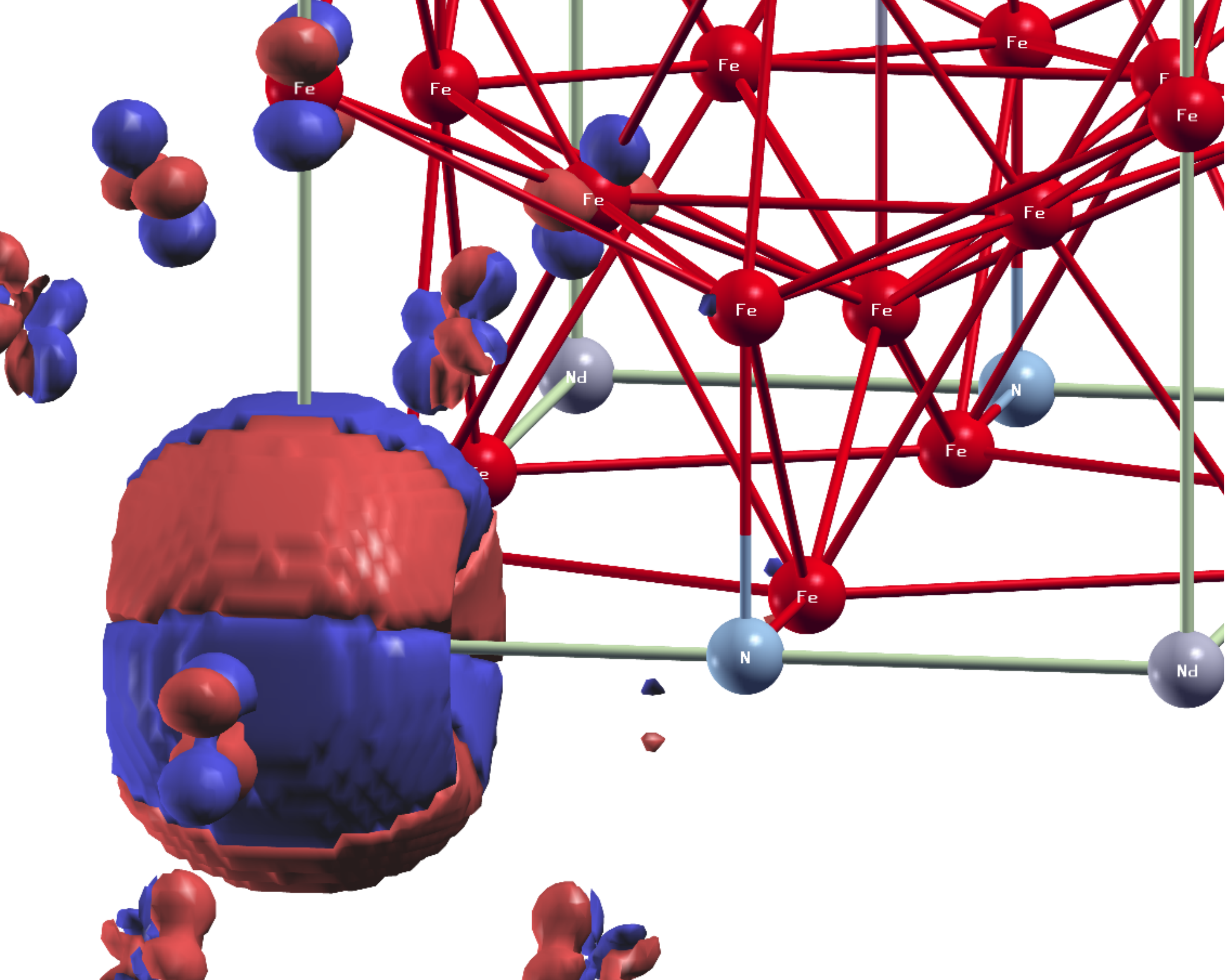}}
\subfloat[NdFe$_{12}$N, $m=0$]{\includegraphics[width=0.5\linewidth]{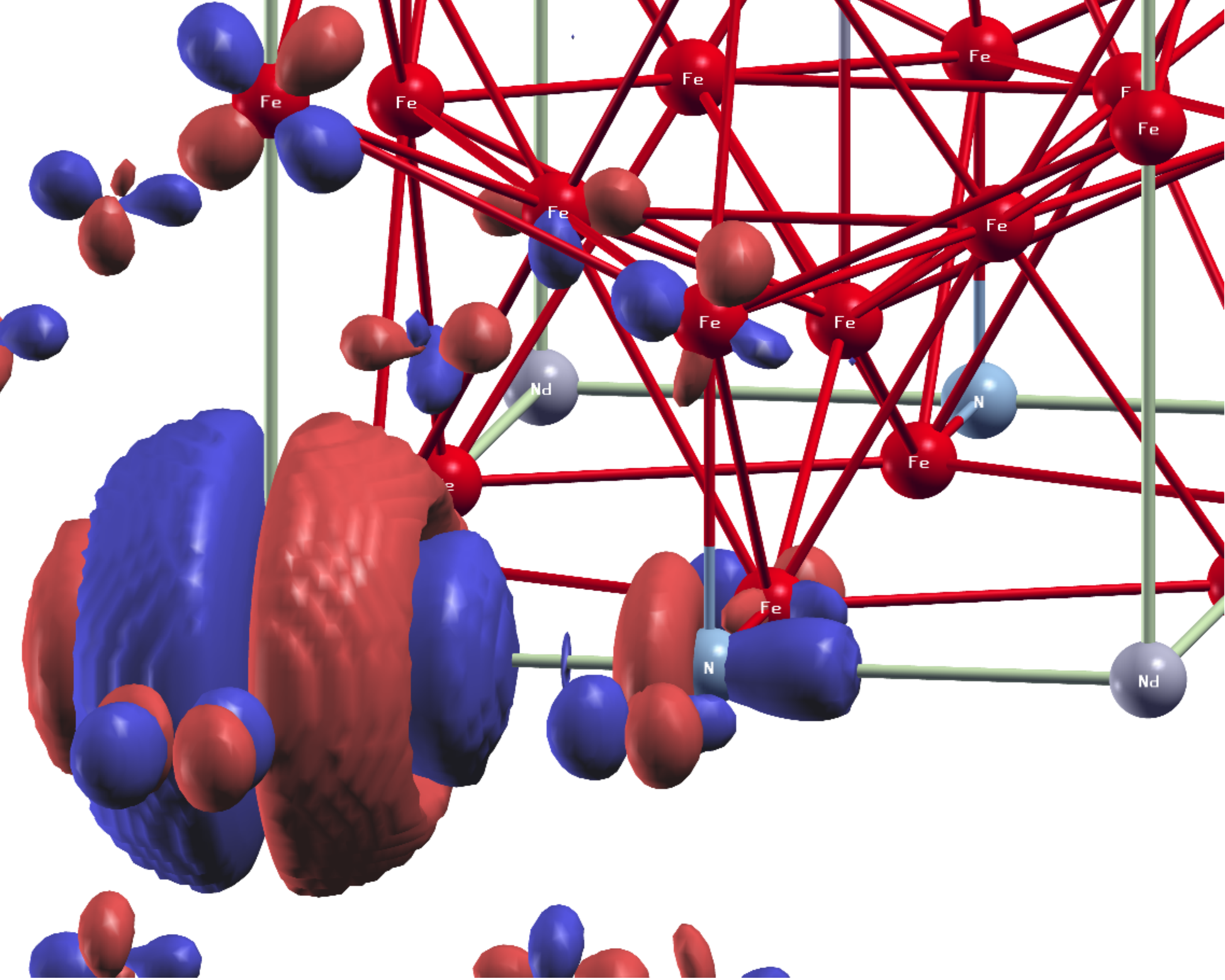}}
\end{minipage}
\begin{minipage}[b]{0.95\linewidth}
\subfloat[NdFe$_{12}$Li, $m=-3$]{\includegraphics[width=0.5\linewidth]{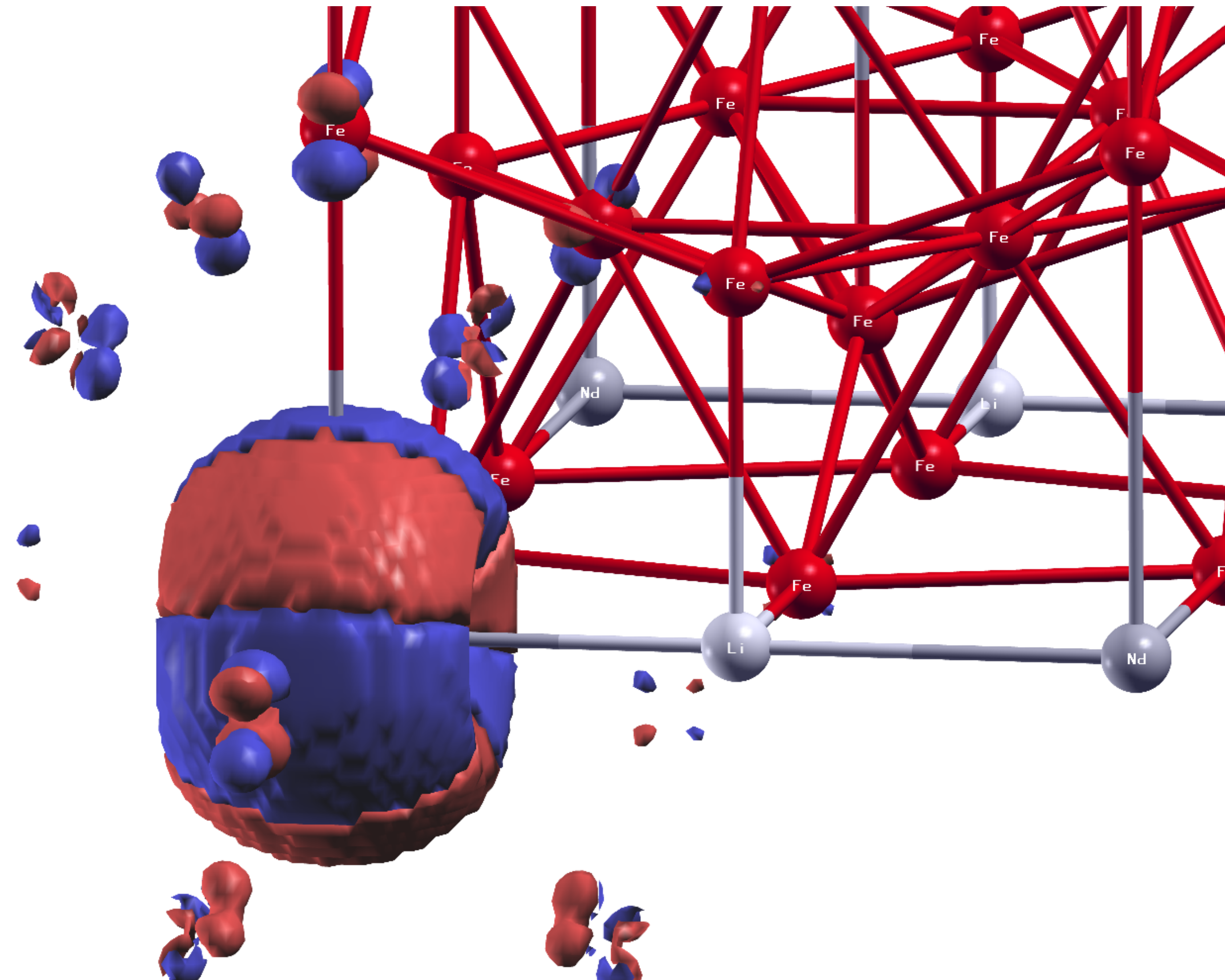}}
\subfloat[NdFe$_{12}$Li, $m=0$]{\includegraphics[width=0.5\linewidth]{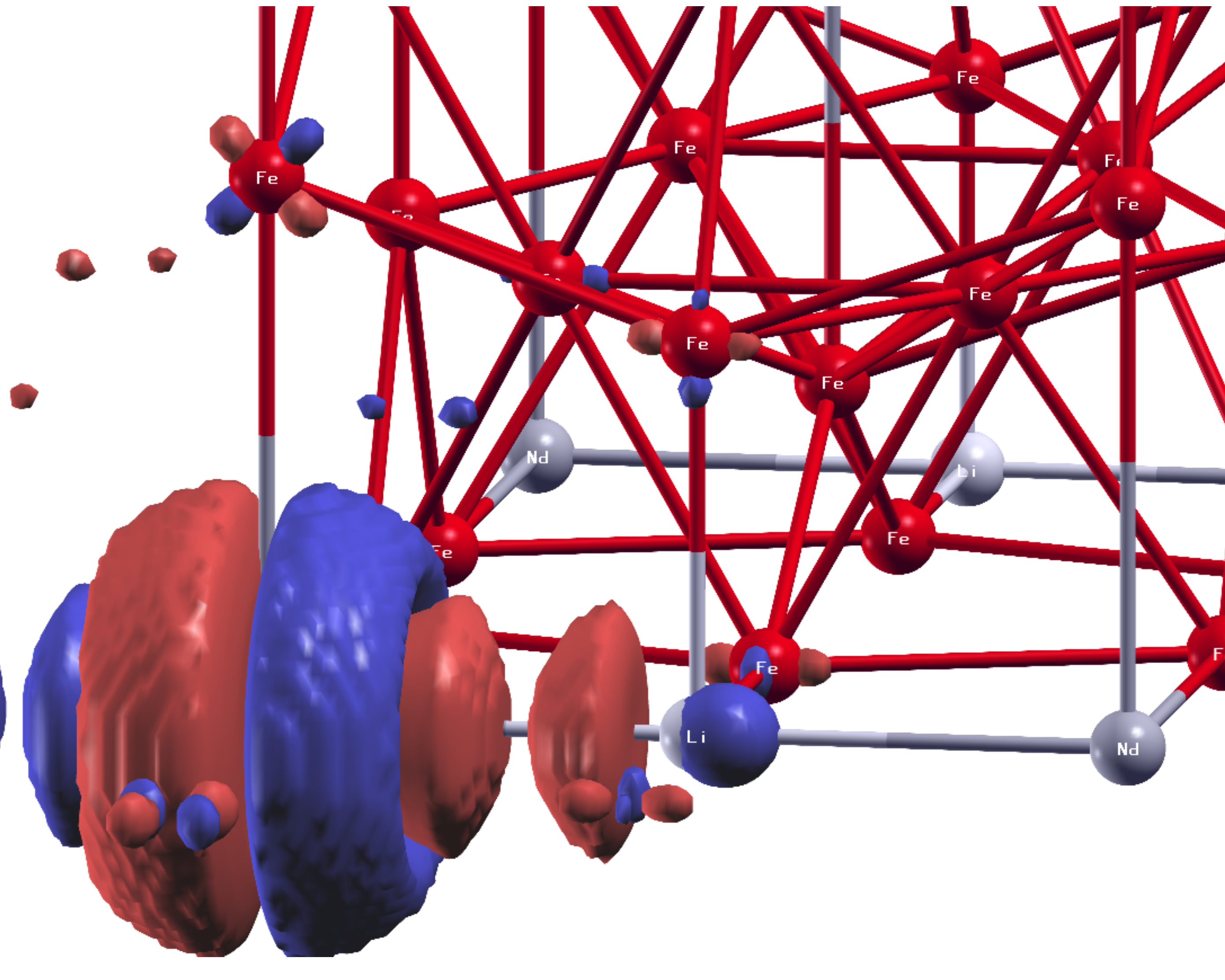}}
\end{minipage}
\caption{$4f$ Wannier orbitals of NdFe$_{12}$N and NdFe$_{12}$Li, for magnetic quantum number $m=-3$ and $m=0$ and window size $\left[-2,2 \right]$~eV. The orbital with $m=0$ points towards and leak to the N or Li sites, while orbitals with $m=\pm3$ do not. All of them leak somewhat to the nearest-neighbor Fe atoms.}
\label{wannier_orbitals}
\end{figure}

In Fig.~\ref{wannier_orbitals} we display the complex Wannier orbitals constructed for Nd 4$f$ states with window size $\omega_{\textrm{win}}=2$~eV with magnetic quantum numbers $m=0$ and $m=-3$, in the presence of interstitial N or Li.  The orbitals with $m=\pm 3$ do not point towards the N or Li atom, and leak only to neighboring Fe atoms. On the other hand, the orbital with $m=0$ (corresponding to $f_{z^3}$ cubic orbital) points towards the interstitial site, and shows strong leakage to the interstitial atom,  particularly in the Li case. The same applies, to a lesser extent, to the orbitals $m=\pm 1$ that are also pointing towards the interstitials.

The N (Li) insertion  has thus two effects on the CFP. The first one is due to the electrostatic interaction between the 4$f$ electrons and the interstitial ions. This interaction with the negative N (positive Li) ion pushes the on-site energies of the $m=-1,0,1$ orbitals, which point towards the interstitial, to higher (lower) energies. 

The second contribution is due to hybridization between the 4$f$ states and the N 2$p$ (Li 2$s$ and 2$p$) bands, which is expected to mainly affect the $m=-1,0,1$ orbitals pointing towards the interstitial. Mixing with the empty Li 2$s$ and 2$p$ bands pushes them to lower energies,  while the opposite shift is induced due to hybridization with mostly filled N 2$p$ located well below rare-earth 4$f$ states, see Fig.~\ref{DOS_spectral_function}. Hence, one sees that both the electrostatic and hybridization effects act in the same direction, raising the on-site energies of the $m=-1,0,1$ orbitals in the case of N and lowering them in the case of Li.

This analysis explains the effect of interstitials on the CFP $A_2^0\langle r^2\rangle$. 
Indeed, the contribution due to $A_2^0\langle r^2\rangle$ into the CF Hamiltonian \ref{H_cf} is $A_2^0\langle r^2\rangle\hat{T}_2^0/\lambda_2^0$, where the matrix of the one-electron operator $\hat{T}_2^0/\lambda_2^0$ reads
$$ \hat{T}_2^0/\lambda_2^0 = 2
\quad
\begin{pmatrix} 
-\frac{1}{3} & & & &  \\
  & 0 & & & & (0)   \\
  & & \frac{1}{5} & & \\
  & & & \frac{4}{15} &  \\
  & & & & \frac{1}{5} & \\
  &  (0)  & & & & 0 & \\
  & & & & & & -\frac{1}{3}
\end{pmatrix}
\quad
$$
in the basis of complex 4$f$ orbitals. Hence, the energy level of 4$f$ orbitals $m=\pm3$ is negatively correlated with $A_2^0\langle r^2\rangle$, while the energy levels of the orbitals with $m=-1,0,1$ are positively correlated with $A_2^0\langle r^2\rangle$ (orbitals with $m=\pm2$ are unaffected by $\hat{O}_l^m$). Thus, the effect of N (Li) insertion is to enhance (reduce) the value of $A_2^0\langle r^2\rangle$.


One may argue that the Hubbard-I approximation neglects the hybridization function in solving the quantum impurity problem, hence, hybridization to the bath is not included explicitly
when solving for the self-energy $\Sigma$ in the DMFT (Hubbard-I) step of our DFT+Hub-I calculations.
However, our Wannier orbitals constructed within the ``small'' energy window do contain the effect of hybridization implicitly, which is  evidenced by their ``leakage'' to neighboring sites due to mixing of rare-earth 4$f$ states with Fe 3$d$, N 2$p$ and Li 2$s$ bands. The real-space Wannier functions of Fig.~\ref{wannier_orbitals} thus represent a convenient visualization of hybridization between rare-earth and other orbitals.  In order to quantify the amount of this admixture of the conduction band states we also expand those extended small-window Wannier orbitals $ |w^{\sigma}_m (\vk)\rangle$ in the basis of localized Wannier functions $|\tilde{w}^{\alpha\sigma'}_{lm'}(\vk)\rangle$ (labeled by spin $\sigma'$, orbital $l$ and magnetic $m'$ quantum numbers, as well as atomic site $\alpha$) constructed within a large energy window for all relevant bands. In Appendix~\ref{app_projection} we derive the corresponding projection operators relating $ |w^{\sigma}_m (\vk)\rangle$ and  $|\tilde{w}^{\alpha\sigma'}_{lm'}(\vk)\rangle$.  We employ it to extract the corresponding  contribution $\tilde{\rho}^{m\sigma}_{\alpha l}(\omega)$  of the shell $l$ on the site $\alpha$
 into the spectral function  of the ``small-window'' 4$f$ orbital $\sigma m$.

\begin{figure}
\begin{minipage}[b][12cm]{0.9\linewidth}
\subfloat{\includegraphics[width=\linewidth]{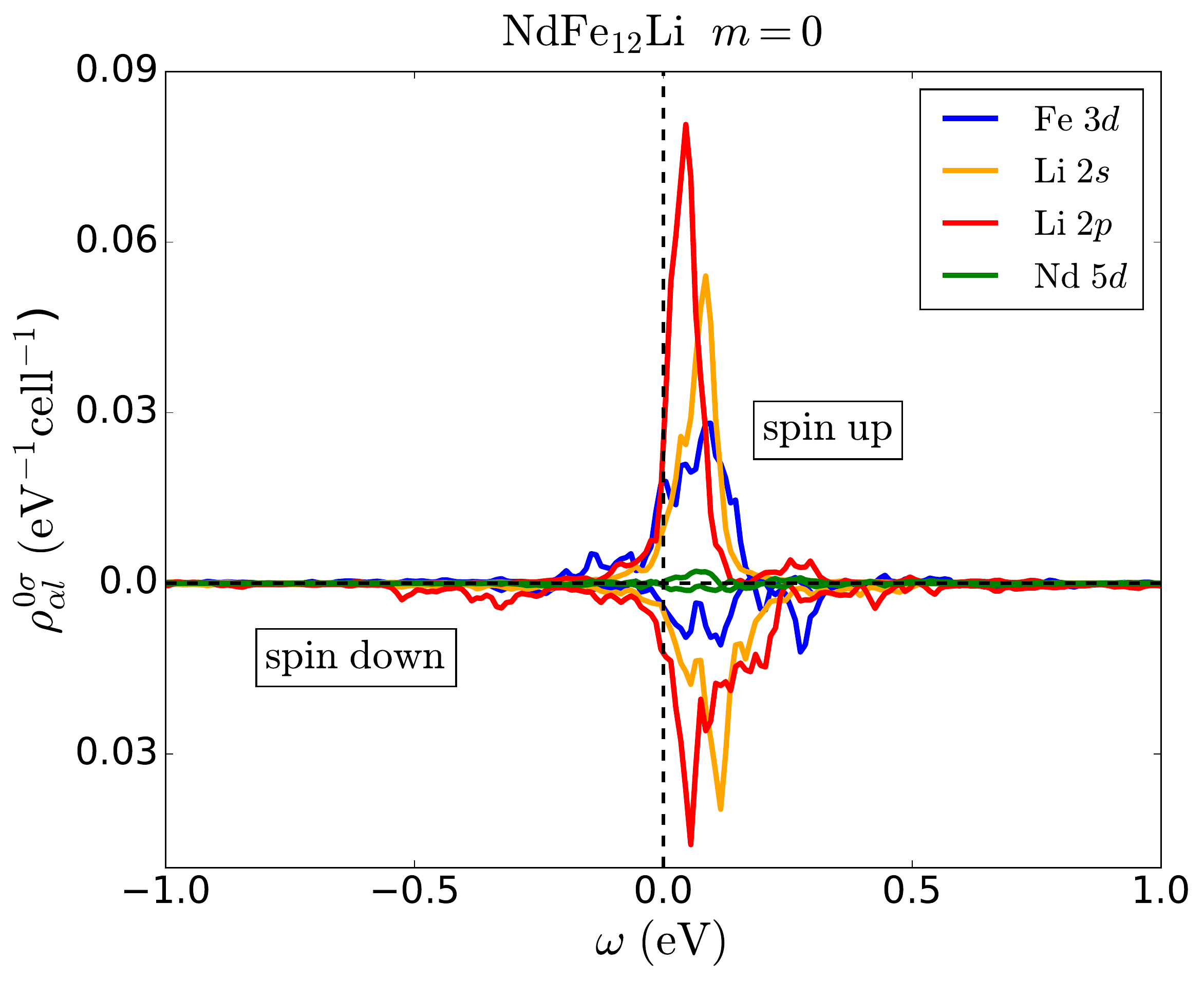}\label{sfig:projected_DOS_Li_0}}
\vfill{}
\subfloat{\includegraphics[width=\linewidth]{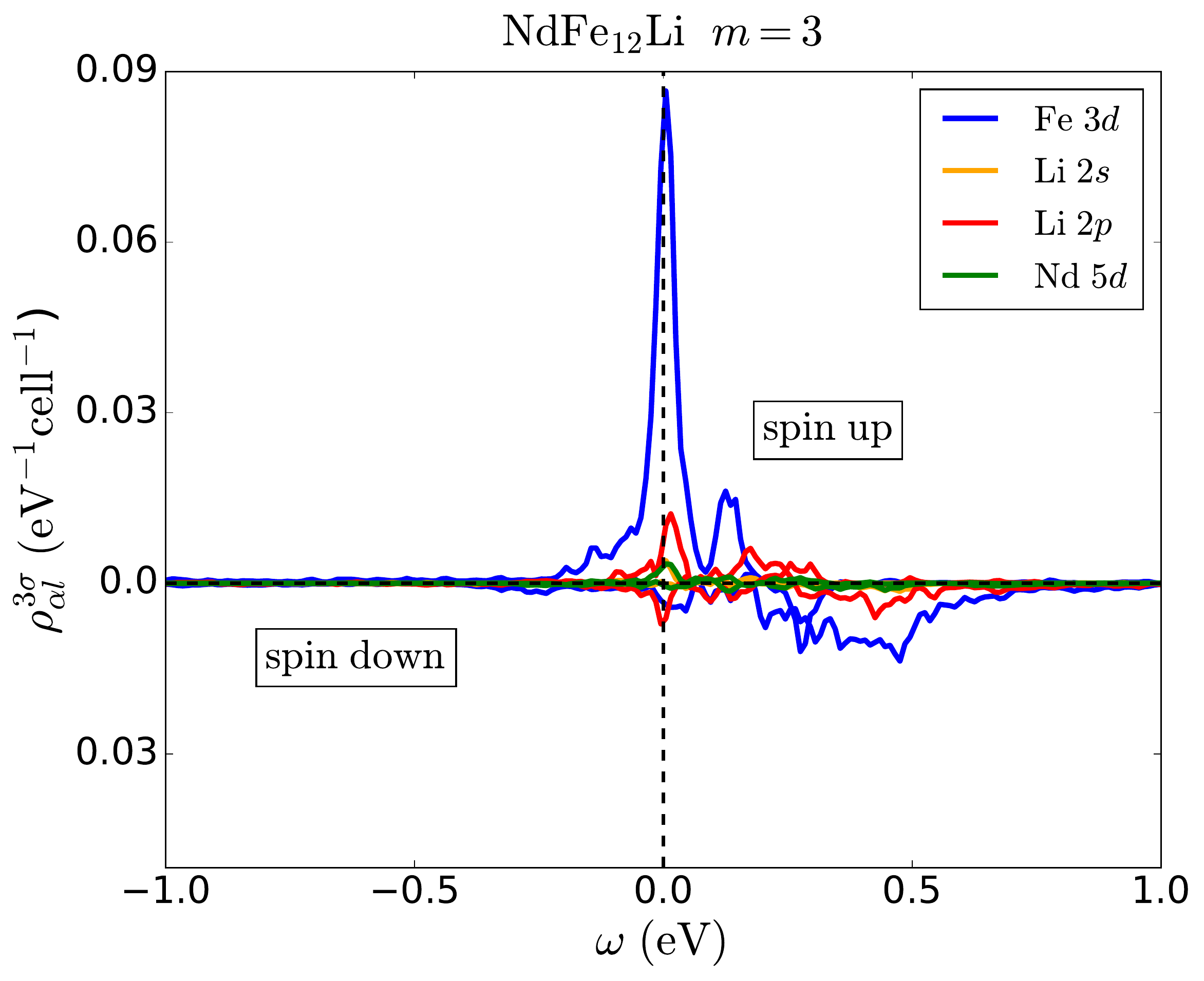}\label{sfig:projected_DOS_Li_3}}
\end{minipage}
\caption{a. Projected spectral functions $\rho^{0\sigma }_{\alpha l}(\omega)$ for the 4$f$ orbital with $m=$0 in NdFe$_{12}$Li, where the atom $\alpha$ and shell $l$ are given in the legend. The magnitude $\rho^{m\sigma }_{\alpha l}(\omega)$ indicates the amount of admixture of the character $\alpha l$ into a given 4$f$ orbital, for its precise formulation see the text. b. The same for the 4$f$ orbital $m=$3 }
\label{projected_DOS_Li}
\end{figure}

\begin{figure}
\begin{minipage}[b][12cm]{0.9\linewidth}
\subfloat{\includegraphics[width=\linewidth]{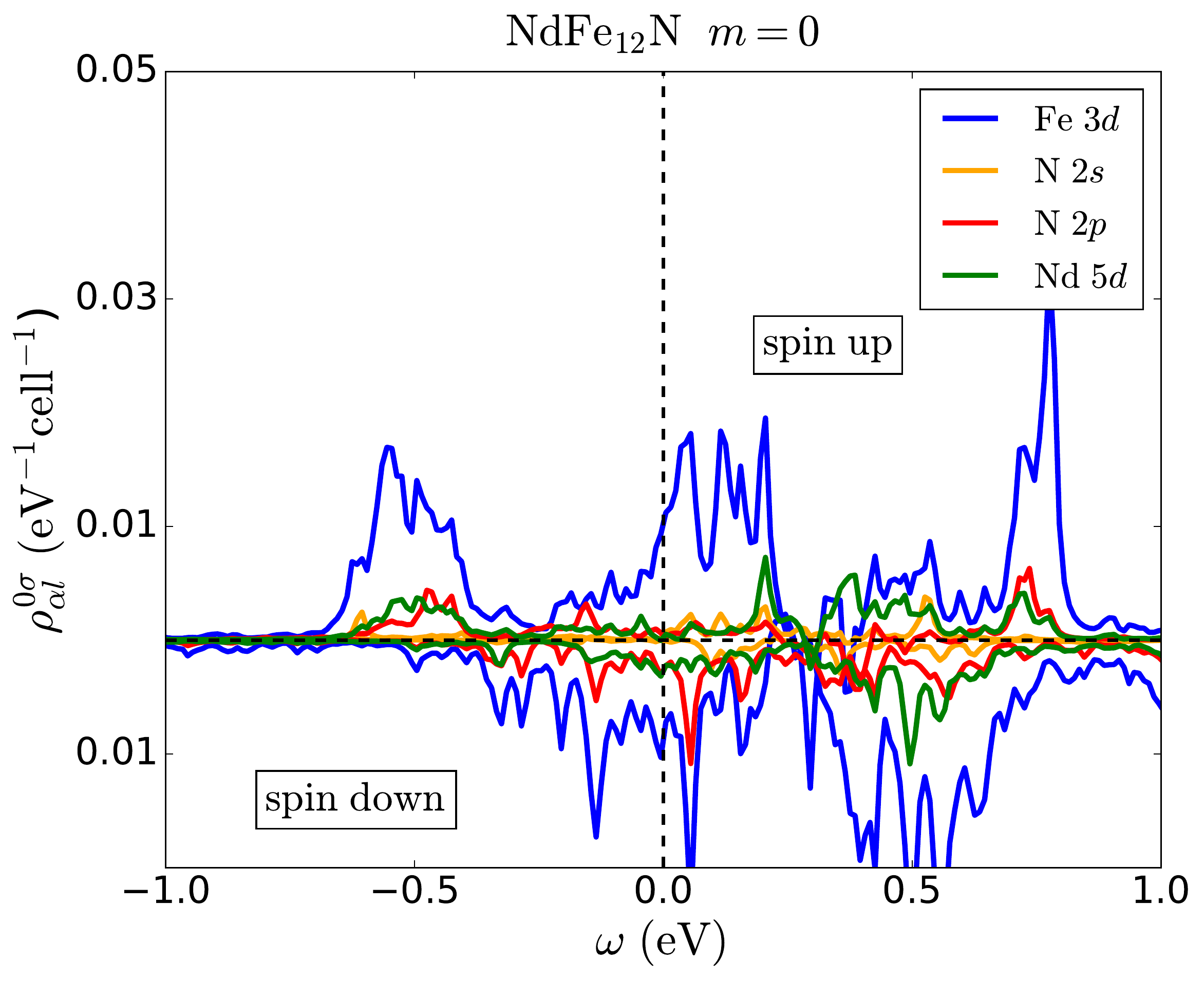}\label{sfig:NdFe12N_0}}
\vfill{}
\subfloat{\includegraphics[width=\linewidth]{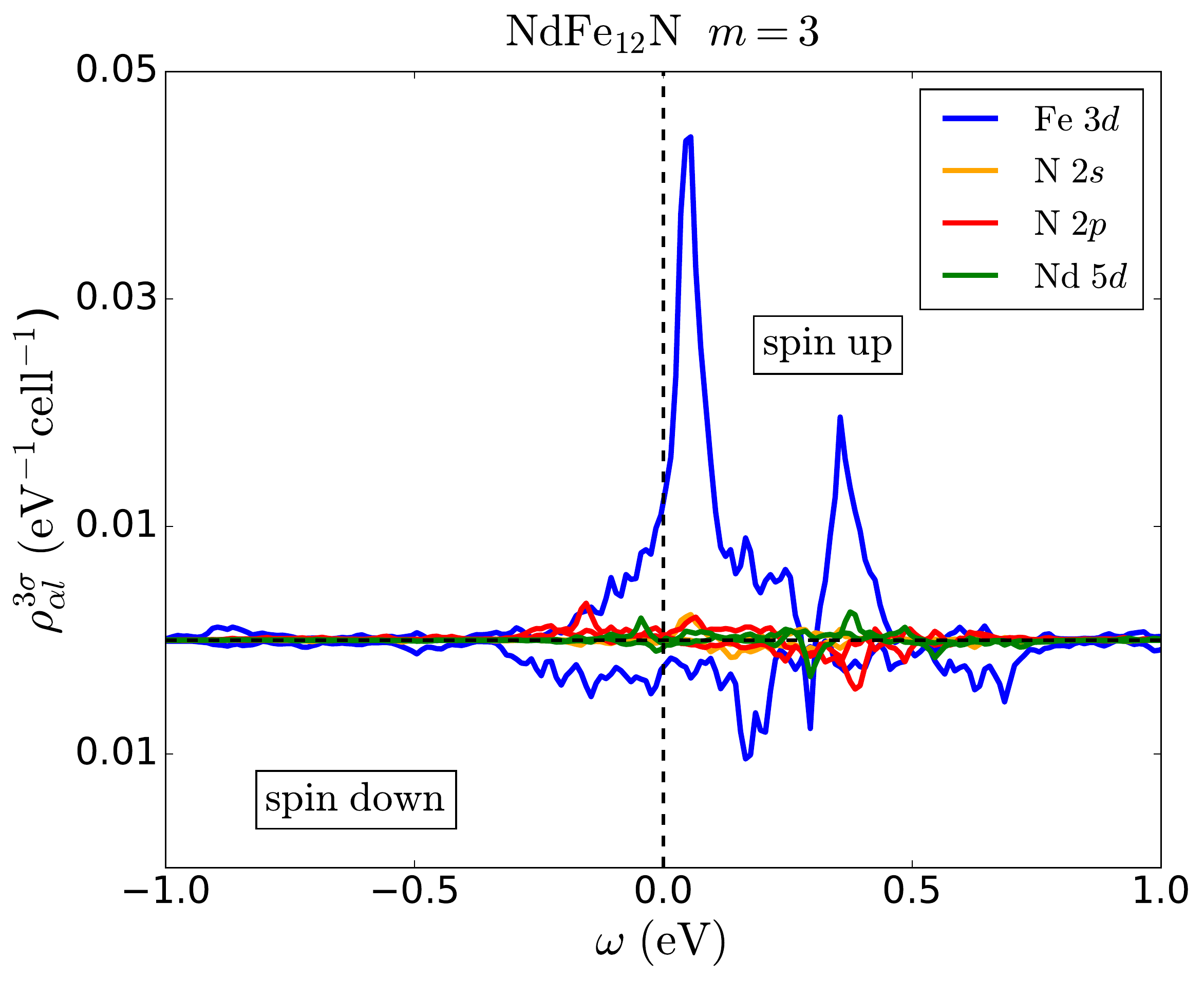}\label{sfig:NdFe12N_3}}
\end{minipage}
\captionsetup[minipage]{justification=raggedright}
\caption{a. Projected spectral functions $\rho^{0\sigma }_{\alpha l}(\omega)$ for the 4$f$ orbital with $m=$0 in NdFe$_{12}$N. For the notation see caption of Fig.~\ref{projected_DOS_Li}. b. The same for the 4$f$ orbital $m=$3, $\rho^{3\sigma }_{\alpha l}(\omega)$  }
\label{projected_DOS_N}
\end{figure}

The comparison of $\rho^{m\sigma }_{\alpha l}$ for the orbital $m=0$ and  $m=3$ are shown in Figs. \ref{projected_DOS_N} and \ref{projected_DOS_Li} for NdFe$_{12}$Li and NdFe$_{12}$N, respectively. One may notice in Fig.~\ref{sfig:projected_DOS_Li_0} that Nd $f_{z^3}$ ($m=0$) in NdFe$_{12}$Li exhibits a strong hybridization with Li 2$s$ and 2$p$; their contribution is significantly larger than the admixture of Fe 3$d$ states. We further observe that spin up states are hybridizing more strongly than spin down states. In contrast, in the same compound for $m=3$ (Fig.~\ref{sfig:projected_DOS_Li_3}), there is a peak of hybridization with Fe states but barely any with the Li 2$s$ and 2$p$ ones. The same difference, but much less pronounced, is noticeable in the case of NdFe$_{12}$N, see Fig.~\ref{projected_DOS_N}. Hence, one may conclude, that the effect of the hybridization with the interstitial on the CF is much larger for Li than for N. In the latter case the electrostatic shift due to the negative charge on N seems to play the leading role.

\section{Conclusion}
In conclusion, we propose a novel first-principles approach for calculating crystal and exchange fields in rare-earth systems. This approach is formulated within the DFT+DMFT framework with local correlations on the rare-earth 4$f$ shell treated within the quasi-atomic Hubbard-I approximation. The 4$f$ states are represented by Wannier functions constructed from a narrow energy range of Kohn-Sham states of mainly 4$f$ character. We employ a charge-density averaging that suppresses the contribution due to the self-interaction of the 4$f$ orbitals to the one-electron Kohn-Sham potential. We thus reduce the effect of this unphysical self-interaction from the crystal-field splitting, while keeping non-spherical contributions to CFP from other bands. Similarly, by removing the contribution due to the 4$f$ magnetic density from the exchange-correlation potential we suppress its unphysical contribution to the exchange field at the rare-earth site.

The present approach is effectively free from adjustable parameters and can be applied to evaluate CFP in any localized lanthanide compound.
While in the present work we chose the value for the on-site interaction parameters $U$ and $J$, they can in principle be evaluated using constrained local-density or random-phase approximation\cite{aryasetiawan2004}. Moreover, we show that the crystal-field splitting exhibits a rather weak dependence on the value of $U$ chosen within a reasonable range for lanthanide 4$f$ shells (4 to 8 eV). Our choice for the local basis representing 4$f$ orbitals, namely, that we construct it from a narrow range of Kohn-Sham bands with heavy 4$f$ character, is physically motivated as it allows for the impact of the hybridization on the CFP being included within DFT+Hub-I.  

We apply this approach to evaluate the crystal  and exchange-field potentials as well as the resulting single-ion magnetic anisotropies in several rare-earth hard-magnetic intermetallics. First, we verify that our {\it ab initio } scheme reproduces the measured crystal-field parameters (CFP) in the well-known hard magnet SmCo$_5$. We subsequently apply it to prospective rare-earth hard magnetic intermetallics of the $R$Fe$_{12}X$ family (where $R=$Nd, Sm and $X$ can be N, Li or vacancy). Our calculations reproduce the strong out-of-plane anisotropy of NdFe$_{12}$N due to a large positive value of the key CFP $A_2^0\langle r^2\rangle$ induced by insertion of N. Interestingly, we find that  interstitial Li has a strong opposite effect, leading to a large negative value of $A_2^0\langle r^2\rangle$. 
We thus predict a strong out-of-plane anisotropy in the hypothetical compound  SmFe$_{12}$Li. 
We also find the anisotropy in SmFe$_{12}$Li to persist to higher temperatures as compared to NdFe$_{12}$N. 
Hence, Sm-based compounds may represent interesting candidates for hard-magnetic applications.
Of course, the thermodynamic stability of SmFe$_{12}$Li and technological feasibility of Li doping still need to be demonstrated by future studies. 

We analyze the effect of N and Li interstitials on $A_2^0\langle r^2\rangle$ by evaluating the Bader charges as well as by studying the leakage of 4$f$ Wannier orbitals to interstitial sites and quantifying the 4$f$ hybridization with N 2$p$ and Li 2$s$ states.

Extensions of the present approach beyond the Hubbard-I approximation are promising for applications to other rare-earth intermetallics. In particular, a similar DFT+DMFT technique suppressing subtle self-interaction and double-counting contributions to the Kohn-Sham potential might be necessary to study, for example, the impact of a spin-polarized transition-metal sublattice on heavy-fermion behavior in Yb-based intermetallics\cite{Mazet2013,Mazet2015}. 

\begin{acknowledgements}
This work was supported by IDRIS/GENCI Orsay under project t2017091393, 
the ECOS-Sud grant A13E04,
the French Agence Nationale de la Recherche in the framework of the international collaborative DFG-ANR project RE-MAP
and the European Research Council under its Consolidator Grant scheme (project 617196).
LP acknowledges the financial support of the Ministry of Education and Science of the Russian Federation (Grant No. 14.Y26.31.0005).
TM acknowledges the financial support by the Ministry of Education, Culture, Sports, Science and Technology (MEXT) of Japan as a social and scientific priority issue (Creation of new functional devices and high-performance materials to support next-generation industries, CDMSI) to be tackled by using post-K computer, and also by MEXT KAKENHI Grant Number 16H06345.
We thank Tilmann Hickel, Halil Soezen, Antoine Georges and Dominique Givord  for useful discussions.
\end{acknowledgements}

\appendix  
\section{crystal-field parameters in SmCo$_5$}
\label{SmCo5}

SmCo$_5$ has been studied more extensively than other hard magnetic rare-earth intermetallics, so ample experimental data is available in this case. 
In particular, several groups estimated the CF parameters using inelastic neutron scattering or magnetization measurements. 
Therefore, this compound is a good benchmark to test our approach. 
SmCo$_5$ has already been studied within DFT+Hub I to evaluate its ground state magnetization and photoemission spectra\cite{graanas2012charge}, but the CF parameters were not calculated in this work.

\begin{figure}
\includegraphics[width=0.95\linewidth]{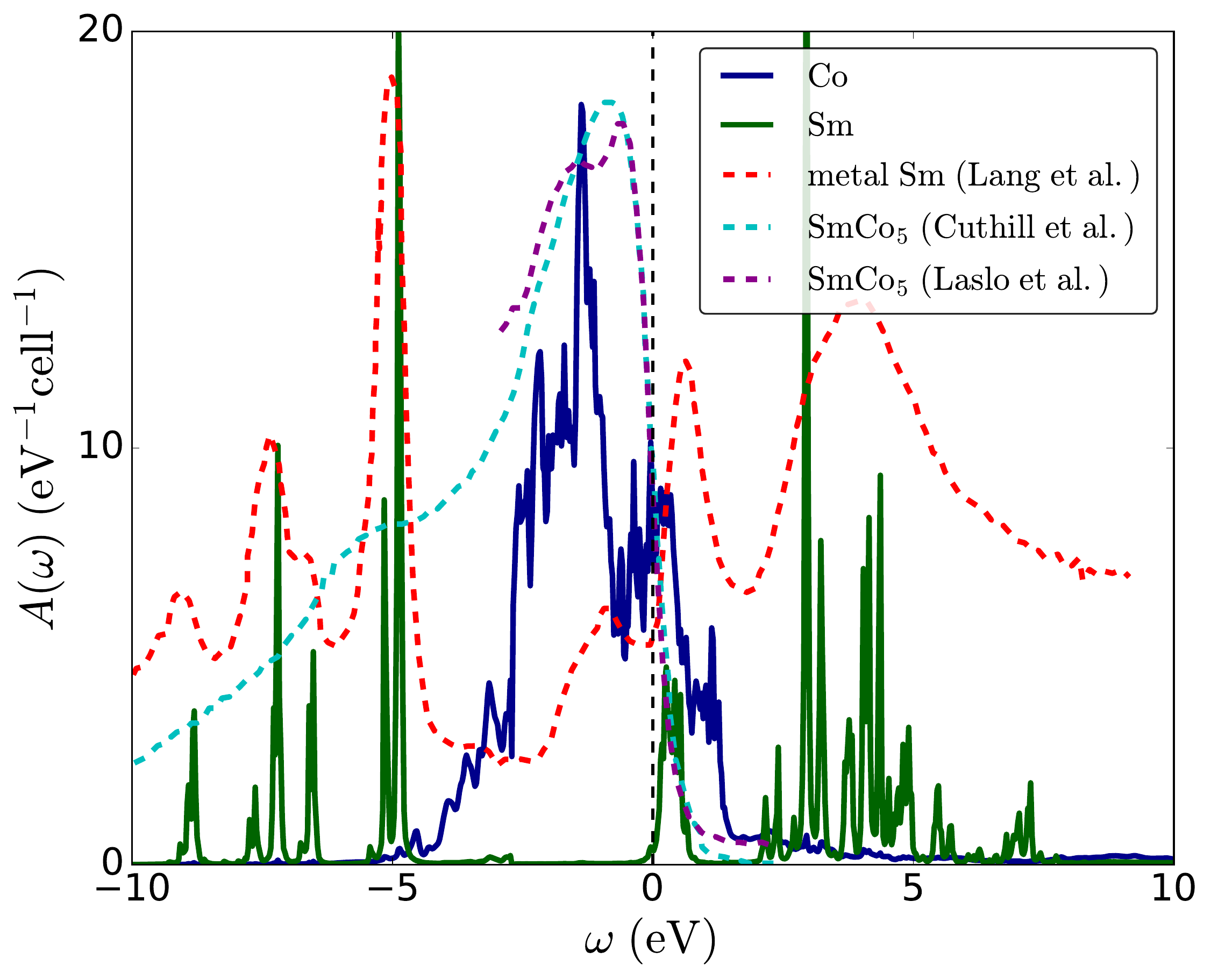}
\caption{Atom-resolved spectral function of SmCo$_5$ obtained within self-consistent spin-polarized DFT+Hubbard I (full lines).
The two inequivalent Co types are summed to give the total Co 3$d$ spectral function.
The occupied part of the experimental spectrum of SmCo$_5$ (light blue from Ref.~\onlinecite{cuthill1974x} and purple from Ref.~\onlinecite{laslo2014effects}) and the full experimental spectrum of metal Sm (red, from Ref.~\onlinecite{lang1981study}) are shown for comparison in dotted lines.
}
\label{fig:spectral_function_smco5}
\end{figure}

The calculated spectral function of SmCo$_5$ is shown in Fig.~\ref{fig:spectral_function_smco5}.
We find a total magnetic moment on the 4$f$ shell of Sm of 0.42~$\mu_B$, antiparallel with the Co moments.
This compares well with the measured value of 0.38~$\mu_B$ at 4.2~K\cite{givord1979}.

The calculated CFP and exchange fields for SmCo$_5$ are listed in Table~\ref{tab:smco5}, together with experimental data. 
The calculations on SmCo$_5$ are done at the experimental lattice constants.

\begin{table*}
	\centering
	\resizebox{\linewidth}{!}{
	\begin{tabular}{ c | c | c  c | c | c | c | c | c | c | c | c |}
		\cline{2-12}
	         & PM & \multicolumn{2}{c|}{FM} & Tils {\it et al.} & Zhao {\it et al.} & Givord {\it et al.}  & Sankar {\it et al.} &  Bushow {\it et al.} & Richter {\it et al.} &  Hummler {\it et al.} &  Novak {\it et al.} \\ 
	         &        &   $\uparrow$  & $\downarrow$ & $^\dagger$Ref. \onlinecite{tils1999}  & $^\dagger$Ref. \onlinecite{tie1991}  &  $^\dagger$Ref. \onlinecite{givord1979} &  $^\dagger$Ref. \onlinecite{sankar1975} &  $^\dagger$Ref. \onlinecite{buschow1974} & Ref. \onlinecite{richter1995} & Ref. \onlinecite{Hummler1996} & Ref. \onlinecite{novak1994} \\
	         \hline
		    \multicolumn{1}{| c|}{$A_2^0\langle r^2\rangle$} & -140  & -313 & -262 &  -326 & -330 & -200  & -420  &  -180 &  -755 & -509 & -160 \\
		    \multicolumn{1}{| c|}{$A_4^0\langle r^4\rangle$} &  -40 & -40 & -55 & -  & -45 & 0 &  25 & 0 & -37 & -20 & -33 \\
		    \multicolumn{1}{| c|}{$A_6^0\langle r^6\rangle$} &  33  & 35  &  25  & -  & 0 & 50 & 0 & 0 & 11 & 2  & 40 \\
		    \multicolumn{1}{| c|}{$A_6^6\langle r^6\rangle$} &  -684 &  -731 & -593 & - & 0 & 0 & 6 & 0 & 290 & -55 & 168\\
		    \hline
		    \multicolumn{1}{| c|}{$B_{\mathrm{ex}}$ ($T$)} & -  &  \multicolumn{2}{c | }{227} &  260 & 327.5 & 260.5 & 357 & 298 & -  & 279 & - \\
		    \hline
	\end{tabular}}
	\caption{\label{tab:smco5} Calculated CF parameters in ferromagnetic (FM) and paramagnetic (PM)  SmCo$_5$ in kelvin. For the FM case we list the CF parameters for each spin direction. The exchange field in the FM phase (in tesla) is also listed. For comparison, measured and calculated values from several groups are also given. 
References corresponding to an experimental work are denoted by the symbol $^\dagger$.}
\end{table*}

One may notice that the CF parameter $A_2^0\langle r^2\rangle$  exhibits a strong dependence on the spin polarization; it is about twice larger in the FM phase. For other CF parameters this dependence is small.

Our results for $A_2^0\langle r^2\rangle$ are in good agreement with the experimental (rather wide) range from about -180 to -420 K. The calculated $B_{\mathrm{ex}}$ also agrees rather well with the experimental range from 260 to 360 T. One may notice that the experimental measurements were performed at room temperature, hence, in ferromagnetic SmCo$_5$. Also, the most recent experimental values\cite{tils1999} of $A_2^0\langle r^2\rangle$  are in very good agreement with our results for the FM phase.

The main discrepancy between our theoretical and experimental CFP lies in  the large value that we find for $A_6^6\langle r^6\rangle$. The high-order CF parameters are usually assumed to be rather small in SmCo$_5$. However, as noted in Ref. \onlinecite{Hummler1996}, experimental inelastic neutron and susceptibility data are not particularly sensitive to those high-order parameters. Hence, they are often assumed to be small from the onset and neglected in the fitting procedure.

In order to facilitate the reproducibility of our calculations, we provide below the one-electron Hamiltonian of Eq.~\ref{H1el_HI} for a converged, ferromagnetic calculation of SmCo$_5$, used to obtain the CFP of Table~\ref{tab:smco5}.
\begin{widetext}
$$
\hat{H}_{\mathrm{1el}}^{\uparrow\uparrow} = 
\quad
\begin{pmatrix}
-26.5763 & 0 & 0 & 0 & 0 & 0 & 0.0235 \\
0 & -26.5003 & 0 & 0 & 0 & 0 & 0  \\
0 & 0 & -26.4465 & 0 & 0 & 0 & 0  \\
0 & 0 & 0 & -26.3511 & 0 & 0 & 0  \\
0 & 0 & 0 & 0 & -26.2796 & 0 & 0  \\
0 & 0 & 0 & 0 & 0 & -26.1675 & 0  \\
0.0235 & 0 & 0 & 0 & 0 & 0 & -26.0800
\end{pmatrix}
\quad
$$
$$
\hat{H}_{\mathrm{1el}}^{\downarrow\downarrow} = 
\quad
\begin{pmatrix}
-26.1093 & 0 & 0 & 0 & 0 & 0 & 0.0191 \\
0 & -26.1926 & 0 & 0 & 0 & 0 & 0  \\
0 & 0 & -26.3023 & 0 & 0 & 0 & 0  \\
0 & 0 & 0 & -26.3802 & 0 & 0 & 0  \\
0 & 0 & 0 & 0 & -26.4688 & 0 & 0  \\
0 & 0 & 0 & 0 & 0 & -26.5252 & 0  \\
0.0191 & 0 & 0 & 0 & 0 & 0 & -26.6069
\end{pmatrix}
\quad
$$
$$\hat{H}_{\mathrm{1el}}^{\uparrow\downarrow} = (\hat{H}_{\mathrm{1el}}^{\downarrow\uparrow})^\dagger =
\quad
\begin{pmatrix}
0 & 0.2032 & 0 & 0 & 0 & 0 & 0 \\
0 & 0 & 0.2632 & 0 & 0 & 0 & 0 \\
0 & 0 & 0 & 0.2893 & 0 & 0 & 0 \\
0 & 0 & 0 & 0 & 0.2886 & 0 & 0 \\
0 & 0 & 0 & 0 & 0 & 0.2633 & 0 \\
-0.0002 & 0 & 0 & 0 & 0 & 0 & 0.2036 \\
0 & - 0.0003 & 0 & 0 & 0 & 0 & 0
\end{pmatrix}
\quad
$$
\end{widetext}
where
$$
\hat{H}_{\mathrm{1el}}= 
\begin{pmatrix}
\hat{H}_{\mathrm{1el}}^{\uparrow\uparrow} & \hat{H}_{\mathrm{1el}}^{\uparrow\downarrow} \\
\hat{H}_{\mathrm{1el}}^{\downarrow\uparrow} & \hat{H}_{\mathrm{1el}}^{\downarrow\downarrow}
\end{pmatrix}
$$

\section{Eigenenergies and eigenstates of  the 4$f$ shells in SmCo$_5$ and $R$Fe$_{12}X$}
\label{WF_tables}

In this Appendix we present the actual converged eigenfunctions and eigenstates of the 4$f$ shell  obtained within our DFT+HubI approach. 
In tables \ref{tab:atomic_levels_Nd} and \ref{tab:atomic_levels_Sm}, we list those eigenenergies and the corresponding wavefunctions for the ground-state multiplet, as well as for the lowest-energy state of the first exited multiplet, in all the materials considered. The eigenenergies are given with respect to the ground state.  The eigenvalues are expanded in the basis of total angular momentum $J$ as $\sum_{J,m_J} a(J,m_J) |J;m_J\rangle$; we include all contributions with $|a(J,m_J)| > 0.03$.

One sees that the eigenstates of Sm belonging to the ground-state multiplet feature a rather significant admixture of the exited $J=7/2$ multiplet. The inter-multiplet mixing is markedly lower in the case of Nd.

To our awareness, only 4$f$ eigenstates in SmCo$_5$ have been measured to date. Our calculated intra- and inter-multiplet splittings are in good agreement with the results of  of Tils \emph{et al.} (Ref.~\onlinecite{tils1999}) and Givord \emph{et al.} (Refs.~\onlinecite{givord1979}), see the lowest panel of Table~\ref{tab:atomic_levels_Sm}.  Moreover, the actual eigenstates and their order are also in very good agreement with the magnetic form-factor measurements\cite{givord1979,Laforest1981}, especially for the lowest-energy states.
\footnote{The opposite sign of the $J=7/2$ contributions in Refs.~\onlinecite{givord1979} and ~\onlinecite{Laforest1981} with respect to ours is due to their choice for the coordination frame, with the $z$ axis aligned to the spin moment of Sm. 
We align $z$ with the spin of Co sublattice, hence opposite to the Sm spin. 
Flipping the direction of the $z$-axis leads to the change in the relative sign between the $J=5/2$ and  $J=7/2$ contributions.
There is a mismatch between the signs of $m_J$ in $| J,m_J \rangle$ reported in Refs.~\onlinecite{givord1979} and ~\onlinecite{Laforest1981}.
The correct signs are reported in Ref.~\onlinecite{Laforest1981} (D. Givord, private communication).}

\renewcommand{\arraystretch}{2}

\begin{table}
	\centering
	\scriptsize
	\tiny
	\begin{tabularx}{1\linewidth}{  c | c |X |}
		\hline
		\multicolumn{3}{| c|}{{\small NdFe$_{12}$}}  \\
		\hline
		\multicolumn{1}{| c|}{GSM} &
		0 & $0.996|9/2;+9/2\rangle+0.081|11/2;+9/2\rangle$ \\
		\multicolumn{1}{| c|}{} &       
		8 & $0.995|9/2;+5/2\rangle+0.094|11/2;+5/2\rangle$ \\
		\multicolumn{1}{| c|}{} & 
		18 & $0.989|9/2;+3/2\rangle+0.145|11/2;+3/2\rangle-0.037|9/2;-5/2\rangle$  \\
		\multicolumn{1}{| c|}{} & 
		18 & $0.999|9/2;+7/2\rangle+0.042|11/2;+7/2\rangle$  \\
		\multicolumn{1}{| c|}{} & 
		39 & $0.989|9/2;+1/2\rangle+0.145|11/2;+1/2\rangle-0.032|9/2;-7/2\rangle$  \\
		\multicolumn{1}{| c|}{} & 		      
		51 & $0.996|9/2;-3/2\rangle+0.085|11/2;-3/2\rangle$  \\
		\multicolumn{1}{| c|}{} & 		      
		51 & $0.994|9/2;-1/2\rangle+0.105|11/2;-1/2\rangle$  \\
		\multicolumn{1}{| c|}{} & 		      
		56 & $0.992|9/2;-5/2\rangle+0.121|11/2;-5/2\rangle+0.038|9/2;+3/2\rangle$   \\
		\multicolumn{1}{| c|}{} &		      
		82 & $0.985|9/2;-7/2\rangle+0.166|11/2;-7/2\rangle+0.033|9/2;+1/2\rangle$  \\
		\multicolumn{1}{| c|}{} & 		      
		88 & $0.998|9/2;-9/2\rangle+0.057|11/2;-9/2\rangle$  \\
		\hline
		\multicolumn{1}{| c|}{FEM} & 
		309 & $0.987|11/2;+5/2\rangle+0.126|13/2;+5/2\rangle-0.093|9/2;+5/2\rangle$ \\  
		\multicolumn{1}{| c|}{ } &  & $-0.046|11/2;-3/2\rangle$ \\
		\hline
		\hline
		\multicolumn{3}{ |c|}{{\small NdFe$_{12}$N}}  \\
		\hline
		\multicolumn{1}{| c|}{GSM} &
		 0 & $0.994|9/2;+9/2\rangle+0.092|11/2;+9/2\rangle-0.054|9/2;+1/2\rangle$  \\
		 \multicolumn{1}{| c|}{} & 		 
		 32 & $0.998|9/2;+5/2\rangle+0.065|11/2;+5/2\rangle$  \\
		 \multicolumn{1}{| c|}{} & 		 
		 38 & $0.994|9/2;+3/2\rangle+0.106|11/2;+3/2\rangle$ \\
		 \multicolumn{1}{| c|}{} &  		 
		 41 & $0.997|9/2;+7/2\rangle-0.065|9/2;-1/2\rangle$ \\
		 \multicolumn{1}{| c|}{} & 
		 56 & $0.993|9/2;+1/2\rangle+0.101|11/2;+1/2\rangle+0.055|9/2;+9/2\rangle$ \\
		 \multicolumn{1}{| c|}{} & 
		 66 & $0.998|9/2;-3/2\rangle+0.052|11/2;-3/2\rangle$ \\
		 \multicolumn{1}{| c|}{} & 
		 66 & $0.966|9/2;-1/2\rangle-0.242|9/2;-9/2\rangle+0.064|9/2;+7/2\rangle+0.061|11/2;-1/2\rangle$ \\
		 \multicolumn{1}{| c|}{} & 
		 67 & $0.996|9/2;-5/2\rangle+0.077|11/2;-5/2\rangle$  \\
		 \multicolumn{1}{| c|}{} & 
		 73 & $0.970|9/2;-9/2\rangle+0.241|9/2;-1/2\rangle$ \\
		 \multicolumn{1}{| c|}{} & 
		 80 & $0.996|9/2;-7/2\rangle+0.093|11/2;-7/2\rangle$ \\
		\hline
		\multicolumn{1}{| c|}{FEM} & 		 		 
		 301 & $0.994|11/2;+11/2\rangle+0.091|13/2;+11/2\rangle-0.051|11/2;+3/2\rangle$  \\
		\hline
		\hline
		\multicolumn{3}{ |c|}{{\small NdFe$_{12}$Li}}  \\
		\hline
		\multicolumn{1}{| c|}{GSM} &
      0 & $0.993|9/2;+7/2\rangle+0.116|11/2;+7/2\rangle$  \\
      \multicolumn{1}{| c|}{} &       
      16 & $0.990|9/2;+5/2\rangle+0.137|11/2;+5/2\rangle$  \\
      \multicolumn{1}{| c|}{} &
      24 & $0.994|9/2;+9/2\rangle+0.105|9/2;+1/2\rangle$   \\
      \multicolumn{1}{| c|}{} &
      31 & $0.989|9/2;+3/2\rangle+0.137|11/2;+3/2\rangle-0.041|9/2;-5/2\rangle$  \\
      \multicolumn{1}{| c|}{} &
      40 & $0.983|9/2;+1/2\rangle+0.146|11/2;+1/2\rangle-0.107|9/2;+9/2\rangle$   \\
      \multicolumn{1}{| c|}{} &
      51 & $0.986|9/2;-1/2\rangle+0.163|11/2;-1/2\rangle$   \\
      \multicolumn{1}{| c|}{} &
      65 & $0.985|9/2;-3/2\rangle+0.166|11/2;-3/2\rangle$    \\
      \multicolumn{1}{| c|}{} &
      80 & $0.988|9/2;-5/2\rangle+0.148|11/2;-5/2\rangle+0.042|9/2;+3/2\rangle$   \\
      \multicolumn{1}{| c|}{} &
      93 & $0.990|9/2;-7/2\rangle+0.135|11/2;-7/2\rangle$   \\
      \multicolumn{1}{| c|}{} &
      126 & $0.984|9/2;-9/2\rangle+0.179|11/2;-9/2\rangle$    \\
	\hline
 		\multicolumn{1}{| c|}{FEM} & 	
      319 & $0.983|11/2;+7/2\rangle+0.134|13/2;+7/2\rangle-0.116|9/2;+7/2\rangle$ \\  
      \multicolumn{1}{| c|}{ } &  &$+0.042|11/2;-1/2\rangle$ \\ 
		\hline
	\end{tabularx}
	\caption{\label{tab:atomic_levels_Nd} Energies (in meV) and wavefunctions (expanded in the total angular momentum $J$ basis) of the atomic eigenstates in the ground state multiplet (GSM) and of the lowest eigenstate of the first excited multiplet (FEM) for NdFe$_{12}X$. }
\end{table}

\begin{table}
	\centering
	\tiny
	\begin{tabularx}{1\linewidth}{  c | c |X |}		
		\hline
		\multicolumn{3}{ |c|}{{\small SmFe$_{12}$}}  \\
		\hline
		\multicolumn{1}{| c|}{GSM} &
      0 & $0.986|5/2;+5/2\rangle+0.164|7/2;+5/2\rangle$     \\
      \multicolumn{1}{| c|}{} &           
      27 & $0.983|5/2;+3/2\rangle+0.183|7/2;+3/2\rangle$   \\
      \multicolumn{1}{| c|}{} &            
      47 & $0.968|5/2;+1/2\rangle+0.245|7/2;+1/2\rangle+0.043|9/2;+1/2\rangle$  \\
      \multicolumn{1}{| c|}{} &        
      69 & $0.977|5/2;-1/2\rangle+0.209|7/2;-1/2\rangle+0.041|9/2;-1/2\rangle$  \\
      \multicolumn{1}{| c|}{} &        
      85 & $0.976|5/2;-3/2\rangle+0.218|7/2;-3/2\rangle$   \\
      \multicolumn{1}{| c|}{} &       
      99 & $0.987|5/2;-5/2\rangle+0.159|7/2;-5/2\rangle+0.032|9/2;-5/2\rangle$   \\
	\hline
	\multicolumn{1}{| c|}{FEM}   &    
      191 & $0.987|7/2;+7/2\rangle+0.154|9/2;+7/2\rangle+0.034|7/2;-1/2\rangle$ \\ 
		\hline
		\hline
		\multicolumn{3}{ |c|}{{\small SmFe$_{12}$N}}  \\
		\hline
		\multicolumn{1}{| c|}{GSM} &
      0 & $0.997|5/2;+5/2\rangle+0.079|7/2;+5/2\rangle$\\
      \multicolumn{1}{| c|}{} &        
      2 & $0.990|5/2;+3/2\rangle+0.143|7/2;+3/2\rangle$ \\
      \multicolumn{1}{| c|}{} &      
      13 & $0.976|5/2;+1/2\rangle+0.216|7/2;+1/2\rangle$  \\
      \multicolumn{1}{| c|}{} &   
      31 & $0.982|5/2;-1/2\rangle+0.186|7/2;-1/2\rangle+0.035|9/2;-1/2\rangle$  \\
      \multicolumn{1}{| c|}{} &      
      48 & $0.978|5/2;-3/2\rangle+0.209|7/2;-3/2\rangle$  \\
      \multicolumn{1}{| c|}{} &      
      75 & $0.971|5/2;-5/2\rangle+0.238|7/2;-5/2\rangle$   \\
     	\hline
     	\multicolumn{1}{| c|}{FEM}   &   
      176 & $0.988|7/2;+5/2\rangle+0.129|9/2;+5/2\rangle-0.078|5/2;+5/2\rangle$  \\
		\hline
		\hline
		\multicolumn{3}{ |c|}{{\small SmFe$_{12}$Li}}  \\
		\hline
		\multicolumn{1}{| c|}{GSM} &
		0 & $0.977|5/2;+5/2\rangle+0.212|7/2;+5/2\rangle+0.037|9/2;+5/2\rangle$ \\
		\multicolumn{1}{| c|}{} &       		
		39 & $0.965|5/2;+3/2\rangle+0.256|7/2;+3/2\rangle+0.047|9/2;+3/2\rangle$ \\
		\multicolumn{1}{| c|}{} &       		
		67 & $0.960|5/2;+1/2\rangle+0.278|7/2;+1/2\rangle+0.042|9/2;+1/2\rangle$  \\
		\multicolumn{1}{| c|}{} &      		
		90 & $0.946|5/2;-1/2\rangle+0.318|7/2;-1/2\rangle+0.051|9/2;-1/2\rangle$  \\
		\multicolumn{1}{| c|}{} &       	
		116 & $0.941|5/2;-3/2\rangle+0.330|7/2;-3/2\rangle+0.068|9/2;-3/2\rangle$ \\
		\multicolumn{1}{| c|}{} &      		
		137 & $0.974|5/2;-5/2\rangle+0.221|7/2;-5/2\rangle+0.053|9/2;-5/2\rangle$ 	 \\
     	\hline
     	\multicolumn{1}{| c|}{FEM}   &       	
		202 & $0.977|7/2;+7/2\rangle+0.211|9/2;+7/2\rangle+0.035|11/2;+7/2\rangle$ \\ 
		\hline
		\hline
		\multicolumn{3}{ |c|}{{\small SmCo$_{5}$}}  \\
		\hline
		\multicolumn{1}{| c|}{GSM} &
        0 (0 / 0) & $0.984|5/2;+5/2\rangle+0.171|7/2;+5/2\rangle$\\
\multicolumn{1}{| c|}{} &       		  
        33 (31 / 28) & $0.983|5/2;+3/2\rangle+0.181|7/2;+3/2\rangle$  \\
\multicolumn{1}{| c|}{} &       		
        52 ( - / 47)& $0.973|5/2;+1/2\rangle+0.225|7/2;+1/2\rangle+0.033|9/2;+1/2\rangle$ \\
\multicolumn{1}{| c|}{} &       		
        71 ( - / 73)& $0.977|5/2;-1/2\rangle+0.209|7/2;-1/2\rangle+0.035|9/2;-1/2\rangle$ \\
\multicolumn{1}{| c|}{} &       		
        86 ( - / 91)& $0.977|5/2;-3/2\rangle+0.211|7/2;-3/2\rangle-0.032|9/2;+9/2\rangle$  \\
\multicolumn{1}{| c|}{} &       		
        95 ( - / 109) & $0.989|5/2;-5/2\rangle+0.122|7/2;-5/2\rangle+0.073|7/2;+7/2\rangle$\\
        \multicolumn{1}{| c|}{} & &  $+0.033|9/2;-5/2\rangle$ \\
     	\hline
     	\multicolumn{1}{| c|}{FEM}   &  
        188 (166 / - ) & $0.963|7/2;+7/2\rangle+0.187|7/2;-5/2\rangle+0.164|9/2;+7/2\rangle$\\
        \multicolumn{1}{| c|}{} & &  $-0.093|5/2;-5/2\rangle$  \\
		\hline
	\end{tabularx}
	\caption{\label{tab:atomic_levels_Sm} Same as Table~\ref{tab:atomic_levels_Nd}, for SmFe$_{12}$X and SmCo$_5$.
	For SmCo$_5$, between brackets next to the calculated energies: energies of the atomic eigenstates measured by Tils \emph{et al.} (left, Ref.~\onlinecite{tils1999}) and Givord \emph{et al.} (right, Ref.~\onlinecite{givord1979}). Note that only Tils \emph{et al.} directly measure the eigenenergies, while Givord  \emph{et al.} obtain them from an atomic Hamiltonian fitted to reproduce the measured magnetic form factor.}
\end{table}

\renewcommand{\arraystretch}{1.2}

\section{Importance of the charge averaging}
\label{appendix:imp_of_averaging}
In this Appendix we 
explicitly demonstrate the effect of averaging of 4$f$ charge density (eq.~\ref{eq_Gat}) by comparing the CFP calculated with and without this averaging (but in both cases the 4$f$ magnetic density is suppressed following Eq.~\ref{Nk_av}) in two materials, NdFe$_{12}$N and SmCo$_5$, that are known to have an out-of-plane magnetic anisotropy.

The corresponding values are displayed in Table \ref{tab:imp_of_averaging}.
One sees that the difference is largest for the lowest-order CFP $A_2^0\langle r^2\rangle$, where calculations without averaging lead to the wrong sign with respect to experiment (suggesting in-plane anisotropy in both cases). Hence, the proper averaging of 4$f$ charge density is crucial for a correct description of the single-ion anisotropy. For the higher order terms the difference between two approaches is smaller.
This suggests that the self-interaction contribution in the CFP has predominantly $l=2$ symmetry.

\begin{table}
	\centering
	\caption{\label{tab:imp_of_averaging}  crystal-field parameters and exchange field in NdFe$_{12}$N and SmCo$_5$ in the ferromagnetic phase, calculated with and without averaging over the ground state multiplet. 
	}
	\begin{tabularx}{1\linewidth}{c|  Y Y  |Y Y   |Y Y   |Y Y   | }
		\cline{2-9}
		&  \multicolumn{4}{c|}{NdFe$_{12}$N } & \multicolumn{4}{c |}{SmCo$_{5}$}   \\
		\cline{2-9}
		& \multicolumn{2}{c|}{with} & \multicolumn{2}{c|}{without}  & \multicolumn{2}{c|}{with} & \multicolumn{2}{c|}{without}   \\
		&  $\uparrow$  & $\downarrow$ &      $\uparrow$  & $\downarrow$ &  $\uparrow$  & $\downarrow$ &      $\uparrow$  & $\downarrow$   \\
		\hline
		\multicolumn{1}{ |c|}{$A_2^0\langle r^2\rangle$} & 477  & 653   & -190  & 26    &  -313   & -262     & 278   & 331    \\
		\multicolumn{1}{ |c|}{$A_4^0\langle r^4\rangle$} & 75    & 112   & 30      & 82    &  -40     & -55       & -30   & -37     \\
                 \multicolumn{1}{ |c|}{$A_4^4\langle r^4\rangle$} & -105 & -141 & -65     & -124 &  0        & -0        &   0     &  0        \\
		\multicolumn{1}{ |c|}{$A_6^0\langle r^6\rangle$} &  32   & 63     & 27      & 64    &  35      & 25        &  38   &  25      \\
		\multicolumn{1}{ |c|}{$A_6^4\langle r^6\rangle$} &  -65  &  -91   & -61    & -112  & 0         & 0         &  0      &  0        \\
		\multicolumn{1}{ |c|}{$A_6^6\langle r^6\rangle$} &      0  &  -0    & 0       & 0       & -731    & -593    &  -945 &  -806   \\
		\hline
		\multicolumn{1}{ |c|}{$B_{\mathrm{ex}}$ ($T$)} & \multicolumn{2}{c|}{217} & \multicolumn{2}{c|}{206} & \multicolumn{2}{c|}{227} & \multicolumn{2}{c|}{235}   \\
		\hline
	\end{tabularx}
\end{table}

\section{Dependence of results on Coulomb $U$ and Hund's $J_H$}
\label{app_U_J}
To perform DFT+DMFT calculations, we have to choose a value for the on-site screened Coulomb interaction parameter $U$ and for the Hund's coupling parameter $J_H$. Several methods have been developed in order to compute those parameters from first principles, most notably the constrained local density approximation\cite{dederichs1984ground} and, more recently, the constrained random phase approximation\cite{aryasetiawan2004}. 

\begin{figure}
\includegraphics[width=0.9\linewidth]{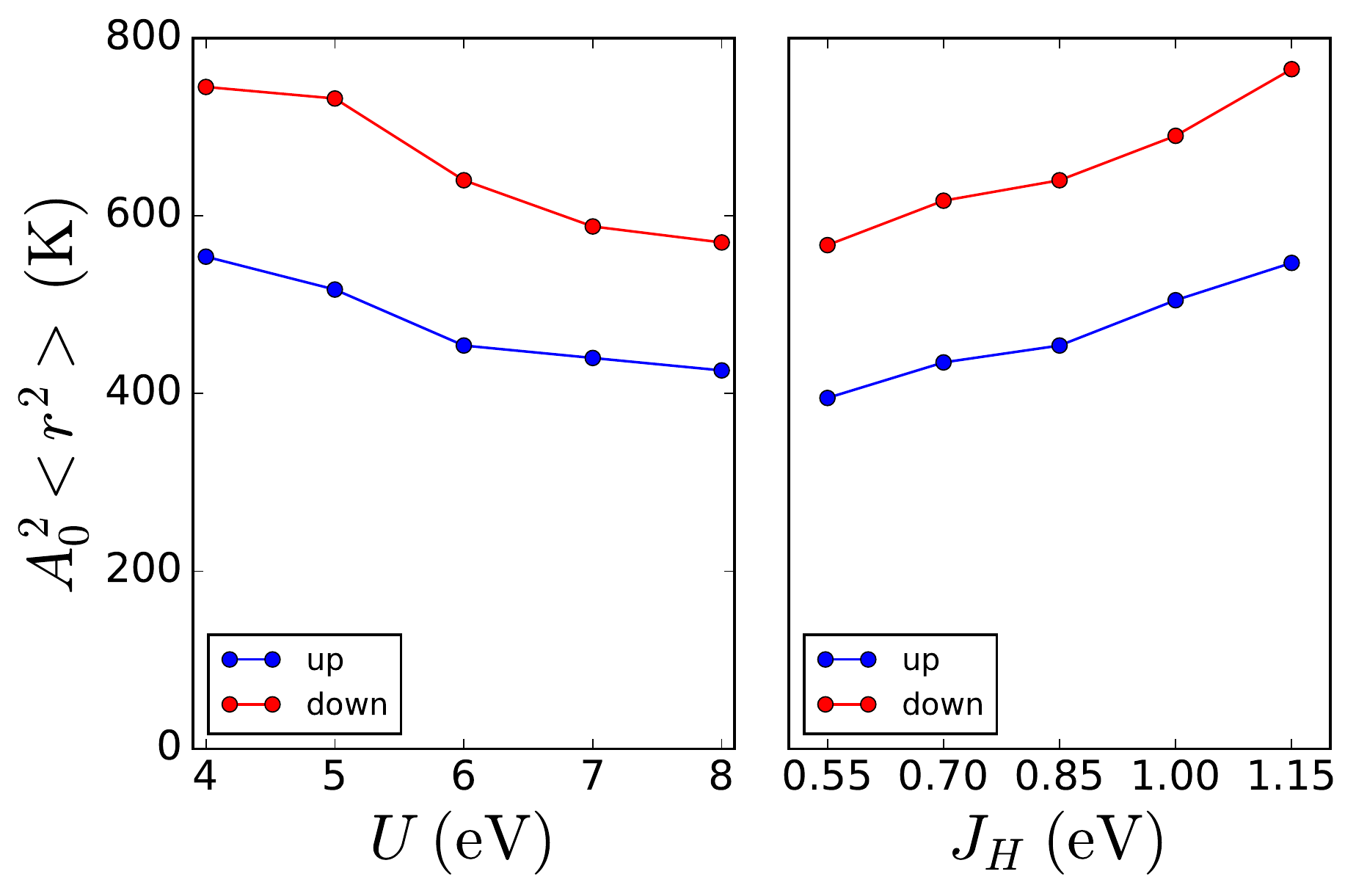}
\caption{CFP $A_2^0\langle r^2\rangle$ in NdFe$_{12}$N as a function of $U$ for $J_H = 0.85$~eV (left-hand panel), and as a function of $J_H$ for $U=6$~eV (right-hand panel). Our reference values are ($U=6$~eV, $J_H = 0.85$~eV).}
\label{U_J_dep}
\end{figure}

In the present work, however, we do not attempt a first principles determination. We use $U=6$~eV and $J_H=0.85$~eV because these values have given satisfactory results in other calculations on rare-earth materials\cite{Pourovskii2007,Locht2016}.
They are also in line with reported values calculated from first principles\cite{Nilsson2013}. Nevertheless, it is preferable that results obtained by our calculation scheme do not depend too strongly on the value of $U$ and $J_H$. In Fig.~\ref{U_J_dep}, we show that the dependence of the CFP $A_2^0\langle r^2\rangle$ in NdFe$_{12}$N is very moderate, as long as the values of U and J are chosen within a reasonable ranges for rare-earth ions. 

Furthermore, we observe that smaller values of $U$ lead to slightly larger values of $A_2^0\langle r^2\rangle$: this is not surprising if we keep in mind that a large $U$ is favorable to a strong localization of the 4$f$ electrons, hence to a weaker coupling to the crystal-field. 

\section{Dependence of results on window size}
\label{app_window_size}
\begin{figure}
\begin{minipage}[t]{0.95\linewidth}
\subfloat[{$\left[-20,20\right]$~eV window} ]{\includegraphics[width=0.5\linewidth]{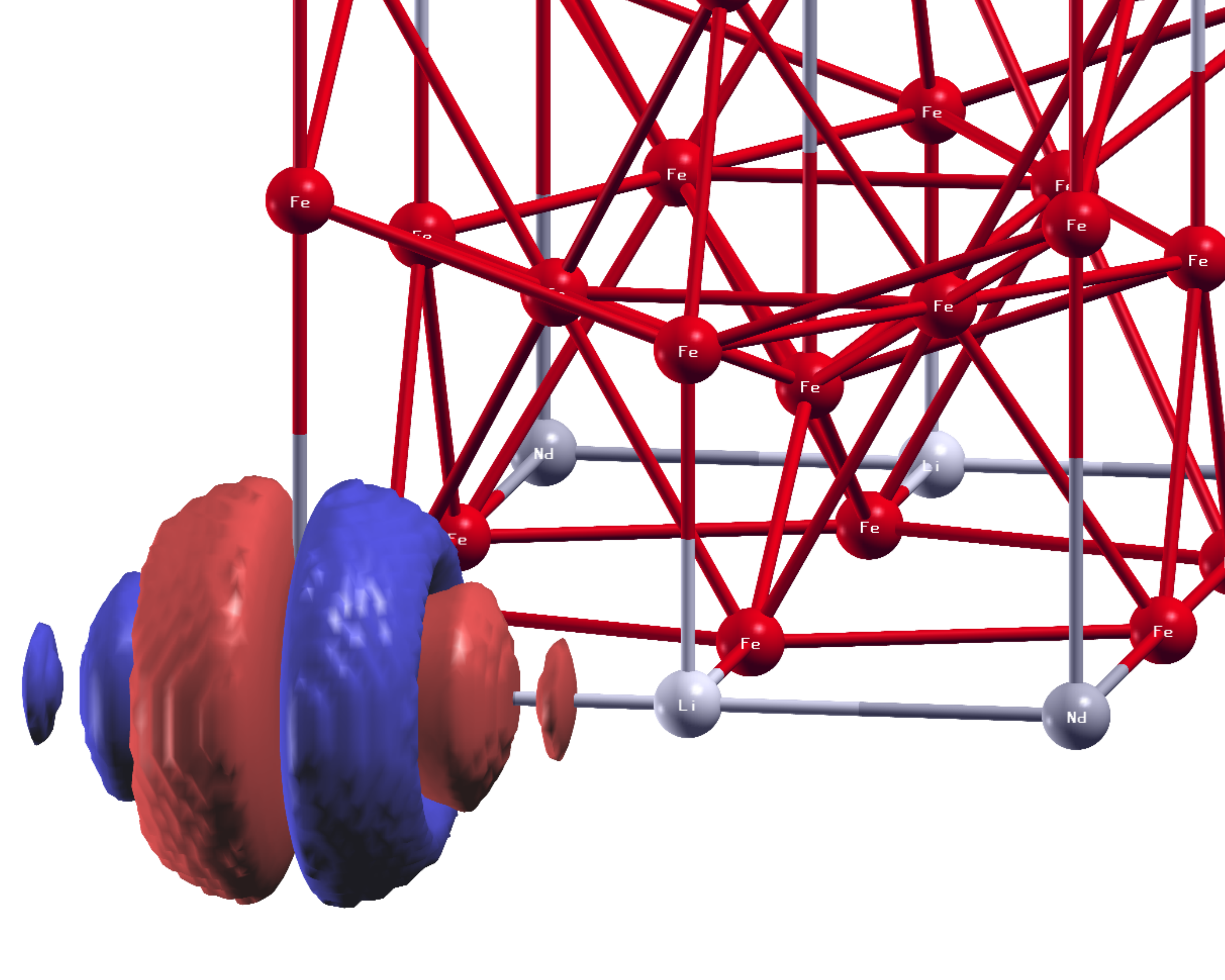}\label{sfig:large_wannier}}
\subfloat[{$\left[-2,2\right]$~eV window} ]{\includegraphics[width=0.5\linewidth]{NdFe12Li_4_small}\label{sfig:small_wannier}}
\end{minipage}
\caption{NdFe$_{12}$Li $4f$ Wannier orbital $m=0$ constructed with a large window $\left[-20,20\right]$~eV (left) and a small window $\left[-2,2\right]$~eV (right). The use of a large window essentially removes all hybridization between the rare-earth and neighboring atoms.}
\label{wanniers_window_size}
\end{figure}

Another important parameter of our calculations is the size of the window around the Fermi level that we use to construct the $4f$ Wannier functions. 
In Fig.~\ref{wanniers_window_size} we compare the Wannier orbitals constructed for the same orbital $m=0$ in NdFe$_{12}$Li for two different window sizes: a small window with $\omega_{\textrm{win}} = 2$~eV, and a large one with $\omega_{\textrm{win}} = 20$~eV. For the large window, the Wannier orbital (WO) takes essentially pure Nd 4$f$ orbital character, while the small-window WO leaks significantly to neighboring sites, in particular, to Li.  

\begin{figure}
\includegraphics[width=0.9\linewidth]{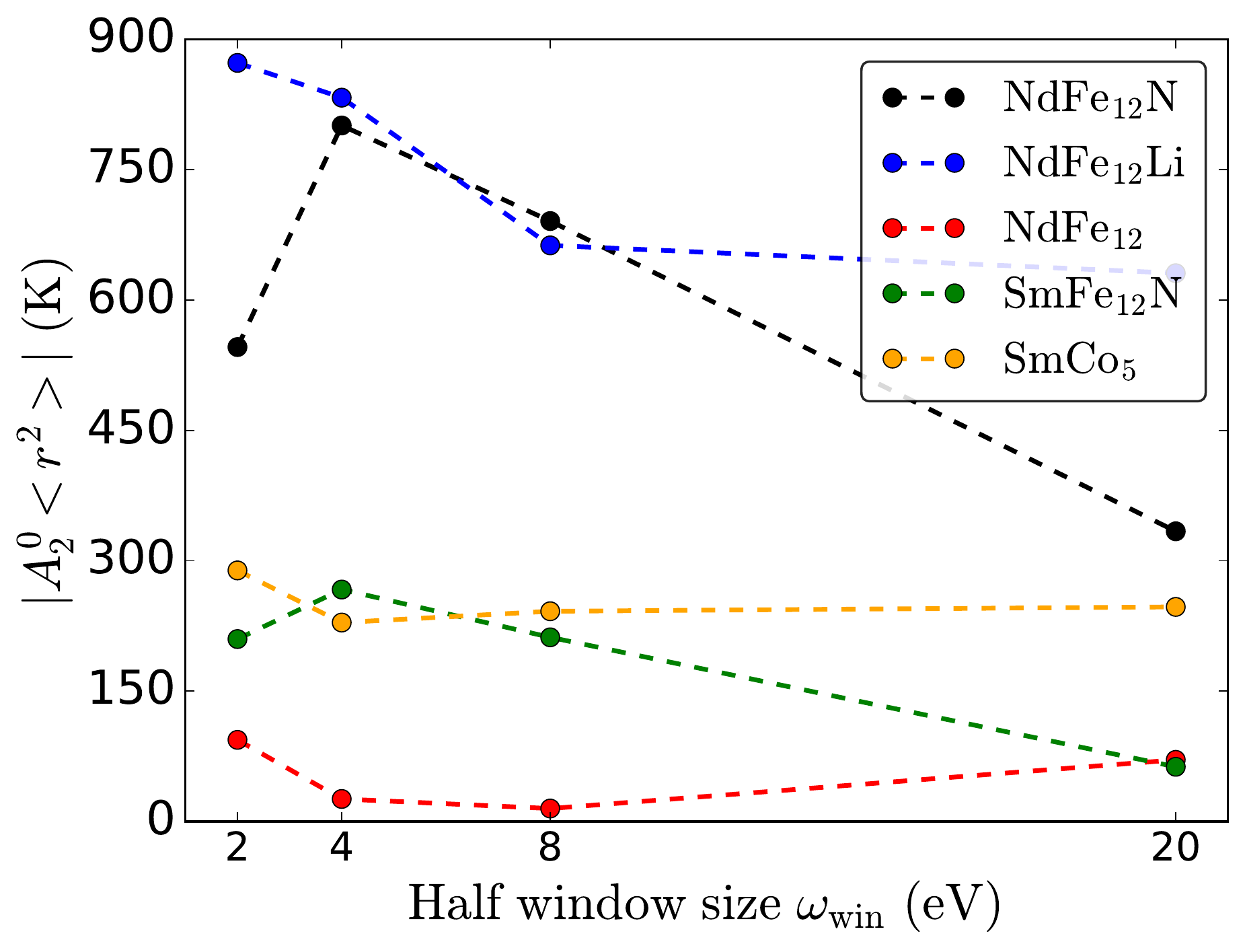}
\caption{Absolute value of the CFP $A_2^0\langle r^2\rangle$ as a function of window size $\omega_{\textrm{win}}$ for NdFe$_{12}$, NdFe$_{12}$N, NdFe$_{12}$Li and SmCo$_5$.}
\label{window_size_dep}
\end{figure}

The effect of the window size on the CF parameters is shown more quantitatively in Fig.~\ref{window_size_dep}, which displays those parameters computed for several window choices $[-\omega_{\mathrm{win}},\omega_{\mathrm{win}}]$ for different materials. 
The smallest window size of $\omega_{\mathrm{win}}=$2~eV is required to enclose all the 4$f$-like bands, increasing it to 4 eV includes most of the Fe states and part of the N or Li states inside the window. 
The largest size of 20 eV gives Wannier functions with essentially pure orbital character. 
One may notice a relatively mild dependence of the CFP on the choice of the window up to $\omega_{\mathrm{win}}=$8~eV.

\section{Projection of extended Wannier orbitals to localized Wannier basis}\label{app_projection}

In this Appendix we derive the projection operator between localized and extended Wannier spaces. 
A set of Wannier-like functions $| \tilde{w}^{\alpha\sigma}_{lm}(\vk)\rangle$ is constructed for an atom $\alpha$ of the unit cell 
and quantum numbers $(lm\sigma)$ as a combinations of  Kohn-Sham Bloch waves for a range of bands within the chosen energy window $\tilde{\cal{W}}$:
\begin{equation}
| \tilde{w}^{\alpha\sigma}_ {lm}(\vk)\rangle = \sum_{\nu\in \cal{\tilde{W}}} \tilde{P}^{\alpha\sigma}_{lm\nu}(\vk) |\phi^k_\nu\rangle
\label{wannier_k_space}
\end{equation}
where $\phi^k_\nu$ are the Bloch functions and $\tilde{P}^{\alpha\sigma}_{lm\nu}(\vk) $ is the corresponding matrix element of the projector constructed as described in Refs.~\onlinecite{Amadon2008} and \onlinecite{Aichhorn2009}.
The corresponding real-space Wannier functions are then obtained by a Fourier transformation 
\begin{equation}
 \tilde{w}^{\alpha\sigma}_{lm}(\vr) = \sum_{\vk} e^{-i\vk.\vr} | \tilde{w}^{\alpha\sigma}_{lm}(\vk)\rangle .
\label{wannier_real_space}
\end{equation}

We assume that the window $\tilde{\cal{W}}$ in Eq.~\ref{wannier_k_space} is {\it large}, i.e. that it includes both rare-earth 4$f$ states and all relevant valence bands that are expected to hybridize with them. In result, with such a large-window construction one obtains a set of mutually-orthogonal and rather well localized Wannier orbtials (WO). In particular, the large-window 4$f$ WOs $ \tilde{w}^{\alpha\sigma} _{lm}(\vr)$ almost do not leak onto neighboring sites, as discussed in the previous section, see Fig.~\ref{sfig:large_wannier}.  If one constructs as many WOs as the number of Kohn-Sham bands within $\tilde{\cal{W}}$ then the projection operator $\tilde{P}(\vk) $ is just a unitary transformation, hence, Eq.~\ref{wannier_k_space} can be inverted
\begin{equation}
|\phi^k_\nu\rangle  \approx \sum_{\alpha \sigma lm} \left[\tilde{P}^{\alpha\sigma}_{lm\nu}(\vk)\right]^* | \tilde{w}^{\alpha\sigma}_{lm}(\vk)\rangle,
\label{phi_from_wan}
\end{equation}
where the equality is approximate because high-energy empty bands usually cross and, hence, one cannot generally chose such a window as to have the same number of bands for all $\vk$-points.  However, those  high-energy states are far from the relevant region close to the Fermi level, and if one applies Eq.~\ref{phi_from_wan} to the bands within a {\it small} window $\cal{W}$ around the the Fermi energy the resulting small non-unitarity of $P(\vk)$ can be neglected.

Alternatively, one may construct 4$f$ Wannier orbitals from the bands within that small window $\cal{W}$ enclosing mainly 4$f$-like Kohn-Sham bands:
\begin{equation}
|w^{\sigma}_m (\vk)\rangle = \sum_{\nu\in \cal{W}} P^{\sigma}_{m\nu}(\vk) |\phi^k_\nu\rangle,
\label{small_win_WO}
\end{equation}
where 4$f$ WOs are constructed for the single rare-earth site in the unit cell for the compounds under consideration. Hence, the site and $l$ labels are suppressed in $|w^{\sigma}_m (\vk)\rangle$. The resulting small-window WOs are rather extended in real space, as one sees in Figs.~\ref{wannier_orbitals} and~\ref{sfig:small_wannier}.

Inserting the expansion Eq.~\ref{phi_from_wan} of the KS states $ |\phi^k_\nu\rangle$    into Eq.~\ref{small_win_WO} one obtains 

\begin{align}
|w^{\sigma}_m (\vk)\rangle  &=   \sum_{\nu\in \textrm{\cal{W}}}\sum_{lm'\sigma'} P^{\sigma}_{m\nu}(\vk) \left[\tilde{P}^{\alpha\sigma'}_{lm'\nu}(\vk)\right]^* | \tilde{w}^{\alpha\sigma'}_{lm'}(\vk)\rangle  \nonumber \\
     &=  \sum_{\alpha \sigma' lm'} U^{\sigma,\alpha\sigma'}_{m,lm'} (\vk) | \tilde{w}^{\alpha\sigma'}_{lm'}(\vk)\rangle
\label{WO_expansion}
\end{align}
where 
\begin{equation}
U^{\sigma,\alpha\sigma'}_{m,l'm'} (\vk) = \sum_{\nu \in\cal{W}} P^{\sigma}_{m\nu}(\vk) \left[\tilde{P}^{\alpha\sigma'}_{lm'\nu}(\vk)\right]^* .
\label{U_proj}
\end{equation}

We  use these projectors $U^{\sigma,\alpha\sigma'}_{m,l'm'} (\vk)$ to project the 4$f$ spectral function computed in the small-window WO basis on large-window localized WOs representing other states (Fe 3$d$, N 2$p$, Li 2$s$ and so on).  Namely, having obtained the  real-axis lattice Green's function in the small-window Wannier basis for the orbital ($\sigma m$) of the 4$f$ shell, $G_{m\sigma}(\vk,\omega+i\delta)$, as well as the corresponding partial spectral function  $\rho_{m\sigma}(\omega) = -\frac{1}{\pi}\textrm{Im}G_{m\sigma}(\vk,\omega+i\delta) $,
we compute the different orbital contributions into it as follows:
\begin{align}
\tilde{\rho}^{m\sigma}_{\alpha l}(\omega) = -\frac{1}{\pi}  \textrm{Im} \sum_k \sum_{m'\sigma'}  &\left[U^{\sigma,\alpha\sigma'}_{m,lm'}(\vk)\right]^*   \label{proj_DOS}  \\
           &\times G_{ m\sigma}(\vk,\omega+i\delta) U^{\sigma,\alpha\sigma'}_{m,lm'} (\vk)  \nonumber 
\end{align}
where $\tilde{\rho}^{m\sigma}_{\alpha l }(\omega)$ is the fraction of the 4$f$ spectral function of orbital index ($\sigma m$) with the character $(\alpha l)$. Using Eq.~\ref{phi_from_wan} and the orthonormality of small-window WOs 
$$
\langle w^{\sigma}_m (\vk) |  w^{\sigma'}_{m'} (\vk)\rangle=\delta_{mm'}\delta_{\sigma\sigma'}=\sum_{\nu} \left[P^{\sigma}_{m\nu}(\vk)\right]^*P^{\sigma'}_{m'\nu}(\vk)
$$
one may easily show the completeness of the expansion (\ref{proj_DOS}) 
$$
\sum_{\alpha l} \tilde{\rho}^{m\sigma}_{\alpha l}(\omega)=\rho_{m\sigma}(\omega).
$$

\bibliography{CF}

\begin{thebibliography}{70}
\expandafter\ifx\csname natexlab\endcsname\relax\def\natexlab#1{#1}\fi
\expandafter\ifx\csname bibnamefont\endcsname\relax
  \def\bibnamefont#1{#1}\fi
\expandafter\ifx\csname bibfnamefont\endcsname\relax
  \def\bibfnamefont#1{#1}\fi
\expandafter\ifx\csname citenamefont\endcsname\relax
  \def\citenamefont#1{#1}\fi
\expandafter\ifx\csname url\endcsname\relax
  \def\url#1{\texttt{#1}}\fi
\expandafter\ifx\csname urlprefix\endcsname\relax\def\urlprefix{URL }\fi
\providecommand{\bibinfo}[2]{#2}
\providecommand{\eprint}[2][]{\url{#2}}

\bibitem[{\citenamefont{Strnat}(1970)}]{strnat1970}
\bibinfo{author}{\bibfnamefont{K.}~\bibnamefont{Strnat}},
  \bibinfo{journal}{IEEE Transactions on Magnetics}
  \textbf{\bibinfo{volume}{6}}, \bibinfo{pages}{182} (\bibinfo{year}{1970}).

\bibitem[{\citenamefont{Herbst}(1991)}]{herbst1991}
\bibinfo{author}{\bibfnamefont{J.}~\bibnamefont{Herbst}},
  \bibinfo{journal}{Reviews of Modern Physics} \textbf{\bibinfo{volume}{63}},
  \bibinfo{pages}{819} (\bibinfo{year}{1991}).

\bibitem[{\citenamefont{K{\"o}rner et~al.}(2016)\citenamefont{K{\"o}rner,
  Krugel, and Els{\"a}sser}}]{korner2016}
\bibinfo{author}{\bibfnamefont{W.}~\bibnamefont{K{\"o}rner}},
  \bibinfo{author}{\bibfnamefont{G.}~\bibnamefont{Krugel}}, \bibnamefont{and}
  \bibinfo{author}{\bibfnamefont{C.}~\bibnamefont{Els{\"a}sser}},
  \bibinfo{journal}{Scientific reports} \textbf{\bibinfo{volume}{6}},
  \bibinfo{pages}{24686} (\bibinfo{year}{2016}).

\bibitem[{\citenamefont{Suzuki et~al.}(2016)\citenamefont{Suzuki, Kuno,
  Urushibata, Kobayashi, Sakuma, Washio, Yano, Kato, and Manabe}}]{suzuki2016}
\bibinfo{author}{\bibfnamefont{S.}~\bibnamefont{Suzuki}},
  \bibinfo{author}{\bibfnamefont{T.}~\bibnamefont{Kuno}},
  \bibinfo{author}{\bibfnamefont{K.}~\bibnamefont{Urushibata}},
  \bibinfo{author}{\bibfnamefont{K.}~\bibnamefont{Kobayashi}},
  \bibinfo{author}{\bibfnamefont{N.}~\bibnamefont{Sakuma}},
  \bibinfo{author}{\bibfnamefont{K.}~\bibnamefont{Washio}},
  \bibinfo{author}{\bibfnamefont{M.}~\bibnamefont{Yano}},
  \bibinfo{author}{\bibfnamefont{A.}~\bibnamefont{Kato}}, \bibnamefont{and}
  \bibinfo{author}{\bibfnamefont{A.}~\bibnamefont{Manabe}},
  \bibinfo{journal}{Journal of Magnetism and Magnetic Materials}
  \textbf{\bibinfo{volume}{401}}, \bibinfo{pages}{259} (\bibinfo{year}{2016}).

\bibitem[{\citenamefont{Harashima
  et~al.}(2015{\natexlab{a}})\citenamefont{Harashima, Terakura, Kino,
  Ishibashi, and Miyake}}]{harashima2015}
\bibinfo{author}{\bibfnamefont{Y.}~\bibnamefont{Harashima}},
  \bibinfo{author}{\bibfnamefont{K.}~\bibnamefont{Terakura}},
  \bibinfo{author}{\bibfnamefont{H.}~\bibnamefont{Kino}},
  \bibinfo{author}{\bibfnamefont{S.}~\bibnamefont{Ishibashi}},
  \bibnamefont{and} \bibinfo{author}{\bibfnamefont{T.}~\bibnamefont{Miyake}},
  \bibinfo{journal}{Physical Review B} \textbf{\bibinfo{volume}{92}},
  \bibinfo{pages}{184426} (\bibinfo{year}{2015}{\natexlab{a}}).

\bibitem[{\citenamefont{Hirayama et~al.}(2015)\citenamefont{Hirayama,
  Takahashi, Hirosawa, and Hono}}]{hirayama2015}
\bibinfo{author}{\bibfnamefont{Y.}~\bibnamefont{Hirayama}},
  \bibinfo{author}{\bibfnamefont{Y.}~\bibnamefont{Takahashi}},
  \bibinfo{author}{\bibfnamefont{S.}~\bibnamefont{Hirosawa}}, \bibnamefont{and}
  \bibinfo{author}{\bibfnamefont{K.}~\bibnamefont{Hono}},
  \bibinfo{journal}{Scripta Materialia} \textbf{\bibinfo{volume}{95}},
  \bibinfo{pages}{70} (\bibinfo{year}{2015}).

\bibitem[{\citenamefont{Miyake et~al.}(2014)\citenamefont{Miyake, Terakura,
  Harashima, Kino, and Ishibashi}}]{miyake2014}
\bibinfo{author}{\bibfnamefont{T.}~\bibnamefont{Miyake}},
  \bibinfo{author}{\bibfnamefont{K.}~\bibnamefont{Terakura}},
  \bibinfo{author}{\bibfnamefont{Y.}~\bibnamefont{Harashima}},
  \bibinfo{author}{\bibfnamefont{H.}~\bibnamefont{Kino}}, \bibnamefont{and}
  \bibinfo{author}{\bibfnamefont{S.}~\bibnamefont{Ishibashi}},
  \bibinfo{journal}{Journal of the Physical Society of Japan}
  \textbf{\bibinfo{volume}{83}}, \bibinfo{pages}{043702}
  (\bibinfo{year}{2014}).

\bibitem[{\citenamefont{Coey}(2011)}]{coey2011}
\bibinfo{author}{\bibfnamefont{J.}~\bibnamefont{Coey}}, \bibinfo{journal}{IEEE
  Transactions on Magnetics} \textbf{\bibinfo{volume}{47}},
  \bibinfo{pages}{4671} (\bibinfo{year}{2011}).

\bibitem[{\citenamefont{Coey}(1996)}]{coey1996}
\bibinfo{author}{\bibfnamefont{J.~M.~D.} \bibnamefont{Coey}},
  \emph{\bibinfo{title}{Rare-earth iron permanent magnets}},
  \bibinfo{number}{54} (\bibinfo{publisher}{Oxford University Press},
  \bibinfo{year}{1996}).

\bibitem[{\citenamefont{Gutfleisch et~al.}(2011)\citenamefont{Gutfleisch,
  Willard, Br{\"u}ck, Chen, Sankar, and Liu}}]{gutfleisch2011}
\bibinfo{author}{\bibfnamefont{O.}~\bibnamefont{Gutfleisch}},
  \bibinfo{author}{\bibfnamefont{M.~A.} \bibnamefont{Willard}},
  \bibinfo{author}{\bibfnamefont{E.}~\bibnamefont{Br{\"u}ck}},
  \bibinfo{author}{\bibfnamefont{C.~H.} \bibnamefont{Chen}},
  \bibinfo{author}{\bibfnamefont{S.}~\bibnamefont{Sankar}}, \bibnamefont{and}
  \bibinfo{author}{\bibfnamefont{J.~P.} \bibnamefont{Liu}},
  \bibinfo{journal}{Advanced materials} \textbf{\bibinfo{volume}{23}},
  \bibinfo{pages}{821} (\bibinfo{year}{2011}).

\bibitem[{\citenamefont{Matsumoto et~al.}(2016)\citenamefont{Matsumoto, Akai,
  Harashima, Doi, and Miyake}}]{matsumoto2016}
\bibinfo{author}{\bibfnamefont{M.}~\bibnamefont{Matsumoto}},
  \bibinfo{author}{\bibfnamefont{H.}~\bibnamefont{Akai}},
  \bibinfo{author}{\bibfnamefont{Y.}~\bibnamefont{Harashima}},
  \bibinfo{author}{\bibfnamefont{S.}~\bibnamefont{Doi}}, \bibnamefont{and}
  \bibinfo{author}{\bibfnamefont{T.}~\bibnamefont{Miyake}},
  \bibinfo{journal}{Journal of Applied Physics} \textbf{\bibinfo{volume}{119}},
  \bibinfo{pages}{213901} (\bibinfo{year}{2016}).

\bibitem[{\citenamefont{Harashima et~al.}(2016)\citenamefont{Harashima,
  Terakura, Kino, Ishibashi, and Miyake}}]{harashima2016}
\bibinfo{author}{\bibfnamefont{Y.}~\bibnamefont{Harashima}},
  \bibinfo{author}{\bibfnamefont{K.}~\bibnamefont{Terakura}},
  \bibinfo{author}{\bibfnamefont{H.}~\bibnamefont{Kino}},
  \bibinfo{author}{\bibfnamefont{S.}~\bibnamefont{Ishibashi}},
  \bibnamefont{and} \bibinfo{author}{\bibfnamefont{T.}~\bibnamefont{Miyake}},
  \bibinfo{journal}{Journal of Applied Physics} \textbf{\bibinfo{volume}{120}},
  \bibinfo{pages}{203904} (\bibinfo{year}{2016}).

\bibitem[{\citenamefont{Hutchings}(1964)}]{hutchings1964}
\bibinfo{author}{\bibfnamefont{M.~T.} \bibnamefont{Hutchings}},
  \bibinfo{journal}{Solid state physics} \textbf{\bibinfo{volume}{16}},
  \bibinfo{pages}{227} (\bibinfo{year}{1964}).

\bibitem[{\citenamefont{Newman and Ng}(1989)}]{newman1989}
\bibinfo{author}{\bibfnamefont{D.}~\bibnamefont{Newman}} \bibnamefont{and}
  \bibinfo{author}{\bibfnamefont{B.}~\bibnamefont{Ng}},
  \bibinfo{journal}{Reports on Progress in Physics}
  \textbf{\bibinfo{volume}{52}}, \bibinfo{pages}{699} (\bibinfo{year}{1989}).

\bibitem[{\citenamefont{Brecher et~al.}(1967)\citenamefont{Brecher, Samelson,
  Lempicki, Riley, and Peters}}]{brecher1967}
\bibinfo{author}{\bibfnamefont{C.}~\bibnamefont{Brecher}},
  \bibinfo{author}{\bibfnamefont{H.}~\bibnamefont{Samelson}},
  \bibinfo{author}{\bibfnamefont{A.}~\bibnamefont{Lempicki}},
  \bibinfo{author}{\bibfnamefont{R.}~\bibnamefont{Riley}}, \bibnamefont{and}
  \bibinfo{author}{\bibfnamefont{T.}~\bibnamefont{Peters}},
  \bibinfo{journal}{Physical Review} \textbf{\bibinfo{volume}{155}},
  \bibinfo{pages}{178} (\bibinfo{year}{1967}).

\bibitem[{\citenamefont{Buschow et~al.}(1974)\citenamefont{Buschow, Van~Diepen,
  and De~Wijn}}]{buschow1974}
\bibinfo{author}{\bibfnamefont{K.~H.~J.} \bibnamefont{Buschow}},
  \bibinfo{author}{\bibfnamefont{A.~M.} \bibnamefont{Van~Diepen}},
  \bibnamefont{and} \bibinfo{author}{\bibfnamefont{H.~W.}
  \bibnamefont{De~Wijn}}, \bibinfo{journal}{Solid State Communications}
  \textbf{\bibinfo{volume}{15}}, \bibinfo{pages}{903} (\bibinfo{year}{1974}).

\bibitem[{\citenamefont{Sankar et~al.}(1975)\citenamefont{Sankar, Rao, Segal,
  Wallace, Frederick, and Garrett}}]{sankar1975}
\bibinfo{author}{\bibfnamefont{S.}~\bibnamefont{Sankar}},
  \bibinfo{author}{\bibfnamefont{V.}~\bibnamefont{Rao}},
  \bibinfo{author}{\bibfnamefont{E.}~\bibnamefont{Segal}},
  \bibinfo{author}{\bibfnamefont{W.}~\bibnamefont{Wallace}},
  \bibinfo{author}{\bibfnamefont{W.}~\bibnamefont{Frederick}},
  \bibnamefont{and} \bibinfo{author}{\bibfnamefont{H.}~\bibnamefont{Garrett}},
  \bibinfo{journal}{Physical Review B} \textbf{\bibinfo{volume}{11}},
  \bibinfo{pages}{435} (\bibinfo{year}{1975}).

\bibitem[{\citenamefont{Givord et~al.}(1979)\citenamefont{Givord, Laforest,
  Schweizer, and Tasset}}]{givord1979}
\bibinfo{author}{\bibfnamefont{D.}~\bibnamefont{Givord}},
  \bibinfo{author}{\bibfnamefont{J.}~\bibnamefont{Laforest}},
  \bibinfo{author}{\bibfnamefont{J.}~\bibnamefont{Schweizer}},
  \bibnamefont{and} \bibinfo{author}{\bibfnamefont{F.}~\bibnamefont{Tasset}},
  \bibinfo{journal}{Journal of Applied Physics} \textbf{\bibinfo{volume}{50}},
  \bibinfo{pages}{2008} (\bibinfo{year}{1979}).

\bibitem[{\citenamefont{Tie-Song et~al.}(1991)\citenamefont{Tie-Song, Han-Min,
  Guang-Hua, Xiu-Feng, and Hong}}]{tie1991}
\bibinfo{author}{\bibfnamefont{Z.}~\bibnamefont{Tie-Song}},
  \bibinfo{author}{\bibfnamefont{J.}~\bibnamefont{Han-Min}},
  \bibinfo{author}{\bibfnamefont{G.}~\bibnamefont{Guang-Hua}},
  \bibinfo{author}{\bibfnamefont{H.}~\bibnamefont{Xiu-Feng}}, \bibnamefont{and}
  \bibinfo{author}{\bibfnamefont{C.}~\bibnamefont{Hong}},
  \bibinfo{journal}{Physical Review B} \textbf{\bibinfo{volume}{43}},
  \bibinfo{pages}{8593} (\bibinfo{year}{1991}).

\bibitem[{\citenamefont{Tils et~al.}(1999)\citenamefont{Tils, Loewenhaupt,
  Buschow, and Eccleston}}]{tils1999}
\bibinfo{author}{\bibfnamefont{P.}~\bibnamefont{Tils}},
  \bibinfo{author}{\bibfnamefont{M.}~\bibnamefont{Loewenhaupt}},
  \bibinfo{author}{\bibfnamefont{K.~H.~J.} \bibnamefont{Buschow}},
  \bibnamefont{and}
  \bibinfo{author}{\bibfnamefont{R.}~\bibnamefont{Eccleston}},
  \bibinfo{journal}{Journal of alloys and compounds}
  \textbf{\bibinfo{volume}{289}}, \bibinfo{pages}{28} (\bibinfo{year}{1999}).

\bibitem[{\citenamefont{Coehoorn}(1991)}]{coehoorn1991}
\bibinfo{author}{\bibfnamefont{R.}~\bibnamefont{Coehoorn}},
  \bibinfo{journal}{Journal of Magnetism and Magnetic Materials}
  \textbf{\bibinfo{volume}{99}}, \bibinfo{pages}{55} (\bibinfo{year}{1991}).

\bibitem[{\citenamefont{Daalderop et~al.}(1992)\citenamefont{Daalderop, Kelly,
  and Schuurmans}}]{daalderop1992}
\bibinfo{author}{\bibfnamefont{G.}~\bibnamefont{Daalderop}},
  \bibinfo{author}{\bibfnamefont{P.}~\bibnamefont{Kelly}}, \bibnamefont{and}
  \bibinfo{author}{\bibfnamefont{M.}~\bibnamefont{Schuurmans}},
  \bibinfo{journal}{Journal of Magnetism and Magnetic Materials}
  \textbf{\bibinfo{volume}{104}}, \bibinfo{pages}{737} (\bibinfo{year}{1992}).

\bibitem[{\citenamefont{Steinbeck et~al.}(1994)\citenamefont{Steinbeck,
  Richter, Eschrig, and Nitzsche}}]{steinbeck1994}
\bibinfo{author}{\bibfnamefont{L.}~\bibnamefont{Steinbeck}},
  \bibinfo{author}{\bibfnamefont{M.}~\bibnamefont{Richter}},
  \bibinfo{author}{\bibfnamefont{H.}~\bibnamefont{Eschrig}}, \bibnamefont{and}
  \bibinfo{author}{\bibfnamefont{U.}~\bibnamefont{Nitzsche}},
  \bibinfo{journal}{Physical Review B} \textbf{\bibinfo{volume}{49}},
  \bibinfo{pages}{16289} (\bibinfo{year}{1994}).

\bibitem[{\citenamefont{Nov{\'a}k and Kuriplach}(1994)}]{novak1994RNi5}
\bibinfo{author}{\bibfnamefont{P.}~\bibnamefont{Nov{\'a}k}} \bibnamefont{and}
  \bibinfo{author}{\bibfnamefont{J.}~\bibnamefont{Kuriplach}},
  \bibinfo{journal}{Physical Review B} \textbf{\bibinfo{volume}{50}},
  \bibinfo{pages}{2085} (\bibinfo{year}{1994}).

\bibitem[{\citenamefont{Steinbeck et~al.}(1996)\citenamefont{Steinbeck,
  Richter, Nitzsche, and Eschrig}}]{steinbeck1996}
\bibinfo{author}{\bibfnamefont{L.}~\bibnamefont{Steinbeck}},
  \bibinfo{author}{\bibfnamefont{M.}~\bibnamefont{Richter}},
  \bibinfo{author}{\bibfnamefont{U.}~\bibnamefont{Nitzsche}}, \bibnamefont{and}
  \bibinfo{author}{\bibfnamefont{H.}~\bibnamefont{Eschrig}},
  \bibinfo{journal}{Physical Review B} \textbf{\bibinfo{volume}{53}},
  \bibinfo{pages}{7111} (\bibinfo{year}{1996}).

\bibitem[{\citenamefont{Nov{\'a}k}(1996)}]{novak1996}
\bibinfo{author}{\bibfnamefont{P.}~\bibnamefont{Nov{\'a}k}},
  \bibinfo{journal}{physica status solidi (b)} \textbf{\bibinfo{volume}{198}},
  \bibinfo{pages}{729} (\bibinfo{year}{1996}).

\bibitem[{\citenamefont{Ning and Brivio}(2007)}]{Ning2007}
\bibinfo{author}{\bibfnamefont{L.}~\bibnamefont{Ning}} \bibnamefont{and}
  \bibinfo{author}{\bibfnamefont{G.~P.} \bibnamefont{Brivio}},
  \bibinfo{journal}{Phys. Rev. B} \textbf{\bibinfo{volume}{75}},
  \bibinfo{pages}{235126} (\bibinfo{year}{2007}).

\bibitem[{\citenamefont{Hu et~al.}(2011)\citenamefont{Hu, Reid, Duan, Xia, and
  Yin}}]{hu2011}
\bibinfo{author}{\bibfnamefont{L.}~\bibnamefont{Hu}},
  \bibinfo{author}{\bibfnamefont{M.~F.} \bibnamefont{Reid}},
  \bibinfo{author}{\bibfnamefont{C.-K.} \bibnamefont{Duan}},
  \bibinfo{author}{\bibfnamefont{S.}~\bibnamefont{Xia}}, \bibnamefont{and}
  \bibinfo{author}{\bibfnamefont{M.}~\bibnamefont{Yin}},
  \bibinfo{journal}{Journal of Physics: Condensed Matter}
  \textbf{\bibinfo{volume}{23}}, \bibinfo{pages}{045501}
  (\bibinfo{year}{2011}).

\bibitem[{\citenamefont{Nov{\'a}k et~al.}(2013)\citenamefont{Nov{\'a}k,
  Kn{\'\i}{\v{z}}ek, and Kune{\v{s}}}}]{novak2013}
\bibinfo{author}{\bibfnamefont{P.}~\bibnamefont{Nov{\'a}k}},
  \bibinfo{author}{\bibfnamefont{K.}~\bibnamefont{Kn{\'\i}{\v{z}}ek}},
  \bibnamefont{and}
  \bibinfo{author}{\bibfnamefont{J.}~\bibnamefont{Kune{\v{s}}}},
  \bibinfo{journal}{Physical Review B} \textbf{\bibinfo{volume}{87}},
  \bibinfo{pages}{205139} (\bibinfo{year}{2013}).

\bibitem[{\citenamefont{Brooks et~al.}(1997)\citenamefont{Brooks, Eriksson,
  Wills, and Johansson}}]{brooks1997density}
\bibinfo{author}{\bibfnamefont{M.}~\bibnamefont{Brooks}},
  \bibinfo{author}{\bibfnamefont{O.}~\bibnamefont{Eriksson}},
  \bibinfo{author}{\bibfnamefont{J.}~\bibnamefont{Wills}}, \bibnamefont{and}
  \bibinfo{author}{\bibfnamefont{B.}~\bibnamefont{Johansson}},
  \bibinfo{journal}{Physical review letters} \textbf{\bibinfo{volume}{79}},
  \bibinfo{pages}{2546} (\bibinfo{year}{1997}).

\bibitem[{\citenamefont{Haverkort et~al.}(2012)\citenamefont{Haverkort,
  Zwierzycki, and Andersen}}]{haverkort2012}
\bibinfo{author}{\bibfnamefont{M.~W.} \bibnamefont{Haverkort}},
  \bibinfo{author}{\bibfnamefont{M.}~\bibnamefont{Zwierzycki}},
  \bibnamefont{and} \bibinfo{author}{\bibfnamefont{O.~K.}
  \bibnamefont{Andersen}}, \bibinfo{journal}{Phys. Rev. B}
  \textbf{\bibinfo{volume}{85}}, \bibinfo{pages}{165113}
  (\bibinfo{year}{2012}).

\bibitem[{\citenamefont{Liddle and van Slageren}(2015)}]{liddle2015improving}
\bibinfo{author}{\bibfnamefont{S.~T.} \bibnamefont{Liddle}} \bibnamefont{and}
  \bibinfo{author}{\bibfnamefont{J.}~\bibnamefont{van Slageren}},
  \bibinfo{journal}{Chemical Society Reviews} \textbf{\bibinfo{volume}{44}},
  \bibinfo{pages}{6655} (\bibinfo{year}{2015}).

\bibitem[{\citenamefont{Ungur and Chibotaru}(2017)}]{ungur2017ab}
\bibinfo{author}{\bibfnamefont{L.}~\bibnamefont{Ungur}} \bibnamefont{and}
  \bibinfo{author}{\bibfnamefont{L.~F.} \bibnamefont{Chibotaru}},
  \bibinfo{journal}{Chemistry-A European Journal}
  \textbf{\bibinfo{volume}{23}}, \bibinfo{pages}{3708} (\bibinfo{year}{2017}).

\bibitem[{\citenamefont{Zhou et~al.}(2011)\citenamefont{Zhou,
  Ozoli{\c{n}}{\v{s}} et~al.}}]{zhou2011crystal}
\bibinfo{author}{\bibfnamefont{F.}~\bibnamefont{Zhou}},
  \bibinfo{author}{\bibfnamefont{V.}~\bibnamefont{Ozoli{\c{n}}{\v{s}}}},
  \bibnamefont{et~al.}, \bibinfo{journal}{Physical Review B}
  \textbf{\bibinfo{volume}{83}}, \bibinfo{pages}{085106}
  (\bibinfo{year}{2011}).

\bibitem[{\citenamefont{Zhou et~al.}(2009)\citenamefont{Zhou,
  Ozoli{\c{n}}{\v{s}} et~al.}}]{zhou2009obtaining}
\bibinfo{author}{\bibfnamefont{F.}~\bibnamefont{Zhou}},
  \bibinfo{author}{\bibfnamefont{V.}~\bibnamefont{Ozoli{\c{n}}{\v{s}}}},
  \bibnamefont{et~al.}, \bibinfo{journal}{Physical Review B}
  \textbf{\bibinfo{volume}{80}}, \bibinfo{pages}{125127}
  (\bibinfo{year}{2009}).

\bibitem[{\citenamefont{Anisimov et~al.}(1997)\citenamefont{Anisimov,
  Poteryaev, Korotin, Anokhin, and Kotliar}}]{anisimov97}
\bibinfo{author}{\bibfnamefont{V.~I.} \bibnamefont{Anisimov}},
  \bibinfo{author}{\bibfnamefont{A.~I.} \bibnamefont{Poteryaev}},
  \bibinfo{author}{\bibfnamefont{M.~A.} \bibnamefont{Korotin}},
  \bibinfo{author}{\bibfnamefont{A.~O.} \bibnamefont{Anokhin}},
  \bibnamefont{and} \bibinfo{author}{\bibfnamefont{G.}~\bibnamefont{Kotliar}},
  \bibinfo{journal}{J. Phys.: Condens. Matter} \textbf{\bibinfo{volume}{9}},
  \bibinfo{pages}{7359} (\bibinfo{year}{1997}).

\bibitem[{\citenamefont{Lichtenstein and Katsnelson}(1998)}]{lichtenstein1998}
\bibinfo{author}{\bibfnamefont{A.}~\bibnamefont{Lichtenstein}}
  \bibnamefont{and}
  \bibinfo{author}{\bibfnamefont{M.}~\bibnamefont{Katsnelson}},
  \bibinfo{journal}{Physical Review B} \textbf{\bibinfo{volume}{57}},
  \bibinfo{pages}{6884} (\bibinfo{year}{1998}).

\bibitem[{\citenamefont{Leb{\`e}gue et~al.}(2005)\citenamefont{Leb{\`e}gue,
  Santi, Svane, Bengone, Katsnelson, Lichtenstein, and
  Eriksson}}]{lebegue2005electronic}
\bibinfo{author}{\bibfnamefont{S.}~\bibnamefont{Leb{\`e}gue}},
  \bibinfo{author}{\bibfnamefont{G.}~\bibnamefont{Santi}},
  \bibinfo{author}{\bibfnamefont{A.}~\bibnamefont{Svane}},
  \bibinfo{author}{\bibfnamefont{O.}~\bibnamefont{Bengone}},
  \bibinfo{author}{\bibfnamefont{M.}~\bibnamefont{Katsnelson}},
  \bibinfo{author}{\bibfnamefont{A.}~\bibnamefont{Lichtenstein}},
  \bibnamefont{and} \bibinfo{author}{\bibfnamefont{O.}~\bibnamefont{Eriksson}},
  \bibinfo{journal}{Physical Review B} \textbf{\bibinfo{volume}{72}},
  \bibinfo{pages}{245102} (\bibinfo{year}{2005}).

\bibitem[{\citenamefont{Leb{\`e}gue et~al.}(2006)\citenamefont{Leb{\`e}gue,
  Svane, Katsnelson, Lichtenstein, and Eriksson}}]{lebegue2006multiplet}
\bibinfo{author}{\bibfnamefont{S.}~\bibnamefont{Leb{\`e}gue}},
  \bibinfo{author}{\bibfnamefont{A.}~\bibnamefont{Svane}},
  \bibinfo{author}{\bibfnamefont{M.}~\bibnamefont{Katsnelson}},
  \bibinfo{author}{\bibfnamefont{A.}~\bibnamefont{Lichtenstein}},
  \bibnamefont{and} \bibinfo{author}{\bibfnamefont{O.}~\bibnamefont{Eriksson}},
  \bibinfo{journal}{Journal of Physics: Condensed Matter}
  \textbf{\bibinfo{volume}{18}}, \bibinfo{pages}{6329} (\bibinfo{year}{2006}).

\bibitem[{\citenamefont{Pourovskii et~al.}(2009)\citenamefont{Pourovskii,
  Delaney, Van~de Walle, Spaldin, and Georges}}]{Pourovskii2009}
\bibinfo{author}{\bibfnamefont{L.~V.} \bibnamefont{Pourovskii}},
  \bibinfo{author}{\bibfnamefont{K.~T.} \bibnamefont{Delaney}},
  \bibinfo{author}{\bibfnamefont{C.~G.} \bibnamefont{Van~de Walle}},
  \bibinfo{author}{\bibfnamefont{N.~A.} \bibnamefont{Spaldin}},
  \bibnamefont{and} \bibinfo{author}{\bibfnamefont{A.}~\bibnamefont{Georges}},
  \bibinfo{journal}{Phys. Rev. Lett.} \textbf{\bibinfo{volume}{102}},
  \bibinfo{pages}{096401} (\bibinfo{year}{2009}).

\bibitem[{\citenamefont{Locht et~al.}(2016)\citenamefont{Locht, Kvashnin,
  Rodrigues, Pereiro, Bergman, Bergqvist, Lichtenstein, Katsnelson, Delin,
  Klautau et~al.}}]{Locht2016}
\bibinfo{author}{\bibfnamefont{I.~L.~M.} \bibnamefont{Locht}},
  \bibinfo{author}{\bibfnamefont{Y.~O.} \bibnamefont{Kvashnin}},
  \bibinfo{author}{\bibfnamefont{D.~C.~M.} \bibnamefont{Rodrigues}},
  \bibinfo{author}{\bibfnamefont{M.}~\bibnamefont{Pereiro}},
  \bibinfo{author}{\bibfnamefont{A.}~\bibnamefont{Bergman}},
  \bibinfo{author}{\bibfnamefont{L.}~\bibnamefont{Bergqvist}},
  \bibinfo{author}{\bibfnamefont{A.~I.} \bibnamefont{Lichtenstein}},
  \bibinfo{author}{\bibfnamefont{M.~I.} \bibnamefont{Katsnelson}},
  \bibinfo{author}{\bibfnamefont{A.}~\bibnamefont{Delin}},
  \bibinfo{author}{\bibfnamefont{A.~B.} \bibnamefont{Klautau}},
  \bibnamefont{et~al.}, \bibinfo{journal}{Phys. Rev. B}
  \textbf{\bibinfo{volume}{94}}, \bibinfo{pages}{085137}
  (\bibinfo{year}{2016}).

\bibitem[{\citenamefont{Gr{\aa}n{\"a}s
  et~al.}(2012)\citenamefont{Gr{\aa}n{\"a}s, Di~Marco, Thunstr{\"o}m,
  Nordstr{\"o}m, Eriksson, Bj{\"o}rkman, and Wills}}]{graanas2012charge}
\bibinfo{author}{\bibfnamefont{O.}~\bibnamefont{Gr{\aa}n{\"a}s}},
  \bibinfo{author}{\bibfnamefont{I.}~\bibnamefont{Di~Marco}},
  \bibinfo{author}{\bibfnamefont{P.}~\bibnamefont{Thunstr{\"o}m}},
  \bibinfo{author}{\bibfnamefont{L.}~\bibnamefont{Nordstr{\"o}m}},
  \bibinfo{author}{\bibfnamefont{O.}~\bibnamefont{Eriksson}},
  \bibinfo{author}{\bibfnamefont{T.}~\bibnamefont{Bj{\"o}rkman}},
  \bibnamefont{and} \bibinfo{author}{\bibfnamefont{J.}~\bibnamefont{Wills}},
  \bibinfo{journal}{Computational materials science}
  \textbf{\bibinfo{volume}{55}}, \bibinfo{pages}{295} (\bibinfo{year}{2012}).

\bibitem[{\citenamefont{Stevens}(1952)}]{stevens1952}
\bibinfo{author}{\bibfnamefont{K.}~\bibnamefont{Stevens}},
  \bibinfo{journal}{Proceedings of the Physical Society. Section A}
  \textbf{\bibinfo{volume}{65}}, \bibinfo{pages}{209} (\bibinfo{year}{1952}).

\bibitem[{\citenamefont{Wybourne and Meggers}(1965)}]{wybourne1965}
\bibinfo{author}{\bibfnamefont{B.~G.} \bibnamefont{Wybourne}} \bibnamefont{and}
  \bibinfo{author}{\bibfnamefont{W.~F.} \bibnamefont{Meggers}},
  \emph{\bibinfo{title}{Spectroscopic properties of rare earths}}
  (\bibinfo{publisher}{Interscience Publishers, New York},
  \bibinfo{year}{1965}).

\bibitem[{\citenamefont{Mulak and Gajek}(2000)}]{mulak2000}
\bibinfo{author}{\bibfnamefont{J.}~\bibnamefont{Mulak}} \bibnamefont{and}
  \bibinfo{author}{\bibfnamefont{Z.}~\bibnamefont{Gajek}},
  \emph{\bibinfo{title}{The effective crystal field potential}}
  (\bibinfo{publisher}{Elsevier}, \bibinfo{year}{2000}).

\bibitem[{\citenamefont{Newman and Ng}(2007)}]{newman2007}
\bibinfo{author}{\bibfnamefont{D.~J.} \bibnamefont{Newman}} \bibnamefont{and}
  \bibinfo{author}{\bibfnamefont{B.}~\bibnamefont{Ng}},
  \emph{\bibinfo{title}{Crystal field handbook}} (\bibinfo{publisher}{Cambridge
  University Press}, \bibinfo{year}{2007}).

\bibitem[{\citenamefont{Aichhorn et~al.}(2016)\citenamefont{Aichhorn,
  Pourovskii, Seth, Vildosola, Zingl, Peil, Deng, Mravlje, Kraberger, Martins
  et~al.}}]{aichhorn2016}
\bibinfo{author}{\bibfnamefont{M.}~\bibnamefont{Aichhorn}},
  \bibinfo{author}{\bibfnamefont{L.}~\bibnamefont{Pourovskii}},
  \bibinfo{author}{\bibfnamefont{P.}~\bibnamefont{Seth}},
  \bibinfo{author}{\bibfnamefont{V.}~\bibnamefont{Vildosola}},
  \bibinfo{author}{\bibfnamefont{M.}~\bibnamefont{Zingl}},
  \bibinfo{author}{\bibfnamefont{O.~E.} \bibnamefont{Peil}},
  \bibinfo{author}{\bibfnamefont{X.}~\bibnamefont{Deng}},
  \bibinfo{author}{\bibfnamefont{J.}~\bibnamefont{Mravlje}},
  \bibinfo{author}{\bibfnamefont{G.~J.} \bibnamefont{Kraberger}},
  \bibinfo{author}{\bibfnamefont{C.}~\bibnamefont{Martins}},
  \bibnamefont{et~al.}, \bibinfo{journal}{Computer Physics Communications}
  \textbf{\bibinfo{volume}{204}}, \bibinfo{pages}{200} (\bibinfo{year}{2016}).

\bibitem[{\citenamefont{Parcollet et~al.}(2015)\citenamefont{Parcollet,
  Ferrero, Ayral, Hafermann, Krivenko, Messio, and Seth}}]{Parcollet2015}
\bibinfo{author}{\bibfnamefont{O.}~\bibnamefont{Parcollet}},
  \bibinfo{author}{\bibfnamefont{M.}~\bibnamefont{Ferrero}},
  \bibinfo{author}{\bibfnamefont{T.}~\bibnamefont{Ayral}},
  \bibinfo{author}{\bibfnamefont{H.}~\bibnamefont{Hafermann}},
  \bibinfo{author}{\bibfnamefont{I.}~\bibnamefont{Krivenko}},
  \bibinfo{author}{\bibfnamefont{L.}~\bibnamefont{Messio}}, \bibnamefont{and}
  \bibinfo{author}{\bibfnamefont{P.}~\bibnamefont{Seth}},
  \bibinfo{journal}{Computer Physics Communications}
  \textbf{\bibinfo{volume}{196}}, \bibinfo{pages}{398} (\bibinfo{year}{2015}).

\bibitem[{\citenamefont{Blaha et~al.}(2001)\citenamefont{Blaha, Schwarz,
  Madsen, Kvasnicka, and Luitz}}]{Wien2k}
\bibinfo{author}{\bibfnamefont{P.}~\bibnamefont{Blaha}},
  \bibinfo{author}{\bibfnamefont{K.}~\bibnamefont{Schwarz}},
  \bibinfo{author}{\bibfnamefont{G.}~\bibnamefont{Madsen}},
  \bibinfo{author}{\bibfnamefont{D.}~\bibnamefont{Kvasnicka}},
  \bibnamefont{and} \bibinfo{author}{\bibfnamefont{J.}~\bibnamefont{Luitz}},
  \emph{\bibinfo{title}{WIEN2k, An augmented Plane Wave + Local Orbitals
  Program for Calculating Crystal Properties}} (\bibinfo{publisher}{Techn.
  Universitat Wien, Austria, ISBN 3-9501031-1-2.}, \bibinfo{year}{2001}).

\bibitem[{\citenamefont{Amadon et~al.}(2008)\citenamefont{Amadon, Lechermann,
  Georges, Jollet, Wehling, and Lichtenstein}}]{Amadon2008}
\bibinfo{author}{\bibfnamefont{B.}~\bibnamefont{Amadon}},
  \bibinfo{author}{\bibfnamefont{F.}~\bibnamefont{Lechermann}},
  \bibinfo{author}{\bibfnamefont{A.}~\bibnamefont{Georges}},
  \bibinfo{author}{\bibfnamefont{F.}~\bibnamefont{Jollet}},
  \bibinfo{author}{\bibfnamefont{T.~O.} \bibnamefont{Wehling}},
  \bibnamefont{and} \bibinfo{author}{\bibfnamefont{A.~I.}
  \bibnamefont{Lichtenstein}}, \bibinfo{journal}{Phys. Rev. B}
  \textbf{\bibinfo{volume}{77}}, \bibinfo{pages}{205112}
  (\bibinfo{year}{2008}).

\bibitem[{\citenamefont{Aichhorn et~al.}(2009)\citenamefont{Aichhorn,
  Pourovskii, Vildosola, Ferrero, Parcollet, Miyake, Georges, and
  Biermann}}]{Aichhorn2009}
\bibinfo{author}{\bibfnamefont{M.}~\bibnamefont{Aichhorn}},
  \bibinfo{author}{\bibfnamefont{L.}~\bibnamefont{Pourovskii}},
  \bibinfo{author}{\bibfnamefont{V.}~\bibnamefont{Vildosola}},
  \bibinfo{author}{\bibfnamefont{M.}~\bibnamefont{Ferrero}},
  \bibinfo{author}{\bibfnamefont{O.}~\bibnamefont{Parcollet}},
  \bibinfo{author}{\bibfnamefont{T.}~\bibnamefont{Miyake}},
  \bibinfo{author}{\bibfnamefont{A.}~\bibnamefont{Georges}}, \bibnamefont{and}
  \bibinfo{author}{\bibfnamefont{S.}~\bibnamefont{Biermann}},
  \bibinfo{journal}{Phys. Rev. B} \textbf{\bibinfo{volume}{80}},
  \bibinfo{pages}{085101} (\bibinfo{year}{2009}).

\bibitem[{\citenamefont{Lechermann et~al.}(2006)\citenamefont{Lechermann,
  Georges, Poteryaev, Biermann, Posternak, Yamasaki, and
  Andersen}}]{lechermann2006}
\bibinfo{author}{\bibfnamefont{F.}~\bibnamefont{Lechermann}},
  \bibinfo{author}{\bibfnamefont{A.}~\bibnamefont{Georges}},
  \bibinfo{author}{\bibfnamefont{A.}~\bibnamefont{Poteryaev}},
  \bibinfo{author}{\bibfnamefont{S.}~\bibnamefont{Biermann}},
  \bibinfo{author}{\bibfnamefont{M.}~\bibnamefont{Posternak}},
  \bibinfo{author}{\bibfnamefont{A.}~\bibnamefont{Yamasaki}}, \bibnamefont{and}
  \bibinfo{author}{\bibfnamefont{O.}~\bibnamefont{Andersen}},
  \bibinfo{journal}{Physical Review B} \textbf{\bibinfo{volume}{74}},
  \bibinfo{pages}{125120} (\bibinfo{year}{2006}).

\bibitem[{\citenamefont{{Pourovskii} et~al.}(2007)\citenamefont{{Pourovskii},
  {Amadon}, {Biermann}, and {Georges}}}]{Pourovskii2007}
\bibinfo{author}{\bibfnamefont{L.~V.} \bibnamefont{{Pourovskii}}},
  \bibinfo{author}{\bibfnamefont{B.}~\bibnamefont{{Amadon}}},
  \bibinfo{author}{\bibfnamefont{S.}~\bibnamefont{{Biermann}}},
  \bibnamefont{and}
  \bibinfo{author}{\bibfnamefont{A.}~\bibnamefont{{Georges}}},
  \bibinfo{journal}{Phys. Rev. B} \textbf{\bibinfo{volume}{76}},
  \bibinfo{pages}{235101} (\bibinfo{year}{2007}).

\bibitem[{\citenamefont{Aichhorn et~al.}(2011)\citenamefont{Aichhorn,
  Pourovskii, and Georges}}]{Aichhorn2011}
\bibinfo{author}{\bibfnamefont{M.}~\bibnamefont{Aichhorn}},
  \bibinfo{author}{\bibfnamefont{L.}~\bibnamefont{Pourovskii}},
  \bibnamefont{and} \bibinfo{author}{\bibfnamefont{A.}~\bibnamefont{Georges}},
  \bibinfo{journal}{Phys. Rev. B} \textbf{\bibinfo{volume}{84}},
  \bibinfo{pages}{054529} (\bibinfo{year}{2011}).

\bibitem[{\citenamefont{Pourovskii}()}]{pourovskii_unpub}
\bibinfo{author}{\bibfnamefont{L.}~\bibnamefont{Pourovskii}},
  \bibinfo{note}{private communication}.

\bibitem[{\citenamefont{Czy{\.z}yk and Sawatzky}(1994)}]{czyzyk1994}
\bibinfo{author}{\bibfnamefont{M.}~\bibnamefont{Czy{\.z}yk}} \bibnamefont{and}
  \bibinfo{author}{\bibfnamefont{G.}~\bibnamefont{Sawatzky}},
  \bibinfo{journal}{Physical Review B} \textbf{\bibinfo{volume}{49}},
  \bibinfo{pages}{14211} (\bibinfo{year}{1994}).

\bibitem[{\citenamefont{Harashima
  et~al.}(2015{\natexlab{b}})\citenamefont{Harashima, Terakura, Kino,
  Ishibashi, and Miyake}}]{harashima2015_proc}
\bibinfo{author}{\bibfnamefont{Y.}~\bibnamefont{Harashima}},
  \bibinfo{author}{\bibfnamefont{K.}~\bibnamefont{Terakura}},
  \bibinfo{author}{\bibfnamefont{H.}~\bibnamefont{Kino}},
  \bibinfo{author}{\bibfnamefont{S.}~\bibnamefont{Ishibashi}},
  \bibnamefont{and} \bibinfo{author}{\bibfnamefont{T.}~\bibnamefont{Miyake}},
  \bibinfo{journal}{Proceedings of Computational Science Workshop 2014
  (CSW2014)} p. \bibinfo{pages}{011021} (\bibinfo{year}{2015}{\natexlab{b}}).

\bibitem[{\citenamefont{Nilsson et~al.}(2013)\citenamefont{Nilsson, Sakuma, and
  Aryasetiawan}}]{Nilsson2013}
\bibinfo{author}{\bibfnamefont{F.}~\bibnamefont{Nilsson}},
  \bibinfo{author}{\bibfnamefont{R.}~\bibnamefont{Sakuma}}, \bibnamefont{and}
  \bibinfo{author}{\bibfnamefont{F.}~\bibnamefont{Aryasetiawan}},
  \bibinfo{journal}{Phys. Rev. B} \textbf{\bibinfo{volume}{88}},
  \bibinfo{pages}{125123} (\bibinfo{year}{2013}).

\bibitem[{\citenamefont{Lang et~al.}(1981)\citenamefont{Lang, Baer, and
  Cox}}]{lang1981study}
\bibinfo{author}{\bibfnamefont{J.}~\bibnamefont{Lang}},
  \bibinfo{author}{\bibfnamefont{Y.}~\bibnamefont{Baer}}, \bibnamefont{and}
  \bibinfo{author}{\bibfnamefont{P.}~\bibnamefont{Cox}},
  \bibinfo{journal}{Journal of Physics F: Metal Physics}
  \textbf{\bibinfo{volume}{11}}, \bibinfo{pages}{121} (\bibinfo{year}{1981}).

\bibitem[{\citenamefont{Bader}(1990)}]{bader1990atoms}
\bibinfo{author}{\bibfnamefont{R.~F.~W.} \bibnamefont{Bader}},
  \emph{\bibinfo{title}{Atoms in Molecules: a quantum theory}}
  (\bibinfo{publisher}{Clarendon: Oxford University Press},
  \bibinfo{year}{1990}).

\bibitem[{\citenamefont{Aryasetiawan et~al.}(2004)\citenamefont{Aryasetiawan,
  Imada, Georges, Kotliar, Biermann, and Lichtenstein}}]{aryasetiawan2004}
\bibinfo{author}{\bibfnamefont{F.}~\bibnamefont{Aryasetiawan}},
  \bibinfo{author}{\bibfnamefont{M.}~\bibnamefont{Imada}},
  \bibinfo{author}{\bibfnamefont{A.}~\bibnamefont{Georges}},
  \bibinfo{author}{\bibfnamefont{G.}~\bibnamefont{Kotliar}},
  \bibinfo{author}{\bibfnamefont{S.}~\bibnamefont{Biermann}}, \bibnamefont{and}
  \bibinfo{author}{\bibfnamefont{A.~I.} \bibnamefont{Lichtenstein}},
  \bibinfo{journal}{Phys. Rev. B} \textbf{\bibinfo{volume}{70}},
  \bibinfo{pages}{195104} (\bibinfo{year}{2004}).

\bibitem[{\citenamefont{Mazet et~al.}(2013)\citenamefont{Mazet, Malterre,
  Fran\ifmmode~\mbox{\c{c}}\else \c{c}\fi{}ois, Dallera, Grioni, and
  Monaco}}]{Mazet2013}
\bibinfo{author}{\bibfnamefont{T.}~\bibnamefont{Mazet}},
  \bibinfo{author}{\bibfnamefont{D.}~\bibnamefont{Malterre}},
  \bibinfo{author}{\bibfnamefont{M.}~\bibnamefont{Fran\ifmmode~\mbox{\c{c}}\else
  \c{c}\fi{}ois}}, \bibinfo{author}{\bibfnamefont{C.}~\bibnamefont{Dallera}},
  \bibinfo{author}{\bibfnamefont{M.}~\bibnamefont{Grioni}}, \bibnamefont{and}
  \bibinfo{author}{\bibfnamefont{G.}~\bibnamefont{Monaco}},
  \bibinfo{journal}{Phys. Rev. Lett.} \textbf{\bibinfo{volume}{111}},
  \bibinfo{pages}{096402} (\bibinfo{year}{2013}).

\bibitem[{\citenamefont{Mazet et~al.}(2015)\citenamefont{Mazet, Malterre,
  Fran\ifmmode~\mbox{\c{c}}\else \c{c}\fi{}ois, Eichenberger, Grioni, Dallera,
  and Monaco}}]{Mazet2015}
\bibinfo{author}{\bibfnamefont{T.}~\bibnamefont{Mazet}},
  \bibinfo{author}{\bibfnamefont{D.}~\bibnamefont{Malterre}},
  \bibinfo{author}{\bibfnamefont{M.}~\bibnamefont{Fran\ifmmode~\mbox{\c{c}}\else
  \c{c}\fi{}ois}},
  \bibinfo{author}{\bibfnamefont{L.}~\bibnamefont{Eichenberger}},
  \bibinfo{author}{\bibfnamefont{M.}~\bibnamefont{Grioni}},
  \bibinfo{author}{\bibfnamefont{C.}~\bibnamefont{Dallera}}, \bibnamefont{and}
  \bibinfo{author}{\bibfnamefont{G.}~\bibnamefont{Monaco}},
  \bibinfo{journal}{Phys. Rev. B} \textbf{\bibinfo{volume}{92}},
  \bibinfo{pages}{075105} (\bibinfo{year}{2015}).

\bibitem[{\citenamefont{Cuthill et~al.}(1974)\citenamefont{Cuthill, McAlister,
  Erickson, and Watson}}]{cuthill1974x}
\bibinfo{author}{\bibfnamefont{J.}~\bibnamefont{Cuthill}},
  \bibinfo{author}{\bibfnamefont{A.}~\bibnamefont{McAlister}},
  \bibinfo{author}{\bibfnamefont{N.}~\bibnamefont{Erickson}}, \bibnamefont{and}
  \bibinfo{author}{\bibfnamefont{R.}~\bibnamefont{Watson}},
  \bibinfo{journal}{AIP Conference Proceedings} \textbf{\bibinfo{volume}{18}},
  \bibinfo{pages}{1039} (\bibinfo{year}{1974}).

\bibitem[{\citenamefont{Laslo et~al.}(2014)\citenamefont{Laslo, Dudric,
  Neumann, Isnard, Coldea, and Pop}}]{laslo2014effects}
\bibinfo{author}{\bibfnamefont{A.}~\bibnamefont{Laslo}},
  \bibinfo{author}{\bibfnamefont{R.}~\bibnamefont{Dudric}},
  \bibinfo{author}{\bibfnamefont{M.}~\bibnamefont{Neumann}},
  \bibinfo{author}{\bibfnamefont{O.}~\bibnamefont{Isnard}},
  \bibinfo{author}{\bibfnamefont{M.}~\bibnamefont{Coldea}}, \bibnamefont{and}
  \bibinfo{author}{\bibfnamefont{V.}~\bibnamefont{Pop}},
  \bibinfo{journal}{Solid State Communications} \textbf{\bibinfo{volume}{199}},
  \bibinfo{pages}{43} (\bibinfo{year}{2014}).

\bibitem[{\citenamefont{Richter et~al.}(1995)\citenamefont{Richter, Steinbeck,
  Nitzsche, Oppeneer, and Eschrig}}]{richter1995}
\bibinfo{author}{\bibfnamefont{M.}~\bibnamefont{Richter}},
  \bibinfo{author}{\bibfnamefont{L.}~\bibnamefont{Steinbeck}},
  \bibinfo{author}{\bibfnamefont{U.}~\bibnamefont{Nitzsche}},
  \bibinfo{author}{\bibfnamefont{P.~M.} \bibnamefont{Oppeneer}},
  \bibnamefont{and} \bibinfo{author}{\bibfnamefont{H.}~\bibnamefont{Eschrig}},
  \bibinfo{journal}{Journal of Alloys and Compounds}
  \textbf{\bibinfo{volume}{225}}, \bibinfo{pages}{469} (\bibinfo{year}{1995}).

\bibitem[{\citenamefont{Hummler and F\"ahnle}(1996)}]{Hummler1996}
\bibinfo{author}{\bibfnamefont{K.}~\bibnamefont{Hummler}} \bibnamefont{and}
  \bibinfo{author}{\bibfnamefont{M.}~\bibnamefont{F\"ahnle}},
  \bibinfo{journal}{Phys. Rev. B} \textbf{\bibinfo{volume}{53}},
  \bibinfo{pages}{3272} (\bibinfo{year}{1996}).

\bibitem[{\citenamefont{Novak and Kuriplach}(1994)}]{novak1994}
\bibinfo{author}{\bibfnamefont{P.}~\bibnamefont{Novak}} \bibnamefont{and}
  \bibinfo{author}{\bibfnamefont{J.}~\bibnamefont{Kuriplach}},
  \bibinfo{journal}{IEEE Transactions on Magnetics}
  \textbf{\bibinfo{volume}{30}}, \bibinfo{pages}{1036} (\bibinfo{year}{1994}),
  ISSN \bibinfo{issn}{0018-9464}.

\bibitem[{\citenamefont{Laforest}(1981)}]{Laforest1981}
\bibinfo{author}{\bibfnamefont{J.}~\bibnamefont{Laforest}}, Ph.D. thesis,
  \bibinfo{school}{USM-INP}, \bibinfo{address}{Grenoble}
  (\bibinfo{year}{1981}).

\bibitem[{\citenamefont{Dederichs et~al.}(1984)\citenamefont{Dederichs,
  Bl{\"u}gel, Zeller, and Akai}}]{dederichs1984ground}
\bibinfo{author}{\bibfnamefont{P.}~\bibnamefont{Dederichs}},
  \bibinfo{author}{\bibfnamefont{S.}~\bibnamefont{Bl{\"u}gel}},
  \bibinfo{author}{\bibfnamefont{R.}~\bibnamefont{Zeller}}, \bibnamefont{and}
  \bibinfo{author}{\bibfnamefont{H.}~\bibnamefont{Akai}},
  \bibinfo{journal}{Physical review letters} \textbf{\bibinfo{volume}{53}},
  \bibinfo{pages}{2512} (\bibinfo{year}{1984}).

\end{thebibliography}

\end{document}